\documentclass[english,aps,prb,twocolumn,superscriptaddress]{revtex4-1}
\pdfoutput=1
\usepackage{color}
\usepackage{babel}
\usepackage{textcomp}
\usepackage{bm}
\usepackage{amsmath}
\usepackage{amssymb}
\usepackage{stackrel}
\usepackage{esint}
\usepackage{graphicx}
\usepackage{hyperref}
\hypersetup{colorlinks=true,linkcolor=blue,citecolor=blue,urlcolor=blue}

\newcommand{\ph}[1]{\hphantom{#1}}
\newcommand{\pd}{\partial}
\newcommand{\Tr}{\textrm{Tr}}
\newcommand{\Id}{\textrm{Id}}
\newcommand{\re}{\textrm{Re}}
\newcommand{\im}{\textrm{Im}}
\newcommand{\sgn}{\textrm{sgn}}

\newcommand{\Ga}{\Gamma}

\newcommand{\rra}{\rightarrow}

\makeatletter

\@ifundefined{textcolor}{}
{%
 \definecolor{BLACK}{gray}{0}
 \definecolor{WHITE}{gray}{1}
 \definecolor{RED}{rgb}{1,0,0}
 \definecolor{GREEN}{rgb}{0,1,0}
 \definecolor{BLUE}{rgb}{0,0,1}
 \definecolor{CYAN}{cmyk}{1,0,0,0}
 \definecolor{MAGENTA}{cmyk}{0,1,0,0}
 \definecolor{YELLOW}{cmyk}{0,0,1,0}
}

\makeatother

\begin{document}

\title{Multiple negative differential conductance regions and inelastic
phonon assisted tunneling in graphene-hBN-graphene structures}

\author{B. Amorim}
\email[Electronic address: ]{amorim.bac@gmail.com}
\affiliation{International Iberian Nanotechnology Laboratory (INL), 4715-330 Braga,
Portugal}
\affiliation{Center of Physics and Department of Physics, Universidade do Minho,
4710-057 Braga, Portugal}

\author{R. M. Ribeiro}
\affiliation{International Iberian Nanotechnology Laboratory (INL), 4715-330 Braga,
Portugal}

\author{N. M. R. Peres}
\affiliation{International Iberian Nanotechnology Laboratory (INL), 4715-330 Braga,
Portugal}

\begin{abstract}
In this paper we study in detail the effect of the rotational alignment
between a hexagonal boron nitride (hBN) slab and the graphene layers
in the vertical current of a a graphene-hBN-graphene device. We show
how for small rotational angles, the transference of momentum by the
hBN crystal lattice leads to multiple peaks in the I-V curve of the
device, giving origin to multiple regions displaying negative differential
conductance. We also study the effect of scattering by phonons in
the vertical current an see how the opening up of inelastic tunneling
events allowed by spontaneous emission of optical phonons leads to
sharp peaks in the second derivative of the current. 

\end{abstract}

\maketitle

\tableofcontents

\section{Introduction}

Being able to tailor the properties of materials at will, aiming at
unveiling new physics and achieving never though before properties,
is the main goal of condensed matter physics and materials science.
However, the degree of manipulation we can undertake using conventional
materials is somewhat limited. In the last ten years, the advent of
two-dimensional materials \cite{NGM2004,NJS2005} opened new avenues
waiting for being explored. One of the less explored avenue is the
one opened by van der Waals (vdW) hybrid structures\cite{PGZ11},
new systems formed by stacking layers of two-dimensional crystals
on top of each other, have emerged as a new approach for manipulating
and tailoring material properties at will\cite{NC_2012,GG_2013}.
Among the various possible combinations of two dimensional crystals,
graphene - semiconductor/insulator - graphene vdW structures, with
semiconducting transition metal dichalcogenide (STMDC) or hexagonal
boron nitride (hBN) as the semiconductor/insulator, have emerged as
some of the most promising from the point of view of applications.
The possibility of controlling electrostatically the effective barrier
height presented by the insulator/semiconductor to the vertical flow
of electrons between the two graphene layers with a gate voltage has
enabled the operation these devices as transistors\cite{BGJ12,Britnell2012b,GJB13},
with ON/OFF ratios as high as $10^{6}$ being possible in graphene-WS$_{2}$-graphene
devices\cite{GJB13}. It was also shown that graphene-STMDC-graphene
devices can operate as photodectectors with high quantum efficiencies
and fast response times \cite{BRE13,YLZ13,MSV15}. Due to the extreme
high quality and atomically sharp interfaces \cite{Haigh2012} between
different layers in vdW structures , lattice mismatch and relative
alignment between consecutive layers play a fundamental role in determining
the electronic coupling between different layers of the vdW structure,
ultimately determining its electronic and optical properties. Lattice
misalignment between different layers has been known to lead to the
formation of Moiré patterns in rotated graphite layers \cite{Pong2005}.
The effect of lattice misalignment and mismatch has been extensively
studied in the context of twisted graphene bilayers and graphene-on-hBN
structures. It was shown theoretically and experimentally, that misalignment
in a graphene bilayer leads to a renormalization of graphene's Fermi
velocity \cite{dosSantos2007,Luican2011}. It was also found out that
mismatch and misalignment controls the formation of mini Dirac cones
in the band structure of graphene - hBN structures systems \cite{Park_2008,Yankowitz_2012,Ortix_2012,Wallbank2013,Wallbank2013_b,Ponomarenko_2013}..
\begin{figure}[t]
\begin{centering}
\includegraphics[width=8cm]{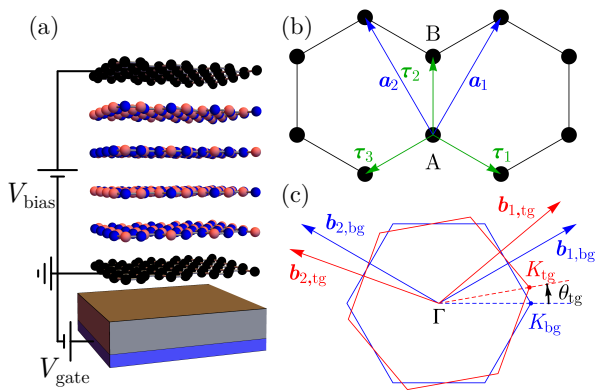}
\par\end{centering}

\protect\caption{\label{fig:Schematic}(a) Schematic of a typical graphene-hBN-graphene
vdW structure with four boron nitride layers, with applied gate, $V_{\text{gate}}$,
and bias, $V_{\text{bias}}$, voltages. (b) Representation of crystalline
structure shared by a graphene/boron nitride monolayer, showing the
lattice basis, $\left\{ \bm{a}_{1},\,\bm{a}_{2}\right\} $, the nearest
neighbour vectors $\bm{\tau}_{i}$, $i=1,2,3$, and the sublattice
A/B sites. (c) Representation of the $1^{\text{st}}$ Brillouin zone
of the rotated bottom and top graphene layers, showing the $K$ points
of both layers and the reciprocal lattice basis vectors $\left\{ \bm{b}_{1,\text{bg,tg}},\,\bm{b}_{2,\text{bg/tg}}\right\} $.}
\end{figure}
The dependence of the vertical current in vdW structures on the rotation
between different layers was first studied in Ref.~\onlinecite{Bistritzer2010}
in the context of twisted bilayer graphene, where it was found that
the current is extremely sensitive to the twist angle. Although this
dependence was not at first completely appreciated, it was soon understood
and verified \cite{MTC14,B__14} that the misalignment between the
graphene layers in graphene-hBN-graphene structures can lead to the
occurrence of negative differential conductance (NDC) regions, with
the I-V curve displaying peaks whose dependence on the bias voltage
depends on the rotation angle between the graphene layers. More recently,
the effect of misalignment on the vertical current in devices formed
by two graphene bilayers \cite{Fallahazad2015,Kang2015,Barrera2015}
and by one graphene monolayer and a graphene bilayer separated by
hBN has also been studied.\cite{Lane2015} Scattering by phonons can
lead to incoherent phonon assisted tunneling between the graphene
layers. This effect has been first theoretically studied for vdW structures
for twisted graphene bilayers \cite{Perebeinos2012}. More recently,
effects of phonon assisted scattering on vertical transport have been
experimentally detected in graphene-hBN-graphite \cite{Jung2015}
and graphene-hBN-graphene structures \cite{Vdovin2015} and have been
proposed as a possible way to probe the phonon spectrum of vdW structures.

In this paper we describe the vertical current in graphene-hBN-graphene
devices with misaligned layers, and for small twist angles, properly
taking into account momentum conservation rules, within the non-equilibrium
Green's function framework and using a tight-binding based continuous
Hamiltonian. We show that the present theory reduces to the ones used
in Refs.~\onlinecite{B__14,MTC14}. By taking into account processes
involving transference of momentum by the hBN crystal lattice to the
tunneling electrons, we find that the vertical current depends sensitively
on the relative alignment between the graphene layers and the hBN
slab and that by carefully controlling this alignment, it is possible
to obtain several peaks in the I-V curve of the device, followed by
regions of NDC, a possibility that has not been considered previously.
We also find out that the structure of graphene wavefunctions manifests
itself in the vertical current, suppressing some of the current peaks
that would be expected with considerations based only on electronic
dispersion relations. We study the effect of resonant disorder in
the graphene layers in the vertical current, treated within the self-consistent
Born approximation (SCBA) which correctly describes the proportionality
of the transport lifetime with the energy\cite{Peres_Rev2010}. We
finally study how phonons and disorder give origin to non-coherent
current between the two graphene layers, deriving an expression for
it.

The paper is organized as follows. In Sec.~\ref{sec:Formalism} we
describe the theoretical framework we employed in this work: in subsection~\ref{sub:Model-Hamiltonian},
we present the Hamiltonian used to model the graphene-hBN-graphene
device and in subsection~\ref{sub:Current-evaluation} we present
the fundamental equations used to treat transport within the non-equilibrium
Green's functions formalism. In Sec.~\ref{sec:Coherent-tunnelling}
we discuss the coherent tunneling flowing through a pristine device
taking into account the lattice mismatch and misalignment between
graphene layers and the hBN slab. The consequences of treating graphene
as part of the device or as external contacts are discussed and the
role of the momentum transferred to the tunneling electrons by the
hBN lattice is analyzed in detail. The effect of an in-plane magnetic
field in the current is also discussed. In Sec.~\ref{sec:Incoherent-tunnelling},
the effects of disorder and phonon scattering into the vertical current
are studied and a expression for the phonon/disorder assisted current
to lowest order in perturbation theory is derived. Finally, in Sec.~\ref{sec:Conclusions}
we conclude. Technical details and longer derivations are include
as Appendices.

\section{Formalism\label{sec:Formalism}}

We want to study the vertical current flowing through a device formed
by two graphene layers (bottom, bg, and top, tg) separated by a few
layers, $\mathcal{N}$, of hBN. The distance between the two graphene
layers is given by $d$. We assume that the top graphene layer and
the hBN slab are rotated with respect to the bottom graphene layer
by an angle of $\theta_{\text{tg}}$ and $\theta_{\text{hBN}}$, respectively.
We assume that layers forming the hBN slab are perfectly aligned with
an $\text{AA}^{\prime}$ stacking\cite{Geick_1966,Ribeiro_2011} (consecutive
honeycomb lattices are perfectly aligned, with each boron/nitrogen
atom of one layer directly on top of the nitrogen/atom of the next
layer). A bias voltage, $V_{\text{bias}}$, can be applied between
the top and bottom graphene layers, which will induce a current between
the two. The doping of the graphene layers can be controlled by application
of a gate voltage to the bottom graphene. A schematic of the typical
device structure is shown in Fig.~\ref{fig:Schematic}.

\subsection{Model Hamiltonian\label{sub:Model-Hamiltonian}}

We model the graphene - hBN - graphene system with the following Hamiltonian
\begin{align}
H= & H_{\text{bg}}+H_{\text{tg}}+H_{\text{hBN}}+\nonumber \\
+ & \left(T_{\text{hBN},\text{bg}}+T_{\text{hBN},\text{tg}}+\text{h.c.}\right),\label{eq:Hamiltonian_total}
\end{align}
where $H_{\text{bg}/\text{tg}}$ is the Hamiltonian describing the
isolated bottom/top graphene layer and $T_{\text{hBN},\text{bg}/\text{tg}}=T_{\text{bg}/\text{tg},\text{hBN}}^{\dagger}$
describes the hopping of electrons from the bottom/top graphene layer
to the hBN slab. The current between the two graphene layers will
be dominated by low energy states. Focusing on the states close to
the $\bm{K}_{\text{bg}}$ and $\bm{K}_{\text{bg}}^{\prime}=-\bm{K}_{\text{bg}}$
points of the bottom graphene layer, we write the Hamiltonian of the
bottom graphene layer in sublattice basis and in term of Bloch states
as the massless Dirac Hamiltonian 
\begin{multline}
H_{\text{bg}}=\sum_{\bm{k},\tau}\bm{c}_{\bm{k}_{\tau},\text{bg}}^{\dagger}\cdot\\
\left[\begin{array}{cc}
V_{\text{bg}} & \tau v_{F}\hbar\left|\bm{k}\right|e^{-\tau i\theta_{\bm{k},\text{bg}}}\\
\tau v_{F}\hbar\left|\bm{k}\right|e^{-\tau i\theta_{\bm{k},\text{bg}}}\cdot & V_{\text{bg}}
\end{array}\right]\cdot\bm{c}_{\bm{k}_{\tau},\text{bg}},\label{eq:Hamiltonian_BG}
\end{multline}
where $v_{F}$ is graphene's Fermi velocity, $V_{\text{bg}}$ is a
on-site potential induced by the applied bias and gate voltages, $\bm{c}_{\bm{k}_{\tau},\text{bg}}^{\dagger}=\left[\begin{array}{cc}
c_{\bm{k}_{\tau},A,\text{bg}}^{\dagger} & c_{\bm{k}_{\tau},B,\text{bg}}^{\dagger}\end{array}\right]$ is the electron creation operator for states localized in the A/B
sublattice, in the $\tau\bm{K}_{\text{bg}}$ valley ($\tau=\pm1$),
with momentum $\tau\bm{K}_{\text{bg}}+\bm{k}$ (measured from the
Brillouin zone center) and $\theta_{\bm{k},\text{bg}}$ is the angle
formed between $\bm{k}$ and $\bm{K}_{\text{bg}}$. We choose the
zero of energy to lie at the Fermi level of the bottom graphene layer,
in which case we have $V_{\text{bg}}=-\epsilon_{\text{F,bg}}$, where
$\epsilon_{\text{F,bg}}$ is the Fermi energy of the bottom graphene
layer measured from its Dirac point. The Hamiltonian in Eq.~(\ref{eq:Hamiltonian_BG})
is diagonalized by the eigenstates $\left|\bm{k},\tau,\lambda\right\rangle _{\text{bg}}=\left[1,\,\lambda\tau e^{-i\tau\theta_{\bm{k},\text{bg}}}\right]^{\dagger}/\sqrt{2}$
with corresponding dispersion relation $\epsilon_{\bm{k},\lambda}=\lambda v_{F}\hbar\left|\bm{k}\right|$,
with $\lambda=\pm1$ for electrons in the conduction/valence band.
Since we will be interested in studying the vertical current to lowest
order in the graphene-hBN coupling, we neglect the effect of the periodic
potential generated by the hBN slab in dispersion relation of graphene
electronic states \cite{Park_2008,Yankowitz_2012,Kidermann2012,Ortix_2012,Wallbank2013,Wallbank2013_b,Jung2014}.
Using the same reference frame in momentum space as in Eq.~(\ref{eq:Hamiltonian_BG}),
the Hamiltonian describing the top graphene layer in the Dirac cone
approximation reads
\begin{multline}
H_{\text{tg}}=\sum_{\bm{k},\tau}\bm{c}_{\bm{k}_{\tau},\text{tg}}^{\dagger}\cdot\\
\left[\begin{array}{cc}
V_{\text{tg}} & \tau v_{F}\hbar\left|\bm{k}^{\prime}\right|e^{-i\tau\theta_{\bm{k}^{\prime},\text{tg}}}\\
\tau v_{F}\hbar\left|\bm{k}^{\prime}\right|e^{i\tau\theta_{\bm{k}^{\prime},\text{tg}}} & V_{\text{tg}}
\end{array}\right]\cdot\bm{c}_{\bm{k}_{\tau},\text{tg}},\label{eq:Hamiltonian_TG}
\end{multline}
where $\bm{k}^{\prime}=\bm{k}+\tau\Delta\bm{K}_{\text{b,t}}$, is
measured from the $\tau\bm{K}_{\text{tg}}$ point of the top graphene
layer, with $\bm{K}_{\text{tg}}=\mathcal{R}(\theta_{\text{tg}})\cdot\bm{K}_{\text{bg}}$
($\mathcal{R}(\theta)$ a rotation matrix), $\Delta\bm{K}_{\text{b,t}}=\bm{K}_{\text{bg}}-\bm{K}_{\text{tg}}$
is the displacement between the Dirac points of the two rotated graphene
layers and $\bm{k}$ is measured with respect to the $\tau\bm{K}_{\text{bg}}$
Dirac point of the bottom graphene graphene layer. $\theta_{\bm{k}^{\prime},\text{tg}}$
is the angle between $\bm{k}^{\prime}$ and $\bm{K}_{\text{tg}}$
and $V_{\text{tg}}$ is an on-site potential, due to the applied bias
and gate voltages, and is given by $V_{\text{tg}}=-\epsilon_{\text{F},\text{tg}}-eV_{\text{bias}}$,
with $\epsilon_{\text{F},\text{tg}}$ the Fermi level of the top graphene
layer measured from its Dirac point and $e>0$ the fundamental electronic
charge. The remaining symbols in Eq.~(\ref{eq:Hamiltonian_TG}) are
similarly defined to the ones in Eq.~(\ref{eq:Hamiltonian_BG}).
Due to the large band gap of boron nitride, we ignore its momentum
dependence, writing the hBN slab Hamiltonian as
\begin{multline}
H_{\text{hBN}}=\\
\sum_{\ell=1}^{\mathcal{N}}\sum_{\bm{k},\tau}\bm{c}_{\bm{k}_{\tau},\ell,\text{hBN}}^{\dagger}\cdot\left[\begin{array}{cc}
E_{\text{B}}+V_{\ell} & 0\\
0 & E_{\text{N}}+V_{\ell}
\end{array}\right]\cdot\bm{c}_{\bm{k}_{\tau},\ell,\text{hBN}}\\
+\sum_{\ell=1}^{\mathcal{N}-1}\sum_{\bm{k},\tau}\bm{c}_{\bm{k}_{\tau},\ell+1,\text{hBN}}^{\dagger}\cdot\left[\begin{array}{cc}
0 & -t_{\perp}\\
-t_{\perp} & 0
\end{array}\right]\cdot\bm{c}_{\bm{k}_{\tau},\ell,\text{hBN}}+\text{h.c},\label{eq:H_hBN_hamiltonian}
\end{multline}
where $\bm{c}_{\bm{k}_{\tau},\ell,\text{hBN}}^{\dagger}=\left[\begin{array}{cc}
c_{\bm{k}_{\tau},\text{B}\ell,\text{hBN}}^{\dagger} & c_{\bm{k}_{\tau},\text{N}\ell,\text{hBN}}^{\dagger}\end{array}\right]$ creates an electron in layer $\ell=1,...,\mathcal{N}$ of the hBN
slab, in the boron ($\text{B}$)/nitrogen ($\text{N}$) site, $\tau$
specifies the valley, $E_{\text{B}}$ and $E_{\text{N}}$ are, respectively,
the on-site energies of boron and nitride sites measured from the
Dirac point of graphene, $t_{\perp}$ is the nearest neighbour interlayer
hoping and $V_{\ell}$ is a potential induced by the applied voltages.
Due to the large energy offset between graphene and hBN sites, the
charge accumulated in the hBN layers will be negligible. In this case
a simple electrostatic calculation (see Appendix~\ref{sec:Capacitor})
gives us $V_{\ell}=-\epsilon_{\text{F,bg}}-\left(\epsilon_{\text{F},\text{tg}}+eV_{\text{bias}}\right)\ell/\left(\mathcal{N}+1\right)$.
For two rotated crystal layers, Bloch states from different layers
can only be coupled provided momentum is conserved modulo any combination
of reciprocal lattice vectors of both layers \cite{Bistritzer2010,Koshino2015},
in a so called generalized \textit{Umklapp} process. Focusing on low
energy states and considering only the three most relevant processes,
the coupling between the graphene layers and the hBN slab is described
by (see Appendix~\ref{sec:Interlayer-hopping-twisted})
\begin{equation}
T_{\text{hBN},\text{X}}=\sum_{\bm{k},\tau}\sum_{n=0}^{2}\bm{c}_{\bm{k}_{\tau}+\tau\bm{g}_{n}^{\text{X},\text{hBN}},\ell_{\text{X}},\text{hBN}}^{\dagger}\cdot\bm{R}_{\frac{2\pi}{3}}^{n}\cdot\hat{\bm{T}}\cdot\bm{R}_{-\frac{2\pi}{3}}^{n}\cdot\bm{c}_{\bm{k}_{\tau},\text{X}},\label{eq:Hamiltonian_graphene_hBN}
\end{equation}
where $\bm{c}_{\bm{k}_{\tau},\text{X}}$ is an annihilation operator
of an electron in the $\text{X}=\text{bg}/\text{tg}$ graphene layer
with momentum $\bm{k}_{\tau}$ measured from the $\tau\bm{K}_{\text{X}}$
point, $\bm{c}_{\bm{k}_{\tau}+\tau\bm{g}_{n}^{\text{X}},\ell_{\text{X}},\text{hBN}}^{\dagger}$
is a creation operator of an electron state in the $\ell_{\text{X}}=1/\mathcal{N}$
layer of the hBN slab, with momentum $\bm{k}_{\tau}+\tau\bm{g}_{n}^{\text{X}}$
measured from $\tau\bm{K}_{\text{X}}$, and with the matrices $\hat{\bm{T}}$
and $\bm{R}_{\theta}$ defined as 
\begin{eqnarray}
\bm{R}_{\theta} & = & \left[\begin{array}{cc}
1 & 0\\
0 & e^{i\theta}
\end{array}\right],\\
\hat{\bm{T}} & = & \left[\begin{array}{cc}
t_{\text{B,C}} & t_{\text{B,C}}\\
t_{\text{N,C}} & t_{\text{N,C}}
\end{array}\right],
\end{eqnarray}
where $t_{\text{B,C}}$ ($t_{\text{N,C}}$) is the hoping parameter
between a carbon site and boron (nitrogen) site and the vectors $\bm{g}_{n}^{\text{X},\text{hBN}}$
are given by (see Appendix~\ref{sec:Interlayer-hopping-twisted})
\begin{eqnarray}
\bm{g}_{0}^{\text{X},\text{hBN}} & = & 0,\\
\bm{g}_{1}^{\text{X},\text{hBN}} & = & \bm{b}_{2,\text{X}}-\bm{b}_{2,\text{hBN}},\\
\bm{g}_{2}^{\text{X},\text{hBN}} & = & -\bm{b}_{1,\text{X}}+\bm{b}_{1,\text{hBN}},
\end{eqnarray}
where $\bm{b}_{i,\text{X}}$ and $\bm{b}_{i,\text{hBN}}$ ($i=1,2$)
are, respectively, the reciprocal lattice vectors of the bottom/top
graphene layer and of the hBN slab (see Fig.~\ref{fig:Schematic}).
Notice that if the boron nitride slab is formed by an even number
of layers, one must replace $\bm{R}_{\frac{2\pi}{3}}^{n}\rra\sigma_{x}\cdot\bm{R}_{\frac{2\pi}{3}}^{n}\cdot\sigma_{x}$
for $T_{\text{hBN},\text{tg}}$, since boron and nitrogen atoms switch
positions in consecutive layers of hBN. Different reciprocal lattice
vectors are related to each other by $\bm{b}_{i,\text{tg}}=\mathcal{R}(\theta_{\text{tg}})\cdot\bm{b}_{i,\text{bg}}$
and $\bm{b}_{i,\text{hBN}}=\left(a_{\text{g}}/a_{\text{hBN}}\right)\mathcal{R}(\theta_{\text{hBN}})\cdot\bm{b}_{i,\text{bg}}$,
where $a_{\text{g}}/a_{\text{hBN}}$ is a scale factor, with $a_{\text{g}}$
($a_{\text{hBN}}$) the lattice parameter of graphene (hBN). Hamiltonians
of the form of Eq.~(\ref{eq:Hamiltonian_graphene_hBN}) have previously
been used to study twisted graphene bilayers \cite{dosSantos2007,Bistritzer2011,dosSantos2012,Moon2013}
and graphene-on-hBN structures \cite{Kidermann2012,Jung2014,Moon2014}.
Considering the three processes coupling the bottom graphene with
hBN and the three processes connecting hBN to the top graphene layer,
described by Eq.~(\ref{eq:Hamiltonian_graphene_hBN}), there are
nine hBN mediated processes coupling the bottom graphene layer to
the top one \cite{B__14}. These nine processes couple a state from
the bottom graphene layer with momentum $\bm{k}$ (measured from $\tau\bm{K}_{\text{bg}}$)
to states of the top graphene layer with momentum $\bm{k}+\tau\bm{\mathcal{Q}}_{n,m}$
(measured from $\tau\bm{K}_{\text{tg}}$) with (see Appendix~\ref{sec:Interlayer-hopping-twisted})
\begin{equation}
\bm{\mathcal{Q}}_{n,m}=\Delta\bm{K}_{\text{b,t}}+\bm{g}_{n}^{\text{bg},\text{hBN}}-\bm{g}_{m}^{\text{tg},\text{hBN}},\, n,m=0,1,2.\label{eq:Momentum_separation}
\end{equation}
The processes with $n\neq m$ involve transfer of momentum by the
hBN lattice, while processes with $n=m$ do not. At zero magnetic
field, the overall three-fold rotational invariance of the graphene-hBN-graphene
structure implies that these nine processes can be organized in three
groups of three, with processes in the same group being related by
$2\pi/3$ rotation and therefore giving the same contribution to the
vertical current. The three groups are 
\begin{align}
\left\{ \bm{\mathcal{Q}}_{0,0},\,\bm{\mathcal{Q}}_{1,1},\,\bm{\mathcal{Q}}_{2,2}\right\} ,\nonumber \\
\left\{ \bm{\mathcal{Q}}_{0,1},\,\bm{\mathcal{Q}}_{1,2},\,\bm{\mathcal{Q}}_{2,0}\right\} ,\label{eq:Momentum_groups}\\
\left\{ \bm{\mathcal{Q}}_{0,2},\,\bm{\mathcal{Q}}_{1,0},\,\bm{\mathcal{Q}}_{2,1}\right\} ,\nonumber 
\end{align}
with length of the vectors in each group being the same. For small
rotation angles and lattice mismatch, $\delta=a_{\text{hBN}}/a_{\text{g}}-1$,
we have %
\footnote{In Ref.~\protect \onlinecite{B__14}, the processes corresponding
to $\bm{\mathcal{Q}}_{0,1}$ and $\bm{\mathcal{Q}}_{0,2}$ where identified
as being equivalent, with $\left|\bm{\mathcal{Q}}_{0,1}\right|=\left|\bm{\mathcal{Q}}_{0,2}\right|$.
This was likely caused by first expanding $\bm{\mathcal{Q}}_{n,m}$
to linear order in $\theta_{\text{tg}}$, $\theta_{\text{hBN}}$ and
$\delta$, and only then evaluating $\left|\bm{\mathcal{Q}}_{n,m}\right|$,
loosing in the process terms involving the product $\delta\theta_{\text{tg}}$
in $\left|\bm{\mathcal{Q}}_{n,m}\right|^{2}$, which lift the equivalence
between the processes associated with $\bm{\mathcal{Q}}_{0,1}$ and
$\bm{\mathcal{Q}}_{0,2}$. %
}
\begin{eqnarray}
\frac{\left|\bm{\mathcal{Q}}_{0,0}\right|^{2}}{K_{\text{g}}^{2}} & \simeq & \theta_{\text{tg}}^{2},\\
\frac{\left|\bm{\mathcal{Q}}_{0,1}\right|^{2}}{K_{\text{g}}^{2}} & \simeq & \theta_{\text{tg}}^{2}+3\left(\theta_{\text{hBN}}^{2}+\delta^{2}-\theta_{\text{tg}}\theta_{\text{hBN}}\right)+\sqrt{3}\delta\theta_{\text{tg}},\\
\frac{\left|\bm{\mathcal{Q}}_{0,2}\right|^{2}}{K_{\text{g}}^{2}} & \simeq & \theta_{\text{tg}}^{2}+3\left(\theta_{\text{hBN}}^{2}+\delta^{2}-\theta_{\text{tg}}\theta_{\text{hBN}}\right)-\sqrt{3}\delta\theta_{\text{tg}},
\end{eqnarray}
with $K_{\text{g}}=4\pi/\left(3a_{\text{g}}\right)$ the length of
$\bm{K}_{\text{bg}/\text{tg}}$.

\subsection{Current evaluation\label{sub:Current-evaluation}}

The standard approach to transport in a mesoscopic device assumes
that the device is attached to external contacts that are in a thermal
equilibrium state with well defined chemical potentials. This is only
an approximation as once a current starts flowing through the system,
the contacts will also be in a non-equilibrium state \cite{Frensley1990,Vogl2010}.
The problem of computing the current flowing through a mesoscopic
device is then analogous to the problem of computing the water flux
through a thin pipe that is connecting two large vessels with different
water levels \cite{Stone1988}. Once water starts flowing through
the pipe, the water levels in each vessel are no longer constants,
however, on short time scales, assuming that the water levels in the
vessels are constant is a reasonable approximation, provided that
these are wide enough with respect to the pipe. In the same way, within
short time scales compared to the depletion time of an external battery,
it is a reasonable approximation to assume that the external contacts
have well defined, constant chemical potentials. Using the non-equilibrium
Green's function technique one can then show that in a mesoscopic
device that is attached to two non-interacting%
\footnote{In mesoscopic transport, the problem of computing the current that
is flowing through the device is reduced to a problem only involving degrees
of freedom in the mesoscopic region by integrating out the external
contacts. This is only done exactly provided the contacts are non-interacting.%
} contacts, bottom and top, in thermal equilibrium state described,
respectively, by the Fermi-Dirac distribution functions $f_{\text{b}}(\omega)=\left[e^{\beta\left(\omega-\mu_{\text{b}}\right)}+1\right]^{-1}$
and $f_{\text{t}}(\omega)=\left[e^{\beta\left(\omega-\mu_{\text{t}}\right)}+1\right]^{-1}$
with $\mu_{\text{b}(\text{t})}$ the chemical potential of the bottom
(top) contact, the current flowing from the bottom to the top contact
is given by \cite{HaugPekka_book} (using a compact notation where
a capital bold face symbols represent matrix elements evaluated in
some one-particle electron basis and omitting the frequency argument
of the different quantities)
\begin{eqnarray}
I_{\text{b}\rra\text{t}} & = & \frac{e}{\hbar}\int\frac{d\omega}{2\pi}f_{\text{b}}(\omega)\Tr\left[\bm{\Gamma}_{\text{b}}\cdot\bm{A}\right]\nonumber \\
 & + & \frac{e}{\hbar}\int\frac{d\omega}{2\pi}i\Tr\left[\bm{\Gamma}_{\text{b}}\cdot\bm{G}^{<}\right],\label{eq:Current_general}
\end{eqnarray}
with the spectral function of the central mesoscopic device given
by 
\begin{eqnarray}
\bm{A} & = & i\left(\bm{G}^{R}-\bm{G}^{A}\right)\nonumber \\
 & = & i\left(\bm{G}^{>}-\bm{G}^{<}\right),\label{eq:def_spectral_function}
\end{eqnarray}
where $\bm{G}^{R/A/</>}$ is the retarded/advanced/lesser/greater
Green's function of the central device (which takes into account coupling
to the external contacts) and $\bm{\Gamma}_{\text{b}(\text{t})}$
is a level width function due to the bottom (top) contact. The level
width function is the density of states of the contacts weighted by
the their coupling to the central device: $\bm{\Gamma}_{\text{b}(\text{t})}=2\pi\bm{\tau}_{\text{b}(\text{t})}\cdot\delta\left(\omega-\bm{H}_{\text{b}(\text{t})}\right)\cdot\bm{\tau}_{\text{b}(\text{t})}^{\dagger}$,
with $\bm{H}_{\text{b}(\text{t})}$ the Hamiltonian describing the
bottom (top) contact and $\bm{\tau}_{\text{b}(\text{t})}$ describing
the coupling between the central device and the contact. The second
equality in Eq.~(\ref{eq:def_spectral_function}) is true by the
very definition of the different Green's functions. A property that
will latter be useful is\cite{Datta_book} 
\begin{eqnarray}
\bm{A} & = & \bm{G}^{R}\cdot\bm{\Gamma}\cdot\bm{G}^{A}\nonumber \\
 & = & \bm{G}^{A}\cdot\bm{\Gamma}\cdot\bm{G}^{R},\label{eq:spectral_RA_AR}
\end{eqnarray}
where the decay rate matrix is defined as $\bm{\Gamma}=-i\left(\left[\bm{G}^{R}\right]^{-1}-\left[\bm{G}^{A}\right]^{-1}\right)$.
This result can be obtained by writing 
\begin{eqnarray}
\bm{G}^{R}-\bm{G}^{A} & = & \bm{G}^{R}\cdot\left(\left[\bm{G}^{A}\right]^{-1}-\left[\bm{G}^{R}\right]^{-1}\right)\cdot\bm{G}^{A}\nonumber \\
 & = & \bm{G}^{A}\cdot\left(\left[\bm{G}^{A}\right]^{-1}-\left[\bm{G}^{R}\right]^{-1}\right)\cdot\bm{G}^{R}.
\end{eqnarray}
Using the Dyson equation for the retarded/advanced Green's function,
$\left[\bm{G}^{R/A}\right]^{-1}=\left[\bm{G}^{0,R/A}\right]^{-1}-\bm{\Sigma}^{R/A}$,
with $\bm{G}^{0}$ indicating the bare Green's function (in the absence
of interactions and coupling to external contacts), and nothing that
$\left[\bm{G}^{0,R}\right]^{-1}$and $\left[\bm{G}^{0,A}\right]^{-1}$
only differ by an infinitesimal constant that is taken to zero, the
decay rate matrix can be written as
\begin{eqnarray}
\bm{\Gamma} & = & i\left[\bm{\Sigma}^{R}-\bm{\Sigma}^{A}\right]\nonumber \\
 & = & i\left[\bm{\Sigma}^{>}-\bm{\Sigma}^{<}\right],\label{eq:def_decay_rate}
\end{eqnarray}
where the last identity is inherited from the second equality in Eq.~(\ref{eq:def_spectral_function}).
The lesser/greater Green's functions obey the Keldysh equation \cite{HaugPekka_book}
\begin{equation}
\bm{G}^{\lessgtr}=\bm{G}^{R}\cdot\bm{\Sigma}^{\lessgtr}\cdot\bm{G}^{A},\label{eq:Keldysh}
\end{equation}
where the lesser/greater self-energy can be split into contributions
from the contacts and interactions as 
\begin{eqnarray}
\bm{\Sigma}^{<} & = & if_{\text{b}}\bm{\Gamma}_{\text{b}}+if_{\text{t}}\bm{\Gamma}_{\text{t}}+\bm{\Sigma}_{\text{int}}^{<},\label{eq:self_energy_split}\\
\bm{\Sigma}^{>} & = & -i\left(1-f_{\text{b}}\right)\bm{\Gamma}_{\text{b}}-i\left(1-f_{\text{t}}\right)\bm{\Gamma}_{\text{t}}+\bm{\Sigma}_{\text{int}}^{>},
\end{eqnarray}
with $\bm{\Sigma}_{\text{int}}^{\lessgtr}$ the contribution from
interactions. In the same manner, the decay rate Eq.~(\ref{eq:def_decay_rate})
can be split into a contribution from external contacts and interactions
\begin{equation}
\bm{\Gamma}=\bm{\Gamma}_{\text{b}}+\bm{\Gamma}_{\text{t}}+\bm{\Gamma}_{\text{int}}.\label{eq:decay_rate_split}
\end{equation}
Using Eqs.~(\ref{eq:spectral_RA_AR}) and (\ref{eq:Keldysh})-(\ref{eq:decay_rate_split})
in Eq.~(\ref{eq:Current_general}), the total current can then be
written as a sum of coherent and incoherent contributions
\begin{equation}
I_{\text{b}\rra\text{t}}=I_{\text{b}\rra\text{t}}^{\text{coh}}+I_{\text{b}\rra\text{t}}^{\text{incoh}},
\end{equation}
with the coherent contribution being given by the Landauer formula
\begin{eqnarray}
I_{\text{b}\rra\text{t}}^{\text{coh}} & = & \frac{e}{\hbar}\int\frac{d\omega}{2\pi}\left(f_{\text{b}}-f_{\text{t}}\right)\mathcal{T},\label{eq:Landauer_formula}
\end{eqnarray}
with the transmission function $\mathcal{T}$ given by
\begin{eqnarray}
\mathcal{T} & = & \Tr\left[\bm{\Gamma}_{\text{b}}\cdot\bm{G}^{R}\cdot\bm{\Gamma}_{\text{t}}\cdot\bm{G}^{A}\right]\label{eq:Transmission:_function}
\end{eqnarray}
and the incoherent contribution, which describes sequential tunneling
processes and plays the same role as vertex corrections in the Kubo
formalism for linear response, being given by 
\begin{multline}
I_{\text{b}\rra\text{t}}^{\text{incoh}}=\frac{e}{\hbar}\int\frac{d\omega}{2\pi}if_{\text{b}}\Tr\left[\bm{\Gamma}_{\text{b}}\cdot\bm{G}^{A}\cdot\bm{\Sigma}_{\text{int}}^{>}\cdot\bm{G}^{R}\right]\\
+\frac{e}{\hbar}\int\frac{d\omega}{2\pi}i\left(1-f_{\text{b}}\right)\Tr\left[\bm{\Gamma}_{\text{b}}\cdot\bm{G}^{R}\cdot\bm{\Sigma}_{\text{int}}^{<}\cdot\bm{G}^{A}\right].\label{eq:Incoherent_current}
\end{multline}
In the following sections we will use these general formalism together
with the model Hamiltonian from Sec.~\ref{sub:Model-Hamiltonian}
to evaluate the vertical current in graphene-hBN-graphene structures.

\section{Current in the non-interacting, pristine limit\label{sec:Coherent-tunnelling}}

\subsection{General discussion}

\begin{figure}
\begin{centering}
\includegraphics[width=8cm]{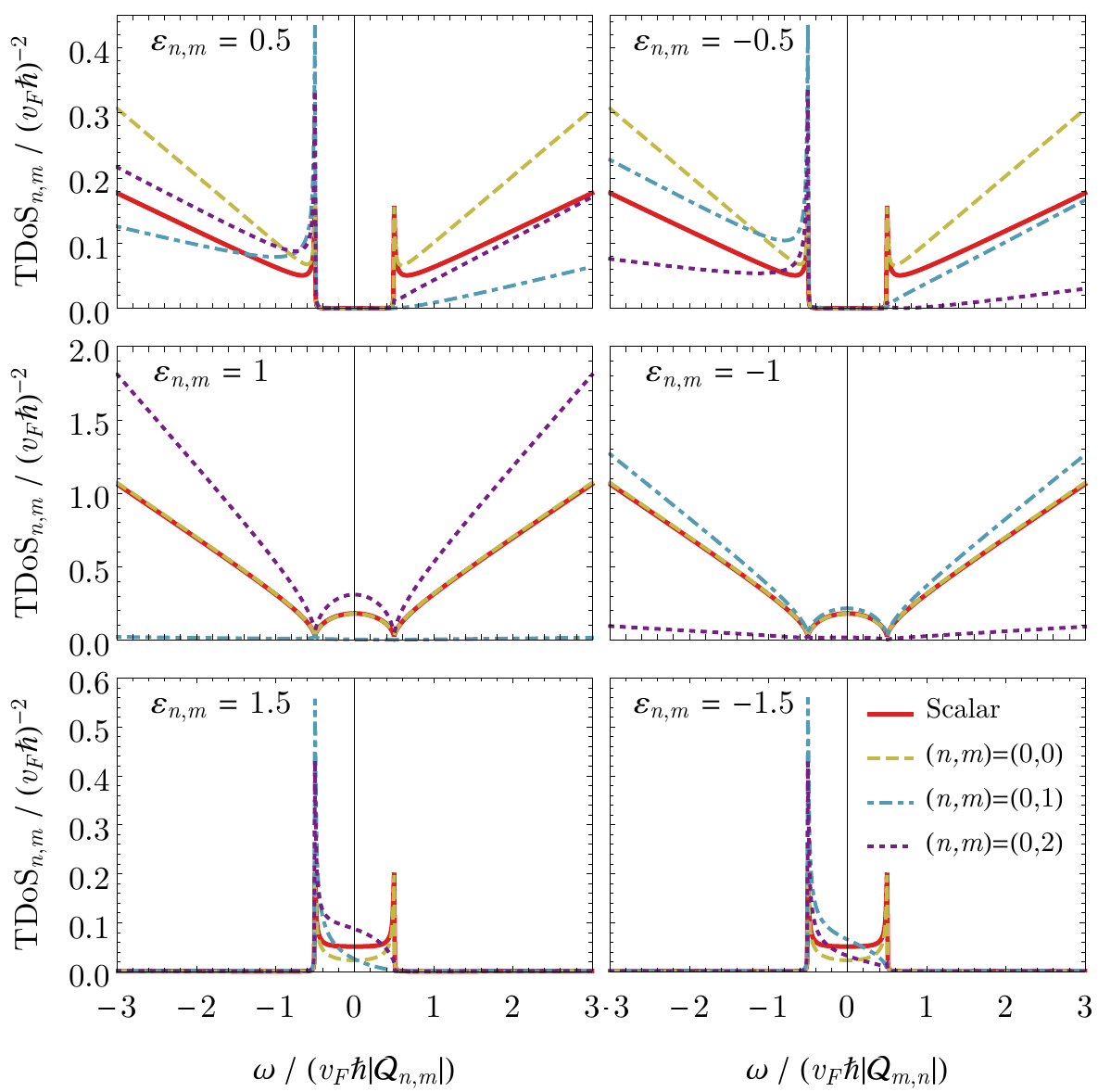}
\par\end{centering}

\protect\caption{\label{fig:TDoS}Plot of the quantity $\text{TDoS}_{m,n}(\omega+\varepsilon_{m,n}v_{F}\hbar\left|\bm{\mathcal{Q}}_{n,m}\right|/2,\omega-\varepsilon_{n,m}v_{F}\hbar\left|\bm{\mathcal{Q}}_{n,m}\right|/2)$
for different values of $\varepsilon_{m,n}$ as a function of the
energy at zero magnetic field and for rotation angles of $\theta_{\text{TG}}=1\text{º}$
and $\theta_{\text{hBN}}=1.5\text{º}$. The solid red line shows the
the tunneling density of states if the wavefunction overlap factors
$\Upsilon_{\bm{k},\lambda}^{B/T,n}$ in Eq.~(\ref{eq:TDoS}) are
set to one. A constant broadening factor of $\gamma=2.5\times v_{F}\hbar\left|\bm{\mathcal{Q}}_{n,m}\right|\times10^{-3}$
was used in all plots.}
\end{figure}
When applying the non-equilibrium Green's function formalism to a
graphene-hBN-graphene device with metal contacts, one is faced with
the issue of how to make the separation between central mesoscopic
region, and the external contacts which are in thermal equilibrium.
Two natural approaches exist: (A) describing the graphene layers as
part of the external contacts, and (B) describing the graphene layers
as part of the central mesoscopic device. In all theoretical analytic
works to date, the graphene layers were assumed to be part of the
external contacts \cite{Feenstra2012,V__13,B__14,MTC14,Barrera2015},
being in equilibrium. However, due to the low density of states of
graphene, it seems more natural to consider graphene as part of the
mesoscopic device. We will start from approach (B) and see that under
certain approximations, it reduces to approach (A). We will first
consider the non-interacting, pristine case. In approach (B), $\bm{G}^{R/A}(\omega)$
in Eq.~(\ref{eq:Transmission:_function}) is the Green's function
of the graphene-hBN-graphene device. We are interested in the matrix
elements of $\bm{G}^{R/A}$ that connect the bottom and the top contacts.
Due to the block diagonal structure of the Hamiltonian (\ref{eq:Hamiltonian_total})
(there is no direct coupling between the two graphene layers), these
can generally be written as
\begin{eqnarray}
\left[\bm{G}^{R}\right]_{\text{b,t}} & = & \bm{G}_{\text{bg}}^{0,R}\cdot\bm{\mathcal{T}}_{\text{bg,tg}}\cdot\bm{G}_{\text{tg}}^{0,R},\\
\left[\bm{G}^{A}\right]_{\text{t,b}} & = & \bm{G}_{\text{tg}}^{0,A}\cdot\bm{\mathcal{T}}_{\text{tg,bg}}\cdot\bm{G}_{\text{bg}}^{0,A},
\end{eqnarray}
where $\bm{G}_{\text{bg}/\text{tg}}^{0,R/A}$ are the Green's function
of the bottom/top graphene layer in the absence of graphene-hBN coupling
(but tacking into account coupling to the external contacts) where
we have defined the hBN mediated tunneling amplitudes
\begin{eqnarray}
\bm{\mathcal{T}}_{\text{bg,tg}} & = & \bm{T}_{\text{bg},\text{hBN}}\cdot\bm{G}_{\text{hBN}}^{R}\cdot\bm{T}_{\text{hBN},\text{tg}},\label{eq:effective_retarded_hoping}\\
\bm{\mathcal{T}}_{\text{tg,bg}} & = & \bm{T}_{\text{tg},\text{hBN}}\cdot\bm{G}_{\text{hBN}}^{A}\cdot\bm{T}_{\text{hBN},\text{bg}},\label{eq:effective_advanced_hoping}
\end{eqnarray}
with $\bm{G}_{\text{hBN}}^{R/A}$ the Green's function of the hBN
slab, which in general takes into account its coupling to the graphene
layers. Therefore, the transmission function Eq.~(\ref{eq:Transmission:_function})
can be written as 
\begin{eqnarray}
\mathcal{T} & = & \Tr\left[\bm{G}_{\text{bg}}^{0,A}\cdot\bm{\Gamma}_{\text{b}}\cdot\bm{G}_{\text{bg}}^{0,R}\bm{\mathcal{T}}_{\text{bg,tg}}\cdot\right.\nonumber \\
 &  & \left.\cdot\bm{G}_{\text{tg}}^{0,R}\cdot\bm{\Gamma}_{\text{t}}\cdot\bm{G}_{\text{tg}}^{0,A}\cdot\bm{\mathcal{T}}_{\text{tg,bg}}\right].\label{eq:transmission_B}
\end{eqnarray}
If we now use Eq.~(\ref{eq:spectral_RA_AR}), we can write the spectral
function of the bottom graphene layer taking into account the coupling
to the bottom metallic contact, $\bm{A}_{\text{bg}}^{0}=i\left(\bm{G}_{\text{bg}}^{0,R}-\bm{G}_{\text{bg}}^{0,A}\right)$,
as $\bm{A}_{\text{bg}}^{0}=\bm{G}_{\text{bg}}^{0,R}\cdot\bm{\Gamma}_{\text{b}}\cdot\bm{G}_{\text{bg}}^{0,A}=\bm{G}_{\text{bg}}^{0,A}\cdot\bm{\Gamma}_{\text{b}}\cdot\bm{G}_{\text{bg}}^{0,R}$
and similarly for the top graphene layer. As such, the transmission
function can be written as
\begin{equation}
\mathcal{T}=\Tr\left[\bm{\mathcal{T}}_{\text{bg,tg}}\cdot\bm{A}_{\text{tg}}^{0}\cdot\bm{\mathcal{T}}_{\text{tg,bg}}\cdot\bm{A}_{\text{bg}}^{0}\right].\label{eq:transmission_A}
\end{equation}
Eq.~(\ref{eq:transmission_A}) is the result that would be directly
obtained, if we followed approach (A) instead, in which case the level
width functions are given by $\bm{\Gamma}_{\text{b}}=\bm{T}_{\text{hBN,bg}}\cdot\bm{A}_{\text{bg}}^{0}\cdot\bm{T}_{\text{hBN},\text{bg}}$
and $\bm{\Gamma}_{\text{t}}=\bm{T}_{\text{hBN,tg}}\cdot\bm{A}_{\text{tg}}^{0}\cdot\bm{T}_{\text{hBN},\text{tg}}$.
As such we have proved that in the non-interacting case both approaches
(A) and (B) coincide. We will leave the discussion for tunneling in
the presence disorder and electron-phonon interactions to the next
section.

In order to make analytic progress, we will employ the wide-band limit
for the metallic contacts, neglecting any frequency dependence of
$\bm{\Gamma}_{\text{b}/\text{t}}$, and assume that the contacts couple
equally to all graphene states, not spoiling translation invariance.
We expect that this last approximation works well for cases where
the metallic contacts are deposited on a small region of the graphene
sample. Within these approximations, the only effect of the metallic
contacts is to introduce a broadening factor of $\gamma_{\text{bg}/\text{tg}}=\Gamma_{\text{b}/\text{t}}/2$
in the Green's function of the bottom/top graphene layer. We will
now write $\mathcal{T}$ for a graphene-hBN-graphene device more explicitly.
Using the graphene-hBN coupling Hamiltonian Eq.~(\ref{eq:Hamiltonian_graphene_hBN}),
the transmission function can be written using the Bloch momentum
basis as (writing explicitly the frequency argument)
\begin{align}
\mathcal{T}(\omega) & =\sum_{\substack{\bm{k},\lambda,\lambda^{\prime}\\
n,m,\tau
}
}\left|_{\text{tg}}\left\langle \bm{k}+\tau\bm{\mathcal{Q}}_{n,m},\tau,\lambda^{\prime}\right|\bm{\mathcal{T}}_{\text{tg,bg}}(\omega)\left|\bm{k},\tau,\lambda\right\rangle _{\text{bg}}\right|^{2}\nonumber \\
 & \times A_{\text{tg},\bm{k}+\tau\bm{\mathcal{Q}}_{n,m},\tau,\lambda^{\prime}}^{0}(\omega_{\text{tg}})A_{\text{bg},\bm{k},\tau,\lambda}^{0}(\omega_{\text{bg}})\label{eq:Transmission_bg_tg}
\end{align}
with the sum on $n,m$ going from $0$ to $2$ and where $\omega_{\text{bg}}=\omega+\epsilon_{\text{F},\text{bg}}$
and $\omega_{\text{tg}}=\omega+\epsilon_{\text{F},\text{tg}}+eV_{\text{bias}}$
are measured from the position of the Dirac point in the bottom and
top graphene layers, respectively. The effective tunneling probability
can be written as 
\begin{multline}
\left|_{\text{tg}}\left\langle \bm{k}+\bm{\mathcal{Q}}_{n,m},\tau,\lambda^{\prime}\right|\bm{\mathcal{T}}_{\text{tg,bg}}(\omega)\left|\bm{k},\tau,\lambda\right\rangle _{\text{bg}}\right|^{2}=\\
=\Upsilon_{\bm{k},\tau,\lambda}^{\text{bg},n}\Upsilon_{\bm{k}+\tau\bm{\mathcal{Q}}_{n,m},\tau,\lambda}^{\text{tg},m}\left|\mathcal{T}_{n,m}(\omega)\right|^{2},
\end{multline}
where $\Upsilon_{\bm{k},\tau,\lambda}^{\text{bg}/\text{tg},n}=1+\tau\lambda\hat{\bm{k}}\cdot\hat{\bm{K}}_{\text{bg}/\text{tg},n}$,
with $\bm{K}_{\text{bg}/\text{tg},n}=\mathcal{R}\left(n2\pi/3\right)\cdot\bm{K}_{\text{bg}/\text{tg}}$,
are graphene wavefunction overlap factors and
\begin{multline}
\mathcal{T}_{n,m}(\omega)=\\
=\frac{1}{2}\text{tr}\left\{ \hat{\bm{T}}^{\dagger}\cdot\bm{R}_{-p\frac{2\pi}{3}}^{m}\cdot\left[\bm{G}_{\text{hBN}}^{A}(\omega)\right]_{\mathcal{N},1}\cdot\bm{R}_{\frac{2\pi}{3}}^{n}\cdot\hat{\bm{T}}\right\} ,\label{eq:T_effective}
\end{multline}
with the trace being performed over the sublattice degrees of freedom.
Neglecting the frequency dependence of $\bm{G}_{\text{hBN}}^{R/A}$
and to lowest order in $t_{\perp}$ we can write
\begin{multline}
\left|\mathcal{T}_{n,m}\right|^{2}\simeq\left(\frac{t_{\perp}^{2}}{E_{\text{B}}E_{\text{N}}}\right)^{\mathcal{N}-1}\times\\
\times\begin{cases}
4\frac{t_{\text{B,C}}^{2}t_{\text{N,C}}^{2}}{E_{\text{B}}E_{\text{N}}}\cos^{2}\left(\frac{\pi}{3}\left(n-m\right)\right) & ,\,\mathcal{N}\text{ is even}\\
\frac{t_{\text{B,C}}^{4}}{E_{\text{B}}^{2}}+\frac{t_{\text{N,C}}^{4}}{E_{\text{N}}^{2}}+2\frac{t_{\text{B,C}}^{2}t_{\text{N,C}}^{2}}{E_{\text{B}}E_{\text{N}}}\cos\left(\frac{2\pi}{3}\left(n-m\right)\right) & ,\,\mathcal{N}\text{ is odd}
\end{cases}\label{eq:Coherent_probability}
\end{multline}
Notice, that in Eq.~(\ref{eq:Transmission_bg_tg}) both valleys give
the same contribution, which can be seen by making a simultaneous
change $\tau\rra-\tau$ and $\bm{k}\rra-\bm{k}$. The transmission
function can then be written as
\begin{equation}
\mathcal{T}(\omega)=Ag_{s}g_{v}\sum_{n,m=0}^{3}\left|\mathcal{T}_{n,m}\right|^{2}\text{TDoS}_{n,m}(\omega_{\text{bg}},\omega_{\text{tg}})
\end{equation}
where $A$ is the area of the device, $g_{s}=g_{v}=2$ are the spin
and valley degeneracies and we have defined the tunneling density
of states as
\begin{align}
\text{TDoS}_{n,m}(\omega_{\text{bg}},\omega_{\text{tg}}) & =\sum_{\lambda,\lambda^{\prime}=\pm1}\int\frac{d^{2}\bm{k}}{\left(2\pi\right)^{2}}\Upsilon_{\bm{k},\lambda}^{\text{bg},n}\Upsilon_{\bm{k}+\bm{\mathcal{Q}}_{n,m},\lambda^{\prime}}^{\text{tg},m}\times\nonumber \\
 & \times A_{\text{bg},\bm{k},\lambda}^{0}(\omega_{\text{bg}})A_{\text{tg},\bm{k}+\bm{\mathcal{Q}}_{n,m},\lambda^{\prime}}^{0}(\omega_{\text{tg}}),\label{eq:TDoS}
\end{align}
which only depends of the graphene's dispersion relation and wavefunctions
(for simplicity we have dropped the valley indice $\tau$). In the
limit of infinite lifetime for graphene electrons, the spectral functions
reduce to $\delta$-functions and it is possible to provide an analytic
expression for $\text{TDoS}_{n,m}(\omega_{\text{bg}},\omega_{\text{tg}})$.
In the presence of a finite, momentum independent, lifetime, it is
still possible to find an approximate analytic expression to Eq.~(\ref{eq:TDoS}).
These analytic expressions lead to a significant speed up in the evaluation
of the current and are presented in Appendix~\ref{sec:Ananlitic-expression-TDoS}.
The presence of the spectral functions for the bottom and top graphene
layers leads to conservation of energy and momentum in the tunneling
process between the two graphene layers. 

\begin{figure}
\begin{centering}
\includegraphics[width=8cm]{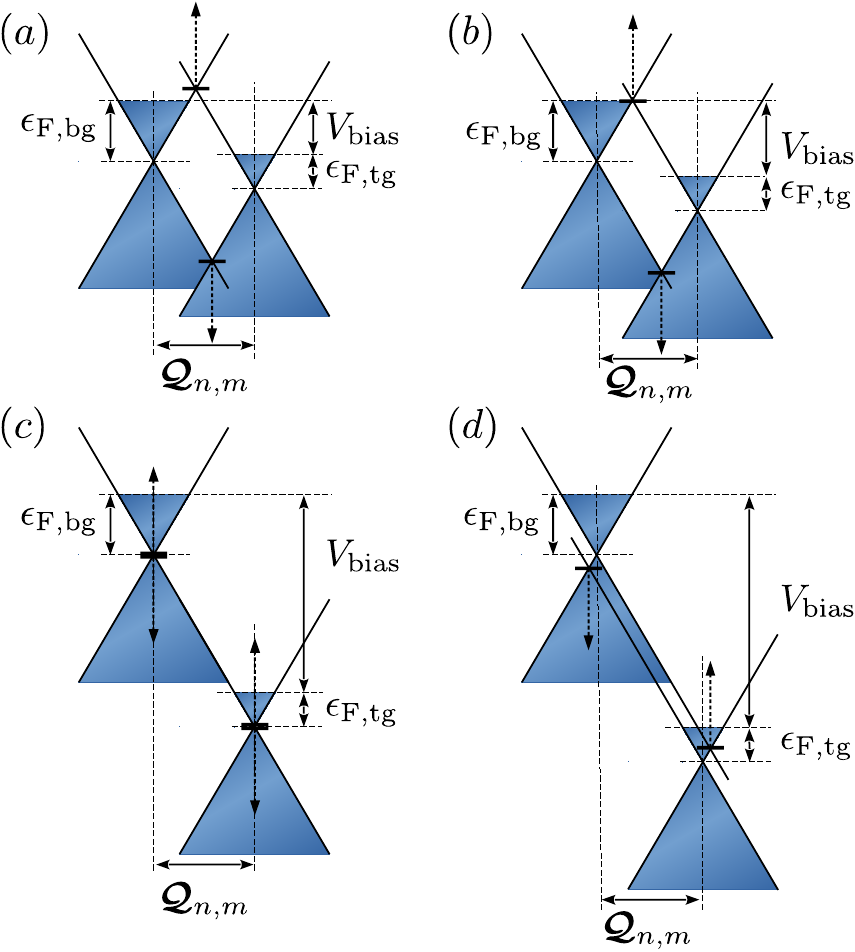}
\par\end{centering}

\protect\caption{\label{fig:Energy-Momentum_diagram}Band diagram representing the
constrains imposed by energy-momentum conservation and Pauli's exclusion
principle in the vertical current of a graphene-hBN-graphene device.
The two cones represent the dispersion relation for electrons of the
bottom and top graphene layers. The shadowed blue regions represent
the occupation of electronic states in both graphene layers. Energy-momentum
conservation is only satisfied when the two shifted Dirac cones intersect
and the energy windows where this occurs are represented by the dashed
arrows. The following cases are represented: (a) Only intraband are
possible, $\varepsilon_{n,m}<1$, these are however Pauli blocked
or there are no states available, therefore in the low temperature
limit, no vertical current flows. (b) Threshold bias voltage above
which intraband processes which satisfy energy-momentum conservation
appear in the energy window where tunneling is allowed by the electronic
occupation factors. (c) Condition which corresponds to the occurrence
of a peak in the current, when $\varepsilon_{n,m}=1$, when both intraband
and interband (conduction-to-valence and valence-to-conduction) processes
are allowed. (d) If one further increases the bias voltage, only interband
tunneling, $\varepsilon_{n,m}>1$, becomes possible and the current
diminishes. }
\end{figure}

\begin{figure}
\begin{centering}
\includegraphics[width=8cm]{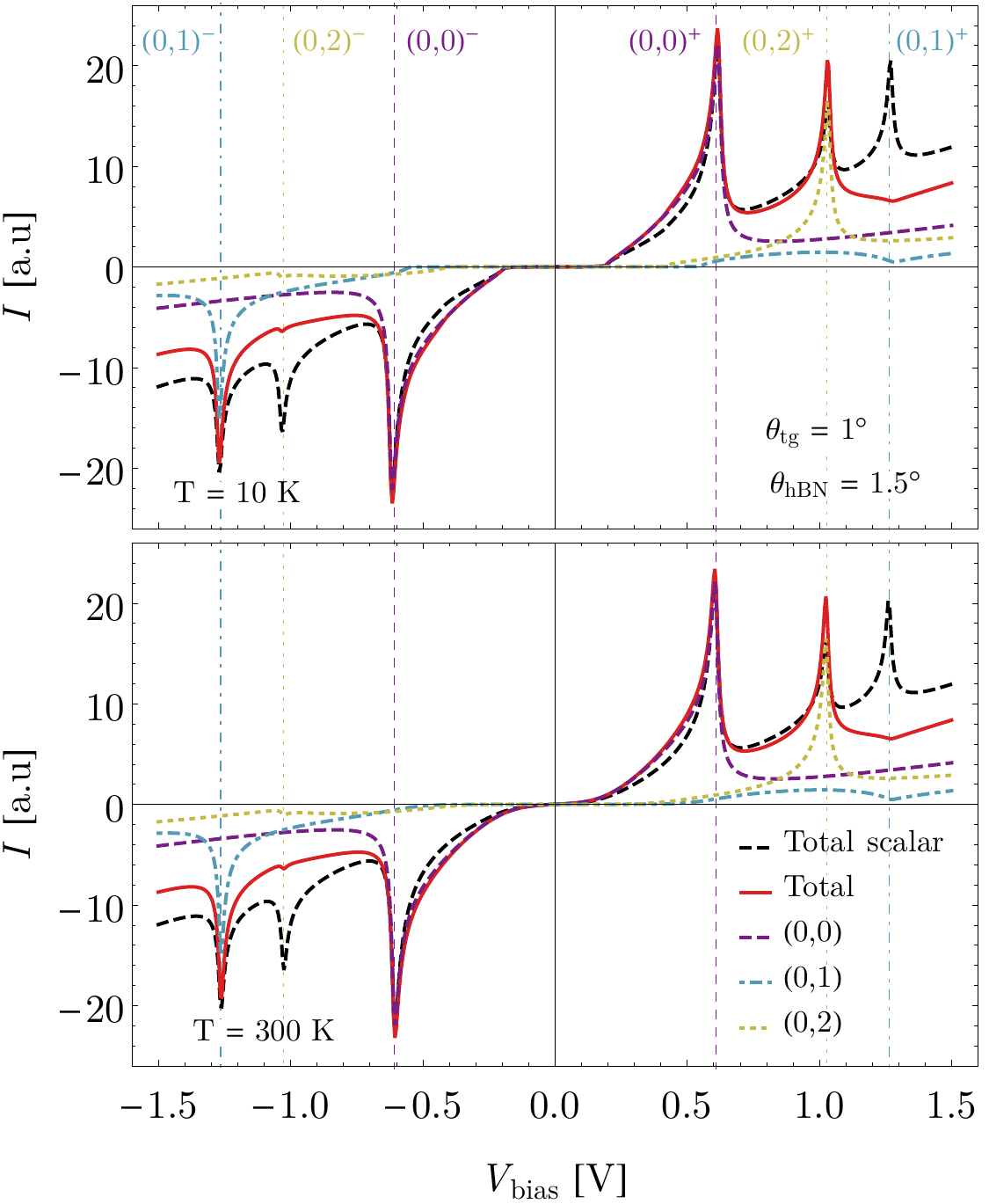}
\par\end{centering}

\protect\caption{\label{fig:I_V_n_m}I-V curves for vertical current in a graphene-hBN-graphene
device with 4 layers of hBN for rotation angles of $\theta_{\text{tg}}=1\text{º}$
and $\theta_{\text{hBN}}=1.5\text{º}$ at gate voltage $V_{\text{gate}}=0$
for two different temperatures. The solid red line indicates the current
due to all the 9 processes coupling both graphene layers, for graphene
electrons, while the dashed black lines represents the total current
for scalar electrons (by setting the wavefunction factors $\Upsilon_{\bm{k},\lambda}^{\text{bg}/\text{tg},n}$
to $1$). The remaining lines represent the contributions to the current
arising from processes involving different $\mbox{\ensuremath{\bm{\mathcal{Q}}}}_{n,m}$
(taking into account the relations imposed by 3-fold rotational invariance,
Eq.~\ref{eq:Momentum_groups}) The dashed vertical lines labeled
by $(n,m)^{\pm}$ mark the bias voltages when the condition $\epsilon_{\text{F,tg}}+eV_{\text{bias}}-\epsilon_{\text{F,bg}}=\pm v_{F}\hbar\left|\bm{\mathcal{Q}}_{n,m}\right|$
is satisfied. Notice that while for scalar electrons all the expected
peaks in the current are present, for Dirac electrons some of them
are absent. Is is due to the suppression by the $\Upsilon_{\bm{k},\lambda}^{\text{bg}/\text{tg},n}$
factors. A constant broadening factor of $\gamma=2.5$ meV was used.}
\end{figure}

\subsection{Results}

\begin{figure}
\begin{centering}
\includegraphics[width=8cm]{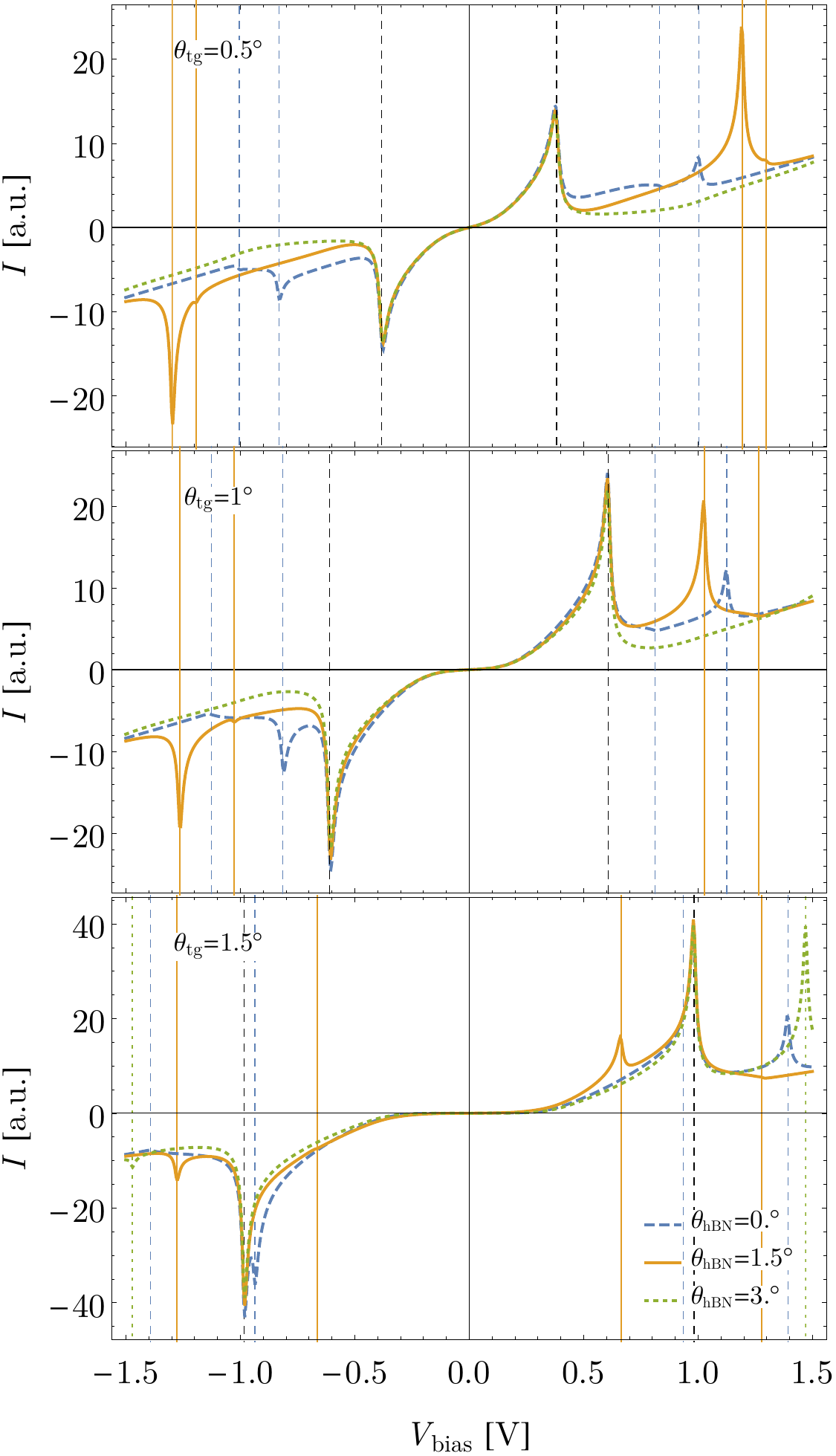}
\par\end{centering}

\protect\caption{\label{fig:I-V_vs_rotation}I-V curves at constant $V_{\text{gate}}=0$
in a graphene-hBN-graphene device with 4 layers of hBN, at $V_{\text{gate}}=0$
and $T=300$ K, for different rotation angles between the top and
bottom graphene layers, and the hBN slab and the bottom graphene layer.
The black dashed line marks the bias voltage when $\varepsilon_{0,0}=\pm1$
(a condition that is independent of $\theta_{\text{hBN}}$). The remaining
vertical lines mark the bias voltages when $\varepsilon_{n,m}=\pm1$
for $n\protect\neq m$ for different values $\theta_{\text{hBN}}$
(the color and type of line match the ones used in the plots).}
\end{figure}

\begin{figure}
\begin{centering}
\includegraphics[width=8cm]{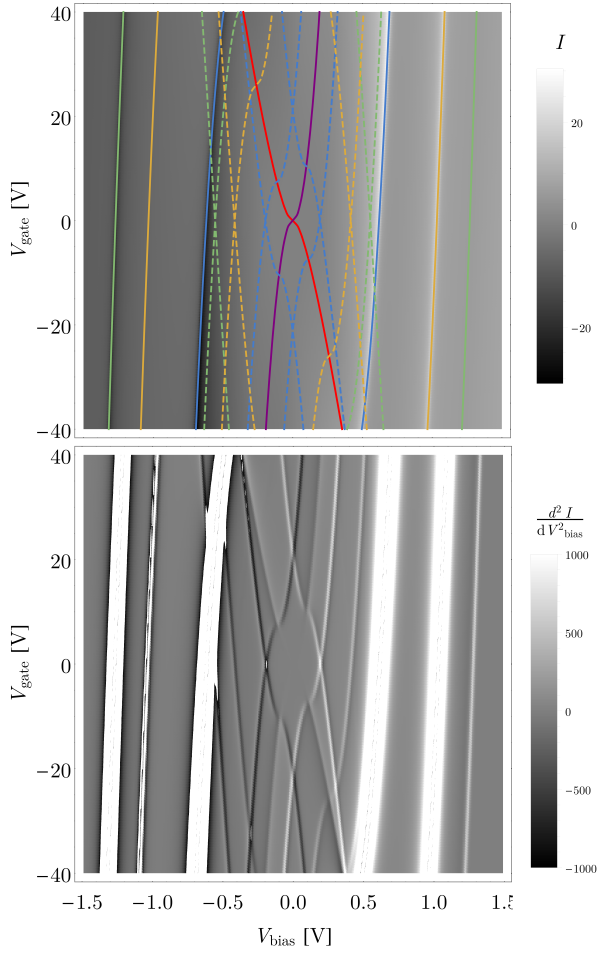}
\par\end{centering}

\protect\caption{\label{fig:Density-plot}Density plot of current, $I$, and its second
derivative with respect to the applied bias voltage, $d^{2}I/dV_{\text{bias}}^{2}$,
as a function of the applied bias and gate voltages at $T=10$ K.
In the current plot, it is also shown the lines defined by the following
conditions: $\epsilon_{\text{F},\text{bg}}=0$ and $\epsilon_{\text{F},\text{tg}}=0$,
represented by the solid lines in red and purple; $\epsilon_{\text{F},\text{tg}}+eV_{\text{bias}}-\epsilon_{\text{F},\text{bg}}=\pm v_{F}\hbar\left|\mathcal{\bm{Q}}_{0,m}\right|$
(Eq.~(\ref{eq:peak_current})) for $m=0$, 1 and $2$, represented
by the solid lines in blue, green and yellow, respectively; $\epsilon_{\text{F},\text{tg}}\pm eV_{\text{bias}}+\epsilon_{\text{F},\text{bg}}=\pm\frac{1}{2}v_{F}\hbar\left|\mathcal{\bm{Q}}_{0,m}\right|$
for $m=0$, 1 and $2$, (Eq.~(\ref{eq:open_currrent})) represented
by the dashed lines in blue, green and yellow. Notice now the guides
lines shown in the current plot match perfectly the sharp features
shown in the $d^{2}I/dV_{\text{bias}}^{2}$ plot. Also, the peaks
expected to occur through channels $(0,1)^{+}$ and $(0,2)^{-}$ are
absent. A constant broadening factor of $\gamma=2.5$ meV was assumed
for both layers.}
\end{figure}

\begin{figure}
\begin{centering}
\includegraphics[width=8cm]{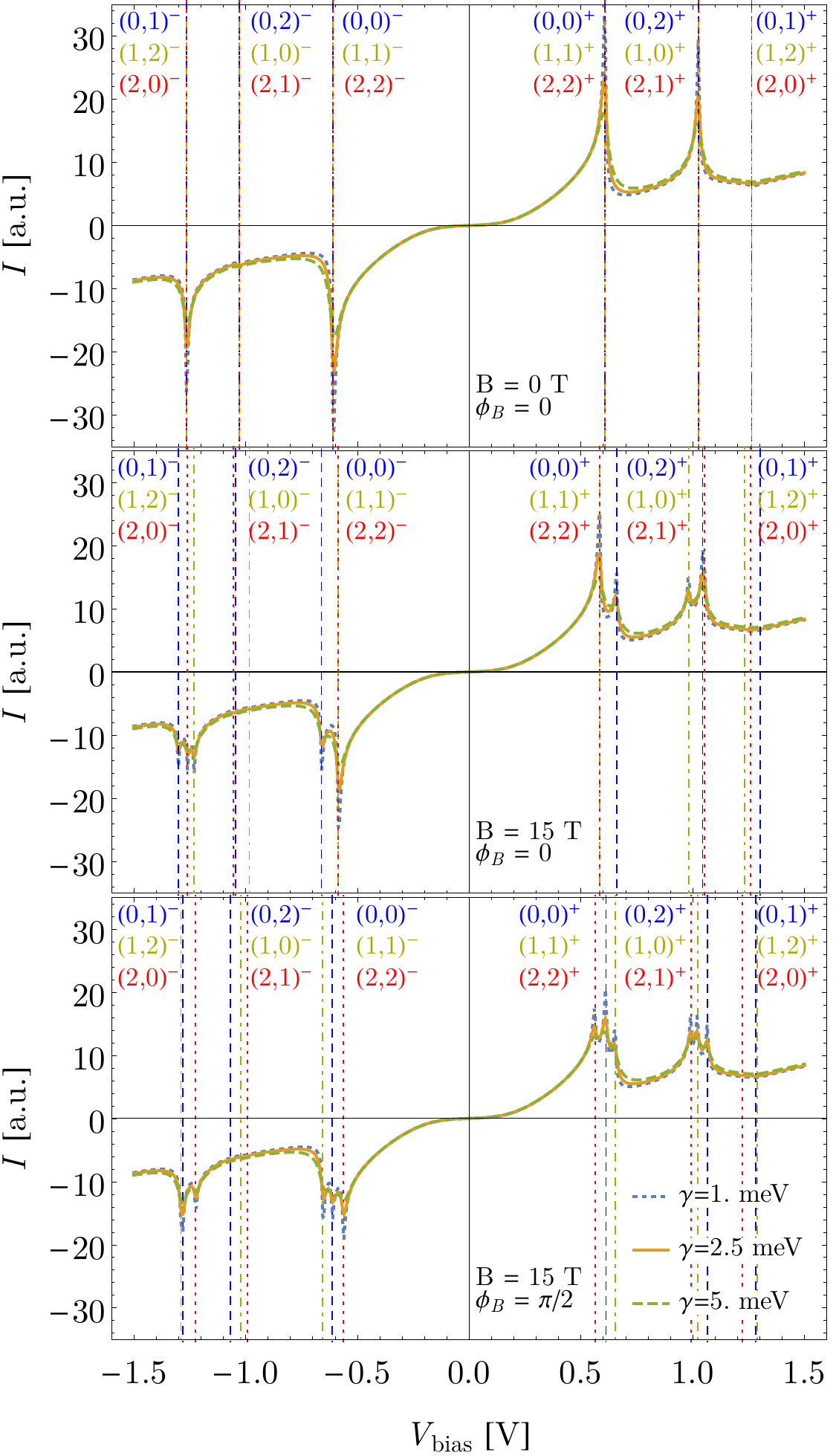}
\par\end{centering}

\protect\caption{\label{fig:I-V_vs_B}I-V curves for a graphene-hBN-graphene device
with 4 layers of hBN, with rotation angles of $\theta_{\text{tg}}=1\text{º}$
and $\theta_{\text{hBN}}=1.5\text{º}$ at constant $V_{\text{gate}}=0$
and $T=300$ K, for different values and orientation of the in-plane
magnetic field and electronic broadening factor. The vertical lines,
labeled by $(n,m)^{\pm}$ mark the bias voltages for which $\varepsilon_{n,m}=\pm1$.
Notice how the applied magnetic field leads to a splitting of the
peaks that occur at zero magnetic field. As the broadening factor
is increased, the peaks become less resolved.}
\end{figure}

The tunneling in a graphene-hBN-graphene structure is controlled both
by energy-momentum conservation and by Pauli's exclusion principle.
The constrains imposed by energy-momentum conservation can be understood
considering that the Dirac cones of the bottom and top graphene layers
are shifted in energy by a value of $\epsilon_{\text{F,tg}}+eV_{\text{bias}}-\epsilon_{\text{F,bg}}$
and in momentum by a value of $\left|\bm{\mathcal{Q}}_{n,m}\right|$,
see Fig.~ \ref{fig:Energy-Momentum_diagram}. The intersection of
the shifted cones allows one to visualize the states which respect
energy-momentum conservation\cite{B__14}. Whenever the bias voltage
is tuned such that 
\begin{equation}
\epsilon_{\text{F,tg}}+eV_{\text{bias}}-\epsilon_{\text{F,bg}}=v_{F}\hbar\left|\bm{\mathcal{Q}}_{n,m}\right|,\label{eq:peak_current}
\end{equation}
there is a complete overlap of the Dirac cones and a maximum in the
current occurs. The information regarding energy-momentum conservation
for an electron tunneling between the two graphene layers is encoded
in the the $\text{TDoS}_{n,m}$ function. In Fig.~\ref{fig:TDoS}
we plot the quantity $\text{TDoS}_{n,m}(\omega-\varepsilon_{n,m}v_{F}\hbar\left|\bm{\mathcal{Q}}_{n,m}\right|/2,\omega+\varepsilon_{n,m}v_{F}\hbar\left|\bm{\mathcal{Q}}_{n,m}\right|/2)$,
for different values of $\varepsilon_{n,m}=\left(\epsilon_{\text{F,tg}}+eV_{\text{bias}}-\epsilon_{\text{F,bg}}\right)/\left(v_{F}\hbar\left|\bm{\mathcal{Q}}_{n,m}\right|\right)$.
For $\varepsilon_{n,m}^{2}<1$, the tunneling is due to intraband
processes (from the conduction/valence band of the bottom graphene
into the conduction/valence band of the top graphene), going to zero
in the pristine limit for $\omega^{2}<\left(v_{F}\hbar\right)^{2}\left|\bm{\mathcal{Q}}_{n,m}\right|^{2}/4$.
For $\varepsilon_{n,m}^{2}>1$, the tunneling is due to interband
processes (from the conduction/valence band of the bottom graphene
layer to the valence/conduction band of the top graphene layer), being
zero in the pristine limit for $\omega^{2}>\left(v_{F}\hbar\right)^{2}\left|\bm{\mathcal{Q}}_{n,m}\right|^{2}/4$.
For $\varepsilon_{n,m}^{2}=1$, $\text{TDoS}_{n,m}(\omega-\varepsilon_{n,m}v_{F}\hbar\left|\bm{\mathcal{Q}}_{n,m}\right|/2,\omega+\varepsilon_{n,m}v_{F}\hbar\left|\bm{\mathcal{Q}}_{n,m}\right|/2)$
diverges in the pristine limit for any value of $\omega$. This divergence
in $\text{TDoS}_{n,m}$ leads to a divergence in the vertical current
\cite{B__14,MTC14}, which is made finite with the introduction of
a finite electronic lifetime. Since for different processes $(n,m)$
with different $\left|\bm{\mathcal{Q}}_{n,m}\right|$ there is a different
effective separation in momentum between the Dirac cones of the bottom
and top graphene layers, one expects the occurrence of multiple peaks
in the I-V curve, followed by subsequent regions of negative differential
conductance. This is indeed the case as shown in Fig.~(\ref{fig:I_V_n_m}).
Based only on energy-momentum conservation, one would expect the occurrence
of three peaks in the I-V curve for positive bias voltage and another
three for negative bias (notice that according to the discussion of
Sec.~\ref{sub:Model-Hamiltonian} from the nine processes coupling
the two graphene layers, only three are independent). This is indeed
the case as shown in Fig.~(\ref{fig:I_V_n_m}). However, the computed
curve only displays two peaks, with the peaks corresponding to the
situations when and $\varepsilon_{0,2}=-1$ being absent. The reason
for the suppression of these peaks is due to the spinorial structure
of graphene electronic wavefunctions, via the overlap factors $\Upsilon_{\bm{k},\lambda}^{\text{bg}/\text{tg},n}$,
that appear in Eq.~(\ref{eq:TDoS}). These overlap factors can severely
suppress the value of $\text{TDoS}_{n,m}$ close to $\varepsilon_{n,m}=\pm1$
and consequently of the height of the peaks in the I-V curve. This
is shown in Fig.~\ref{fig:TDoS} , where it is shown a considerable
suppression of $\text{TDoS}_{n,m}$ for $\varepsilon_{0,1}=1$ and
$\varepsilon_{0,2}=-1$. The effect of the overlap factors can also
be seen in Fig.~(\ref{fig:I_V_n_m}), where it is also shown the
current that would be obtained, if the electronic wavefunction of
graphene where scalars, i.e. by setting $\Upsilon_{\bm{k},\lambda}^{\text{bg}/\text{tg},n}=1$
in (\ref{eq:TDoS}) (see Eq~\ref{eq:TDoS_scalar} in Appendix~\ref{sec:Ananlitic-expression-TDoS}),
displaying the three peaks expected by kinematic considerations. While
the occurrence of NDC in graphene-hBN-graphene has already been experimentally
observed\cite{MTC14}, the occurrence of multiple NDC regions has
not. This might be due to the fact that the position of the current
peaks depends very sensitively in the rotation angles $\theta_{\text{tg}}$
and $\theta_{\text{hBN}}$. This is exemplified in Fig.~\ref{fig:I-V_vs_rotation},
where the computed I-V curves for several rotation angles are shown.
As shown, for a fixed angle of $\theta_{\text{tg}}=1^{\circ}$, changing
$\theta_{\text{hBN}}$ from $1.5^{\circ}$ to $3^{\circ}$ moves the
additional peaks in the current due to the transfer of momentum by
the hBN crystal lattice from a bias voltage of $\sim1$ V to bias
voltages $>1.5$ V. Tunneling processes which satisfy energy-momentum
conservation, can only contribute to the current if these lie in an
energy window between the zero of energy and the bias voltage, as
presented in Fig.~\ref{fig:Energy-Momentum_diagram}. The condition
for which processes allowed by energy-momentum conservation become
allowed by the occupation factors occurs in the limit of zero temperature
when (see Fig.~\ref{fig:Energy-Momentum_diagram}.(b))
\begin{equation}
\epsilon_{\text{F},\text{tg}}\pm eV_{\text{bias}}+\epsilon_{\text{F},\text{bg}}=\pm\frac{1}{2}v_{F}\hbar\left|\bm{\mathcal{Q}}_{n,m}\right|.\label{eq:open_currrent}
\end{equation}
This explains the occurrence of the plateau with nearly zero current
seen at low temperature in Fig.~\ref{fig:I_V_n_m}, and gives origin
to the features in the $d^{2}I/dV_{\text{bias}}^{2}$ as a function
of applied bias and gate voltages as seen in the density plot of Fig.~\ref{fig:Density-plot}.
At higher temperatures, all these sharp features tend to vanish, as
the Fermi-Dirac occupation factors become smoother. 

By applying an in-plane magnetic field, the threefold rotational invariance
of the graphene-hBN-graphene structure is broken, and therefore, the
processes corresponding to the different groups in (\ref{eq:Momentum_groups})
will contribute differently to the current, and one expects that each
peak in the I-V curve will split into three. An in-plane magnetic
field of the form $\bm{B}=B\left(\cos\phi_{B},\sin\phi_{B},0\right)$
can be described by the vector potential $\bm{A}=Bz\left(\sin\phi_{B},-\cos\phi_{B}\right)$.
Neglecting the momentum dependence of $H_{\text{hBN}}$ the effect
of the in-plane magnetic field reduces to an additional transference
of momentum to the tunneling electrons,, which is encoded in a shift
in the $\bm{\mathcal{Q}}_{n,m}$ vectors \cite{Brey1988,Hayden1991,Falko1991,B__14,MTC14}
\begin{equation}
\bm{\mathcal{Q}}_{n,m}\rra\bm{\mathcal{Q}}_{n,m}+\frac{eBd}{\hbar}\left(\sin\phi_{B},-\cos\phi_{B}\right).
\end{equation}
The splitting of the peaks in the I-V curve is shown in Fig.~(\ref{fig:I-V_vs_B}),
where it is also shown the effect of an increasing electronic broadening
factor.

\begin{figure}
\begin{centering}
\includegraphics[width=5cm]{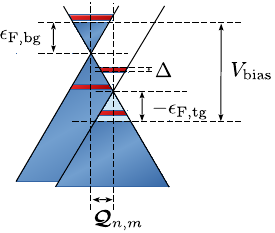}
\par\end{centering}

\protect\caption{\label{fig:Spectrum_reconstruction}Diagram representing the possible
effect of the reconstruction of the graphene Dirac spectrum, due to
the presence of hBN, in the vertical current of a graphene-hBN-graphene
device when $\varepsilon_{n,m}=\pm1$. The red bars represent the
position in energy of the regions, of width $\Delta$, where graphene's
spectrum reconstruction is significant. Provided $eV_{\text{bias}}/\Delta\gg1$,
the peaks in the current will still be present.}
\end{figure}

Finally, we comment the possible effect of the hBN in the electronic
structure of graphene. It is known that the potential modulation with
the periodicity of the Moiré pattern formed by graphene on top of
hBN can lead to a reconstruction of the density of states of graphene
at energies $\sim\pm v_{F}\hbar\left|\bm{g}_{1/2}^{\text{g},\text{hBN}}\right|/2$
measured from the original Dirac cone, where $\left|\bm{g}_{1/2}^{\text{g},\text{hBN}}\right|\simeq4\pi\sqrt{\delta^{2}+\theta_{\text{g},\text{hBN}}^{2}}/(\sqrt{3}a_{\text{g}})$
is the wavevector of the Moiré pattern reciprocal lattice \cite{Park_2008,Yankowitz_2012,Ortix_2012,Wallbank2013,Wallbank2013_b,Ponomarenko_2013}
with $\theta_{\text{g},\text{hBN}}$ is the rotation angle between
the graphene layer and the hBN slab. We have disregarded such effects
in our discussion. As we have seen in Fig.~(\ref{fig:I-V_vs_rotation}),
the additional peaks in the current enabled by the transference of
momentum by the hBN lattice, only appear for reasonable values of
the bias voltage for small twist angles between the graphene layers
and hBN slab. It is precisely in this case that that the reconstruction
of the graphene dispersion relations becomes important at low energy.
The effect of this reconstruction should impact not only the peaks
that involve transference of momentum by the hBN lattice ($n\neq m$),
but also the ones that do not ($n=m$). In this situation one can
question the validity of the results from these section. However,
we argue that the possible reconstruction of the graphene dispersion
relations, should not affect in a profound way the occurrence of peaks
and NDC in the I-V curves of graphene-hBN-graphene devices. The energy
width, $\Delta$, where the reconstruction of the linear dispersion
relation of graphene is significant is of the order of the tens of
meV\cite{Yankowitz_2012,DaSilva2015}, while the total energy window
of states that contribute to the current is, at low temperatures,
of the width of $\sim eV_{\text{bias}}$. Provided the condition $eV_{\text{bias}}/\Delta\gg1$
is satisfied (see Fig.~(\ref{fig:Spectrum_reconstruction})), we
expect that the effect of the dispersion relation reconstruction is
negligible, and apart from a possible reduction of the height of the
peaks, should not affect the current in any drastic way.

\section{Incoherent current: phonon and disorder assisted tunneling\label{sec:Incoherent-tunnelling}}

\subsection{General discussion}

We will now study, in a unified way, the effect of phonons and disorder
in the current of a graphene-hBN-graphene device. We consider a generic
electron-phonon interaction described by the Hamiltonian
\begin{equation}
H_{\text{e-ph}}=\bm{c}^{\dagger}\cdot\bm{M}_{\zeta}\cdot\bm{c}\phi_{\zeta},
\end{equation}
where $\phi_{\zeta}=\left(a_{\zeta}+a_{\zeta}^{\dagger}\right)/\sqrt{2}$
is the phonon field operator, with $a_{\zeta}^{\dagger}$ the creation
operator for a phonon mode $\zeta$, $\bm{M}_{\zeta}$ is an electron-phonon
coupling matrix and $\bm{c}^{\dagger}$ is row vector of electronic
creation operators in an arbitrary basis. For this electron-phonon
interaction, the Fock (or sunset)%
\footnote{We point out that the Hartree (or tadpole) self-energy is local in
time and as such does not give origin to lesser/greater self-energy
terms, contributing only to the retarded/advanced self-energies.%
} contribution to the lesser/greater self-energy is given by (from
now on we write all frequency arguments explicitly)
\begin{equation}
\bm{\Sigma}_{\text{e-ph}}^{\lessgtr}(\omega)=i\sum_{\zeta}\int\frac{d\nu}{2\pi}\bm{M}_{\zeta}\cdot\bm{G}^{\lessgtr}(\omega-\nu)\cdot\bm{M}_{\zeta}^{\dagger}D_{\zeta}^{\lessgtr}(\nu),
\end{equation}
where $D_{\zeta}^{\lessgtr}(\nu)$ is the lesser/greater Green's function
for the phonon field operator, which, assuming the phonons are in
thermal equilibrium, are given by
\begin{equation}
D_{\zeta}^{\lessgtr}(\nu)=\mp i2\pi b(\pm\nu)2\omega_{\zeta}\sgn(\nu)\delta\left(\nu^{2}-\omega_{\zeta}^{2}\right),
\end{equation}
where $b(\nu)=\left(e^{\beta\nu}-1\right)^{-1}$ is the Bose-Einstein
distribution function, which satisfies $1+b(\nu)=-b(-\nu)$, and $\omega_{\zeta}$
is phonon frequency of mode $\zeta$. Therefore, the self-energy due
to electron-phonon interaction reads
\begin{equation}
\bm{\Sigma}_{\text{e-ph}}^{\lessgtr}(\omega)=\sum_{\zeta,s=\pm1}\left(\pm sb(\pm s\omega_{\zeta})\right)\bm{M}_{\zeta}\cdot\bm{G}^{\lessgtr}(\omega-s\omega_{\zeta})\cdot\bm{M}_{\zeta}^{\dagger}.\label{eq:greater_lesser_e-ph_selfenergy}
\end{equation}
We point out that this self-energy can also describe elastic scattering
by impurities by drop the summation over $s$, take $\omega_{\zeta}\rra0$
and set $\pm sb(\pm s\omega_{\zeta})\rra1$, in which case the quantity
$\bm{M}_{\zeta}\bm{M}_{\zeta}^{\dagger}$ is to be interpreted as
the disorder correlator. With this in mind, the following discussion
applies both to inelastic scattering by phonons and elastic scattering
by impurities. Combining Eq.~\ref{eq:greater_lesser_e-ph_selfenergy}
with Eqs.~(\ref{eq:Keldysh}) and (\ref{eq:self_energy_split}),
it is possible to write to lowest order in the electron-phonon interaction
\begin{multline}
\bm{\Sigma}_{\text{e-ph}}^{<}(\omega)\simeq\sum_{\zeta,s=\pm1}if_{\text{b}}\left(\omega-s\omega_{\zeta}\right)sb(s\omega_{\zeta})\times\\
\times\bm{M}_{\zeta}\cdot\bm{G}^{R}(\omega-s\omega_{\zeta})\cdot\bm{\Gamma}_{\text{b }}(\omega-s\omega_{\zeta})\cdot\bm{G}^{A}(\omega-s\omega_{\zeta})\cdot\bm{M}_{\zeta}^{\dagger}.\\
+\sum_{\zeta,s=\pm1}if_{\text{t}}\left(\omega-s\omega_{\zeta}\right)sb(s\omega_{\zeta})\times\\
\times\bm{M}_{\zeta}\cdot\bm{G}^{R}(\omega-s\omega_{\zeta})\cdot\bm{\Gamma}_{\text{t}}(\omega-s\omega_{\zeta})\cdot\bm{G}^{A}(\omega-s\omega_{\zeta})\cdot\bm{M}_{\zeta}^{\dagger},\label{eq:self_energy_lowest_order}
\end{multline}
with $\bm{\Sigma}_{\text{e-ph}}^{>}(\omega)$ obtained by replacing
$f_{\text{b}/\text{t}}\rra1-f_{\text{b}/\text{t}}$ and $b(s\omega_{\zeta})\rra b(-s\omega_{\zeta})$.
Inserting this expression in Eq.~(\ref{eq:Incoherent_current}),
we obtain the lowest order contribution to the non-coherent current
\begin{multline}
I_{\text{b}\rra\text{t}}^{\text{incoh,1-ph}}=\\
=\frac{e}{\hbar}\sum_{\zeta,s}\int\frac{d\omega}{2\pi}f_{\text{b}}(\omega)\left[1-f_{\text{t}}\left(\omega-s\omega_{\zeta}\right)\right]\left(-sb(-s\omega_{\zeta})\right)\mathcal{T}_{\text{b},\text{t}}^{(\zeta,s)}(\omega)\\
-\frac{e}{\hbar}\sum_{\zeta,s}\int\frac{d\omega}{2\pi}\left(1-f_{\text{b}}(\omega)\right)f_{\text{t}}\left(\omega-s\omega_{\zeta}\right)sb(s\omega_{\zeta})\mathcal{T}_{\text{b},\text{t}}^{(\zeta,s)}(\omega)\\
+\frac{e}{\hbar}\sum_{\zeta,s}\int\frac{d\omega}{2\pi}f_{\text{b}}(\omega)\left[1-f_{\text{b}}\left(\omega-s\omega_{\zeta}\right)\right]\left(-sb(-s\omega_{\zeta})\right)\mathcal{T}_{\text{b},\text{b}}^{(\zeta,s)}(\omega)\\
-\frac{e}{\hbar}\sum_{\zeta,s}\int\frac{d\omega}{2\pi}\left(1-f_{\text{b}}(\omega)\right)f_{\text{b}}\left(\omega-s\omega_{\zeta}\right)sb(s\omega_{\zeta})\mathcal{T}_{\text{b},\text{b}}^{(\zeta,s)}(\omega),\label{eq:Incoherent_lowest_order}
\end{multline}
where the 1-phonon (disorder) assisted transmission function is given
by 
\begin{align}
\mathcal{T}_{\ell,\ell^{\prime}}^{(\zeta,s)}(\omega) & =\Tr\left[\bm{\Gamma}_{\ell}(\omega)\cdot\bm{G}^{R}(\omega)\cdot\bm{M}_{\zeta}\cdot\bm{G}^{R}(\omega-s\omega_{\zeta})\right.\cdot\nonumber \\
 & \cdot\left.\bm{\Gamma}_{\ell^{\prime}}(\omega-s\omega_{\zeta})\cdot\bm{G}^{A}(\omega-s\omega_{\zeta})\cdot\bm{M}_{\zeta}^{\dagger}\cdot\bm{G}^{A}(\omega)\right].\label{eq:Tunnelling_assisted}
\end{align}
It is easy to check that 
\begin{multline}
f_{\text{b}}(\omega)\left[1-f_{\text{b}}\left(\omega-s\omega_{\zeta}\right)\right]\left(-sb(-s\omega_{\zeta})\right)=\\
=\left(1-f_{\text{b}}(\omega)\right)f_{\text{b}}\left(\omega-s\omega_{\zeta}\right)sb(s\omega_{\zeta})\label{eq:Fermi_ident}
\end{multline}
and as such the last two terms of Eq.~(\ref{eq:Incoherent_lowest_order})
cancel each other. This cancellation is required since in a steady
state no charge accumulation can occur in the device and therefore,
the current flowing from the top to the bottom contact should satisfy
$I_{\text{t}\rra\text{b}}=-I_{\text{t}\rra\text{b}}$. As such, terms
that involve only the occupation factor of one the contacts must cancel
at the end of any calculation. Processes assisted by a greater number
of phonons can also be obtained. Higher order corrections to Eq.~(\ref{eq:self_energy_lowest_order})
can be obtained by iterating Eq.~(\ref{eq:greater_lesser_e-ph_selfenergy})
using Eqs.~(\ref{eq:Keldysh}) and (\ref{eq:self_energy_split}).
Just as for the lowest order case, contributions involving only occupation
factors from one of the contacts cancel each other. Therefore, the
contribution to the incoherent current assisted by $n$ phonons can
be written as\begin{widetext}
\begin{multline}
I_{\text{b}\rra\text{t}}^{\text{incoh, }n\text{-ph}}=\\
=\frac{e}{\hbar}\sum_{\zeta_{1},s_{1},...,\zeta_{n},s_{n}}\int\frac{d\omega}{2\pi}f_{\text{b}}(\omega)\left(1-f_{\text{t}}\left(\omega+s_{1}\omega_{\zeta_{1}}+...+s_{n}\omega_{\zeta_{n}}\right)\right)\left(s_{1}b(s_{1}\omega_{\zeta_{1}})\right)...\left(s_{n}b(s_{n}\omega_{\zeta_{n}})\right)\mathcal{T}_{\text{b}\rra\text{t}}^{(\zeta_{1},s_{1})....(\zeta_{n},s_{n})}(\omega)\\
-\frac{e}{\hbar}\sum_{\zeta_{1},s_{1},...,\zeta_{n},s_{n}}\int\frac{d\omega}{2\pi}f_{\text{t}}\left(\omega\right)\left(1-f_{\text{b}}(\omega+s_{1}\omega_{\zeta_{1}}+...+s_{n}\omega_{\zeta_{n}})\right)\left(s_{1}b(s_{1}\omega_{\zeta_{1}})\right)...\left(s_{n}b(s_{n}\omega_{\zeta_{n}})\right)\mathcal{T}_{\text{t}\rra\text{b}}^{(\zeta_{1},s_{1})....(\zeta_{n},s_{n})}(\omega),\label{eq:Incoherent_n_order}
\end{multline}
where we have defined the $n$-phonon assisted transmission functions
\begin{multline}
\mathcal{T}_{\text{b}\rra\text{t}}^{(\zeta_{1},s_{1})....(\zeta_{n},s_{n})}(\omega)=\Tr\left[\bm{\Gamma}_{\text{b}}(\omega)\cdot\bm{G}^{R}(\omega)\cdot\bm{M}_{\zeta_{1}}\cdot\bm{G}^{R}(\omega+s_{1}\omega_{\zeta_{1}})...\cdot\bm{M}_{\zeta_{n}}\cdot\bm{G}^{R}(\omega+s_{1}\omega_{\zeta_{1}}+...+s_{n}\omega_{\zeta_{n}})\right.\\
\left.\bm{\Gamma}_{\text{t}}(\omega+s_{1}\omega_{\zeta_{1}}+...+s_{n}\omega_{\zeta_{n}})\cdot\bm{G}^{A}(\omega+s_{1}\omega_{\zeta_{1}}+...+s_{n}\omega_{\zeta_{n}})\cdot\bm{M}_{\zeta_{n}}^{\dagger}\cdot...\cdot\bm{G}^{A}(\omega+s_{1}\omega_{\zeta_{1}})\cdot\bm{M}_{\zeta_{1}}^{\dagger}\bm{G}^{A}(\omega)\right],\label{eq:Transmission_forward_n_ph}
\end{multline}
\begin{multline}
\mathcal{T}_{\text{t}\rra\text{b}}^{(\zeta_{1},s_{1})....(\zeta_{n},s_{n})}(\omega)=\Tr\left[\bm{\Gamma}_{\text{t}}(\omega)\cdot\bm{G}^{A}(\omega)\cdot\bm{M}_{\zeta_{1}}^{\dagger}\cdot\bm{G}^{A}(\omega+s_{1}\omega_{\zeta_{1}})\cdot...\cdot\bm{M}_{\zeta_{n}}^{\dagger}\bm{G}^{A}(\omega+s_{1}\omega_{\zeta_{1}}+...+s_{n}\omega_{\zeta_{n}}).\right.\\
\left.\bm{\Gamma}_{\text{b}}(\omega+s_{1}\omega_{\zeta_{1}}+...+s_{n}\omega_{\zeta_{n}})\cdot\bm{G}^{R}(\omega+s_{1}\omega_{\zeta_{1}}+...+s_{n}\omega_{\zeta_{n}})\cdot\bm{M}_{\zeta_{n}}\cdot...\cdot\bm{G}^{R}(\omega+s_{1}\omega_{\zeta_{1}})\cdot\bm{M}_{\zeta_{1}}\cdot\bm{G}^{R}(\omega)\right].\label{eq:Transmission_backward_n_ph}
\end{multline}
\end{widetext} Notice that with respect to Eq.~(\ref{eq:Incoherent_lowest_order}),
we have made a change of $s_{i}\rra-s_{i}$ in the first line and
made a shift in the frequency variable $\omega\rra\omega+s_{1}\omega_{\zeta_{1}}+...+s_{n}\omega_{\zeta_{n}}$in
the second line of Eq.~(\ref{eq:Incoherent_n_order}). Eqs.~(\ref{eq:Incoherent_n_order}),
(\ref{eq:Transmission_forward_n_ph}) and (\ref{eq:Transmission_backward_n_ph})
have a very simple interpretation. The first/second line of Eq.~(\ref{eq:Incoherent_n_order})
can be understood has the probability of an electron being injected
from the bottom/top contact being collected by the top/bottom contact,
while being scattering by $n$ phonons during the contact to contact
trip, with $s_{i}=\pm1$ representing a phonon absorption/emission
process. We will now use this general formalism to study the effect
of phonon scattering in vertical transport in a graphene-hBN-graphene
device. We will analyze separately the effect of scattering by graphene
and hBN phonons.
\begin{figure}
\begin{centering}
\includegraphics[width=6cm]{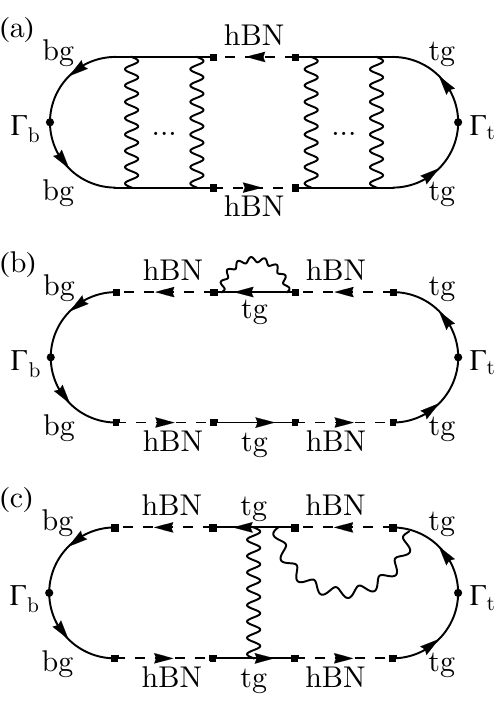}
\par\end{centering}

\protect\caption{\label{fig:Feynman_diag}Diagrammatic representation of contributions
to the current involving phonon scattering. The dots represent the
level width functions due to the bottom and top external metallic
contacts, the squares represent the graphene-hBN coupling, solid lines
represent graphene electronic propagators and dashed lines represent
hBN propagators. The wiggly lines represent phonon propagators. (a)
Ladder diagrams that are resumed in Eq.~(\ref{eq:Current_lowestorder}).
To lowest order in the graphene-hBN coupling these are all the contributions
due to electron-phonon interaction in the graphene layers. (b) Diagram
contributing to the current in higher order in the graphene-hBN coupling,
including the renormalization of the top graphene layer Green's function
by phonons. This kind of diagram can be captured in Eq.~(\ref{eq:Current_lowestorder}),
provided the effect of coupling to the graphene layers is included
into $\bm{G}_{\text{hBN}}$. (c) Higher order diagrams in the graphene-hBN
coupling, including electron-phonon interaction in the graphene layers,
that is not included in Eq.~(\ref{eq:Current_lowestorder}) and more
generically cannot be captured when evaluating the current following
approach (A).}
\end{figure}

\subsubsection{Scattering by phonons in the graphene layers\label{sub:Scattering-graphene}}

We know return to the issue of the consequences of considering graphene
as part of the external contacts or part of the central mesoscopic
region. We will first study the effect of multiple scatterings of
electrons in the graphene layers by phonons (or impurities). We will
first focus on scattering by phonons in the top graphene layer, with
scattering in the bottom layer being treated in the same way. Using
Eq.~(\ref{eq:Transmission_forward_n_ph}), the tunneling amplitude
assisted by $n$ phonon scattering events in the the top graphene
layer can be written to lowest order in the graphene-hBN coupling
as \begin{widetext}
\begin{multline}
\mathcal{T}_{\text{b}\rra\text{t}}^{(\zeta_{1},s_{1})....(\zeta_{n},s_{n})}(\omega)\simeq\Tr\left[\bm{\mathcal{T}}_{\text{tg,bg}}(\omega)\cdot\bm{\Gamma}_{\text{b}}(\omega)\cdot\bm{G}_{\text{bg}}^{R}(\omega)\cdot\bm{\mathcal{T}}_{\text{bg,tg}}(\omega)\cdot\right.\\
\cdot\bm{G}_{\text{tg}}^{R}(\omega)\cdot\bm{M}_{\zeta_{1}}\cdot\bm{G}_{\text{tg}}^{R}(\omega+s_{1}\omega_{\zeta_{1}})...\cdot\bm{M}_{\zeta_{n}}\cdot\bm{G}_{\text{tg}}^{R}(\omega+s_{1}\omega_{\zeta_{1}}+...+s_{n}\omega_{\zeta_{n}})\cdot\bm{\Gamma}_{\text{t}}(\omega+s_{1}\omega_{\zeta_{1}}+...+s_{n}\omega_{\zeta_{n}})\cdot\\
\left.\cdot\bm{G}_{\text{tg}}^{A}(\omega+s_{1}\omega_{\zeta_{1}}+...+s_{n}\omega_{\zeta_{n}})\cdot\bm{M}_{\zeta_{n}}^{\dagger}\cdot...\cdot\bm{G}_{\text{tg}}^{A}(\omega+s_{1}\omega_{\zeta_{1}})\cdot\bm{M}_{\zeta_{1}}^{\dagger}\bm{G}_{\text{tg}}^{A}(\omega)\right],\label{eq:Incoherent_graphene_ph}
\end{multline}
\end{widetext} and similarly for $\mathcal{T}_{\text{t}\rra\text{b}}^{(\zeta_{1},s_{1})....(\zeta_{n},s_{n})}(\omega)$.
These contributions correspond to multiple scatterings of an electron
before leaving the top graphene layer. Summing up all the contributions
of the form of Eq.~(\ref{eq:Incoherent_graphene_ph}), together with
the contribution from the coherent current, we obtain
\begin{align}
I_{\text{b}\rra\text{t}}= & \sum_{n=0}^{\infty}I_{\text{b}\rra\text{t}}^{\text{incoh, }n-\text{ph}}\nonumber \\
= & \frac{e}{\hbar}\sum_{n=0}^{\infty}\int\frac{d\omega}{2\pi}f_{\text{b}}(\omega)\left(1-f_{\text{t}}\left(\omega\right)\right)\times\nonumber \\
\times & \Tr\left[\bm{A}_{\text{bg}}(\omega)\cdot\bm{\mathcal{T}}_{\text{bg,tg}}(\omega)\cdot\bm{A}_{\text{tg}}^{(n)}(\omega)\cdot\bm{\mathcal{T}}_{\text{tg,bg}}(\omega)\right]\nonumber \\
- & \frac{e}{\hbar}\sum_{n=0}^{\infty}\int\frac{d\omega}{2\pi}f_{\text{t}}\left(\omega\right)\left(1-f_{\text{b}}(\omega)\right)\times\nonumber \\
\times & \Tr\left[\bm{A}_{\text{tg}}^{(n)}(\omega)\cdot\bm{\mathcal{T}}_{\text{tg,bg}}(\omega)\cdot\bm{A}_{\text{bg}}(\omega)\cdot\bm{\mathcal{T}}_{\text{bg,tg}}(\omega)\right]
\end{align}
where we have written $\bm{A}_{\text{bg}}(\omega)=\bm{G}_{\text{bg}}^{A}(\omega)\cdot\bm{\Gamma}_{\text{b}}(\omega)\cdot\bm{G}_{\text{bg}}^{R}(\omega)$,
since we are considering only scattering in the top graphene layer.
It can be checked that the different terms $\bm{A}_{\text{tg}}^{(n)}(\omega)$
obey the following recursion relation
\begin{align}
\bm{A}_{\text{tg}}^{(0)}(\omega) & =\bm{G}^{R}(\omega)\cdot\bm{\Gamma}_{\text{t}}(\omega)\cdot\bm{G}^{A}(\omega),\, n=0\label{eq:spectral_iteraction_0}\\
\bm{A}_{\text{tg}}^{(n)}(\omega) & =\sum_{s_{n},\zeta_{n}}\frac{\left[1-f_{\text{t}}\left(\omega-s_{n}\omega_{\zeta_{n}}\right)\right]\left[-s_{n}b(-s_{n}\omega_{\zeta_{n}})\right]}{1-f_{\text{t}}\left(\omega\right)}\times\nonumber \\
\times & \bm{G}_{\text{tg}}^{R}(\omega)\cdot\bm{M}_{\zeta_{n}}\cdot\bm{A}_{\text{tg}}^{(n-1)}(\omega-s_{n}\omega_{\zeta_{n}})\cdot\nonumber \\
 & \cdot\bm{M}_{\zeta_{n}}^{\dagger}\cdot\bm{G}_{\text{tg}}^{A}(\omega),\, n>0.\label{eq:spectral_iteration_n}
\end{align}
This can be compared with the spectral function of the top graphene
layer. Assuming that the top graphene layer is in near equilibrium
with the top contact, then the spectral function can be written as
\begin{equation}
\bm{A}_{\text{tg}}(\omega)\simeq\bm{G}^{R}(\omega)\cdot\left(\bm{\Gamma}_{\text{t}}(\omega)+\bm{\Gamma}_{\text{e-ph,tg}}(\omega)\right)\cdot\bm{G}^{A}(\omega),\label{eq:spectral_near_eq}
\end{equation}
where, under the approximation that the top graphene is in equilibrium
with the top contact, the decay rate due to electron-phonon interaction
can be written as 
\begin{eqnarray}
\bm{\Gamma}_{\text{e-ph,tg}}(\omega) & \simeq & \sum_{s,\zeta}s\left(1-f_{\text{t}}(\omega-s\omega_{\zeta})+b(s\omega_{\zeta})\right)\times\nonumber \\
 & \times & \bm{M}_{\zeta}\cdot\bm{A}_{\text{tg}}(\omega-s\omega_{\zeta})\cdot\bm{M}_{\zeta}^{\dagger}.\label{eq:decay_rate_near_eq}
\end{eqnarray}
It is easy to check that the equilibrium occupation functions satisfy
the equality
\begin{multline}
\frac{\left[1-f\left(\omega-s\omega_{\zeta}\right)\right]\left[-sb(-s\omega_{\zeta})\right]}{1-f\left(\omega\right)}=\\
=s\left(1-f(\omega-s\omega_{\zeta})+b(s\omega_{\zeta})\right).
\end{multline}
Therefore, by inserting Eq.~(\ref{eq:decay_rate_near_eq}) into Eq.~(\ref{eq:spectral_near_eq})
and iterating the equation, we obtain
\begin{equation}
\bm{A}_{\text{tg}}(\omega)\simeq\sum_{n=0}^{\infty}\bm{A}_{\text{tg}}^{(n)}(\omega),
\end{equation}
with the different terms $\bm{A}_{\text{tg}}^{(n)}(\omega)$ coincide
with Eqs.~(\ref{eq:spectral_iteraction_0}) and (\ref{eq:spectral_iteration_n}),
and the $\simeq$ means we are making the approximation that the top
graphene layer is in near equilibrium with the top contact. Therefore,
the sum of all incoherent scattering processes occurring before the
electron leaves the graphene layer and the coherent contribution reproduces
the spectral function of graphene taking into account electron-phonon
interaction / disorder. The same is true for scattering in the bottom
graphene layer. Notice that in Eqs.~(\ref{eq:Transmission_forward_n_ph})
and (\ref{eq:Transmission_backward_n_ph}) retarded/advanced Green's
functions appear to the right/left of $\bm{\Gamma}_{\text{b}}$. Nevertheless,
by using Eq.~(\ref{eq:spectral_RA_AR}), the previous calculation
can also be applied for scattering in the bottom graphene layer. We
have thus arrived to an important conclusion: the expression
\begin{eqnarray}
I_{\text{b}\rra\text{t}} & = & \frac{e}{\hbar}\int\frac{d\omega}{2\pi}\left(f_{\text{b}}(\omega)-f_{\text{t}}(\omega)\right)\times\nonumber \\
 & \times & \Tr\left[\mathcal{\bm{T}}_{\text{bg},\text{tg}}(\omega)\cdot\bm{A}_{\text{tg}}(\omega)\cdot\mathcal{\bm{T}}_{\text{tg},\text{bg}}\cdot\bm{A}_{\text{bg}}(\omega)\right],\label{eq:Current_lowestorder}
\end{eqnarray}
which would be the one obtained if we employed approach (A), actually
already includes the effect of multiple non-coherent scattering processes
in the graphene layers, provided $\bm{A}_{\text{tg}/\text{bg}}(\omega)$
are replaced with the respective expressions in the presence of phonon/disorder
scattering. We also point out that in the case of elastic scattering
due to disorder in the graphene layers, the result from Eq.~(\ref{eq:Current_lowestorder})
can be obtained by performing disorder averages of Eq.~(\ref{eq:transmission_B}),
see Appendix~\ref{sec:Vertex-corrections}. To lowest order in the
graphene-hBN coupling, Eq.~(\ref{eq:Current_lowestorder}) actually
includes all the possible scattering processes of an electron in the
graphene layers. Including the effects of graphene into the Green's
function of hBN that appears in $\mathcal{\bm{T}}_{\text{bg},\text{tg}}(\omega)$,
Eq.~(\ref{eq:Current_lowestorder}) includes only a subclass of all
possible contributions due to electron-phonon interaction, see Fig.~\ref{fig:Feynman_diag}.
Therefore, we conclude that approaches (A) and (B) coincide to lowest
order in the graphene-hBN coupling and to higher order in this coupling,
approach (A) can correctly capture a class of all the possible electron
phonon scatterings.

\subsubsection{Scattering by phonons in the hBN slab}

We will now discuss the effects of scattering by phonons/disorder
in the hBN slab. We will restrict ourselves to the case of tunneling
assisted by one phonon. We write the electron phonon interaction in
a Bloch state basis as
\begin{equation}
H_{\text{e-ph,hBN}}=\frac{1}{\sqrt{N}}\sum_{\bm{k},\bm{q}}\bm{c}_{\bm{k}+\bm{q},\text{hBN}}^{\dagger}\cdot\bm{M}_{\bm{q},\zeta}\cdot\bm{c}_{\bm{k},\text{hBN}}\phi_{\bm{q},\zeta},\label{eq:hBN_electron_phonon}
\end{equation}
where $\bm{c}_{\bm{k},\text{hBN}}^{\dagger}=\left[c_{\bm{k},\text{B},1,\text{hBN}}^{\dagger},\, c_{\bm{k},\text{N},1,\text{hBN}}^{\dagger},\,...,\, c_{\bm{k},\text{B},\mathcal{N},\text{hBN}}^{\dagger},\, c_{\bm{k},\text{N},\mathcal{N},\text{hBN}}^{\dagger}\right]$
and $\phi_{\bm{q},\zeta}=\left(a_{\bm{q},\zeta}+a_{-\bm{q},\zeta}^{\dagger}\right)/\sqrt{2}$
is the phonon field operator and $N$ is the number of unit cells
in the hBN slab. For small rotation angles between the different layers
and assuming only scattering by phonons close to the $\Gamma$ or
$K$ points of hBN, such that only states close to the Dirac points
of each layer are involved, using Eq.~\ref{eq:Incoherent_n_order},
we can write the 1-phonon assisted tunneling current to lowest order
in the graphene-hBN coupling as\begin{widetext} 
\begin{align*}
I_{\text{b}\rra\text{t}}^{\text{incoh, }1\text{-ph}} & =\frac{e}{\hbar N}\sum_{\substack{\bm{k},\lambda,\lambda^{\prime}\\
n,m
}
}\sum_{\bm{q},\zeta,s}\int\frac{d\omega}{2\pi}f_{\text{b}}(\omega)\left(1-f_{\text{t}}\left(\omega+s\omega_{\zeta}\right)\right)\left(sb(s\omega_{\zeta})\right)\left|\mathcal{T}_{n,m,\bm{k},\bm{q}}^{(\zeta,s)1\text{-ph}}(\omega)\right|^{2}\times\\
 & \times\Upsilon_{\bm{k},\lambda}^{\text{bg},n}\Upsilon_{\bm{k}+\bm{\mathcal{Q}}_{n,m}-\bm{q},\lambda^{\prime}}^{\text{tg},m}A_{\text{bg},\bm{k},\lambda}(\omega_{\text{bg}})A_{\text{tg},\bm{k}+\bm{\mathcal{Q}}_{n,m}-\bm{q},\lambda^{\prime}}(\omega_{\text{tg}}+s\omega_{\zeta})\\
 & -\frac{e}{\hbar N}\sum_{\substack{\bm{k},\lambda,\lambda^{\prime}\\
n,m
}
}\sum_{\bm{q},\zeta,s}\int\frac{d\omega}{2\pi}f_{\text{t}}\left(\omega+s\omega_{\zeta}\right)\left(1-f_{\text{b}}(\omega)\right)\left(-sb(-s\omega_{\zeta})\right)\left|\mathcal{T}_{n,m,\bm{k},\bm{q}}^{(\zeta,s)1\text{-ph}}(\omega)\right|^{2}\times\\
 & \times\Upsilon_{\bm{k},\lambda}^{\text{bg},n}\Upsilon_{\bm{k}+\bm{\mathcal{Q}}_{n,m}-\bm{q},\lambda^{\prime}}^{\text{tg},m}A_{\text{bg},\bm{k},\lambda}(\omega_{\text{bg}})A_{\text{tg},\bm{k}+\bm{\mathcal{Q}}_{n,m}-\bm{q},\lambda^{\prime}}(\omega_{\text{tg}}+s\omega_{\zeta}),
\end{align*}
where we have introduced the phonon assisted tunneling amplitude between
the graphene layers
\[
\mathcal{T}_{n,m,\bm{k},\bm{q}}^{(\zeta,s)1\text{-ph}}(\omega)=\frac{1}{2}\text{tr}\left\{ \hat{\bm{T}}^{\dagger}\cdot\bm{R}_{-p\frac{2\pi}{3}}^{m}\cdot\left[\bm{G}_{\text{hBN},\bm{k}+\bm{g}_{n}^{\text{bg}}-\bm{q}}^{A}(\omega+s\omega_{\zeta})\cdot\bm{M}_{\bm{q},\zeta}^{\dagger}\cdot\bm{G}_{\text{hBN},\bm{k}+\bm{g}_{n}^{\text{bg}}}^{A}(\omega)\right]_{\mathcal{N},1}\cdot\bm{R}_{\frac{2\pi}{3}}^{n}\cdot\hat{\bm{T}}\right\} .
\]
Neglecting the momentum and frequency dependence of $\bm{G}_{\text{hBN}}^{A}$and
assuming dispersionless phonons, one can make a shift in the momentum
variable $\bm{q}\rra\bm{k}-\bm{k}^{\prime}+\bm{\mathcal{Q}}_{n,m}$,
such that the summation over $\bm{k}$and $\bm{k}^{\prime}$ factors
and we can write
\begin{align}
I_{\text{b}\rra\text{t}}^{\text{incoh, }1\text{-ph}} & =AA_{\text{cell}}g_{s}g_{v}\frac{e}{\hbar}\sum_{n,m}\sum_{\zeta,s}\int\frac{d\omega}{2\pi}\left[f_{\text{b}}(\omega)\left(1-f_{\text{t}}(\omega+s\omega_{\zeta})\right)sb(s\omega_{\zeta})-f_{\text{t}}(\omega+s\omega_{\zeta})\left(1-f_{\text{b}}(\omega)\right)s\left(1+b(\omega_{\zeta})\right)\right]\times\nonumber \\
 & \times\left|\mathcal{T}_{n,m}^{(\zeta)1\text{-ph}}\right|^{2}\text{DoS}_{\text{bg}}(\omega_{\text{bg}})\text{DoS}_{\text{tg}}(\omega_{\text{tg}}+s\omega_{\zeta}),\label{eq:Incoherent_memory_loss}
\end{align}
\end{widetext}, where $A_{\text{cell}}$ is the area of the unit
cell of hBN and graphene's density of states per spin and valley is
given by
\begin{equation}
\text{DoS}(\omega)=\frac{1}{V}\sum_{\bm{k},\lambda}A_{\bm{k},\lambda}(\omega)=\frac{\left|\omega\right|}{\left(v_{F}\hbar\right)^{2}},
\end{equation}
where the last equality is valid for for non-interacting electrons
in pristine graphene. A similar expression to Eq.~(\ref{eq:Incoherent_memory_loss}),
which included only processes involving spontaneous emission of phonons
(equivalent to assuming that the phonons are at zero temperature),
was recently presented without derivation and used in Ref.~\onlinecite{Vdovin2015}
to model vertical current in graphene-hBN-graphene devices. In the
case of elastic scattering by disorder with short range correlation,
Eq.~(\ref{eq:Incoherent_memory_loss}) becomes, 
\begin{eqnarray}
I_{\text{b}\rra\text{t}}^{\text{incoh, }1\text{-dis}} & = & AA_{\text{cell}}g_{s}g_{v}\frac{e}{\hbar}\sum_{\substack{\zeta,s\\
n,m
}
}\int\frac{d\omega}{2\pi}\left[f_{\text{b}}(\omega)-f_{\text{t}}(\omega)\right]\times\nonumber \\
 & \times & \left|\mathcal{T}_{n,m}^{1\text{-dis}}\right|^{2}\text{DoS}_{\text{bg}}(\omega_{\text{bg}})\text{DoS}_{\text{tg}}(\omega_{\text{tg}}),\label{eq:Incoherent_elastic}
\end{eqnarray}
with $\mathcal{T}_{n,m}^{1\text{-dis}}$ a disorder assisted tunneling
amplitude. Although an expression of the form of Eq.~\ref{eq:Incoherent_elastic}
was previously used to model vertical current in graphene-hBN-graphene
devices \cite{BGJ12,Britnell2012b}, we emphasize that Eq.~\ref{eq:Incoherent_elastic}
only describes processes where there is a complete degradation of
in-plane momentum conservation, something that has been previously
pointed out in Refs.~\onlinecite{B__14,Barrera2015}. The complete
degradation of momentum conservation only occurs for scattering by
dispersionless phonons or for disorder with short distance correlation.

As an example we consider, scattering by optical out-of-plane breathing
modes close to the $\Gamma$ point, with non-zero components of polarization
vector given by
\begin{multline}
\bm{\xi}_{\text{ZB},a,\ell}^{z}=\left(\xi_{\zeta,\text{B},1}^{z},\,\xi_{\zeta,\text{N},1}^{z}\,,\xi_{\zeta,\text{B},2}^{z},\xi_{\zeta,\text{N},2}^{z},\,....\right)\\
=\sqrt{\frac{\mu_{\text{BN}}}{\mathcal{N}}}\left(\frac{1}{\sqrt{m_{\text{B}}}},\,\frac{1}{\sqrt{m_{\text{N}}}},\,\frac{-1}{\sqrt{m_{\text{B}}}},\,\frac{-1}{\sqrt{m_{\text{N}}}},\,....\right),
\end{multline}
where $\mu_{\text{BN}}^{-1}=m_{\text{B}}^{-1}+m_{\text{N}}^{-1}$
is the reduced mass of the hBN phonon mode. We assume that electron-phonon
coupling for this mode can be described as a local change in the value
of the interlayer hoping parameter in Hamiltonian \ref{eq:H_hBN_hamiltonian}.
Considering electrons due close to the $K$ point and phonons close
to the $\Gamma$ point, we derive an electron-phonon Hamiltonian of
the form of Eq.~\ref{eq:hBN_electron_phonon}, with a momentum independent
coupling constant which reads
\begin{equation}
\bm{M}_{\text{ZB}}^{\text{hBN}}=\frac{g_{\text{ZB}}^{\text{hBN}}}{\sqrt{\mathcal{N}}}\left[\begin{array}{cccc}
\bm{0} & \bm{\sigma}_{x}\\
\bm{\sigma}_{x} & \bm{0} & -\bm{\sigma}_{x}\\
 & -\bm{\sigma}_{x} & \bm{0} & \ddots\\
 &  & \ddots & \ddots
\end{array}\right],
\end{equation}
with the electron-phonon coupling constant given by
\begin{equation}
g_{\text{ZB}}^{\text{hBN}}=-\frac{\pd\log t_{\perp}}{\pd\log c_{\text{BN}}}\frac{t_{\perp}}{c_{\text{BN}}}\sqrt{\frac{\hbar}{\mu_{\text{BN}}\omega_{\text{ZB}}^{\text{hBN}}}},
\end{equation}
where $-\pd\log t_{\perp}/\pd\log c_{\text{BN}}\simeq3$ describes
the change of the interlayer hopping, $t$, with the interlayer distance,
$c_{\text{BN}}$, and $\omega_{\text{ZB}}^{\text{hBN}}$ is the out-of-plane
breathing phonon frequency: For this electron-phonon interaction we
obtain to lowest order in $t_{\perp}$ and neglecting the frequency
and momentum dependence
\[
\left|\mathcal{T}_{n,m}^{(\text{ZB})1\text{-ph}}\right|^{2}\simeq\frac{(\mathcal{N}-1)^{2}}{\mathcal{N}}\left|\frac{g_{\text{ZB}}^{\text{hBN}}}{t_{\perp}}\right|^{2}\left|\mathcal{T}_{n,m}\right|^{2},
\]
with $\left|\mathcal{T}_{n,m}\right|^{2}$ given by Eq.~(\ref{eq:Coherent_probability}).

\subsection{Results }

\begin{figure}
\begin{centering}
\includegraphics[width=8cm]{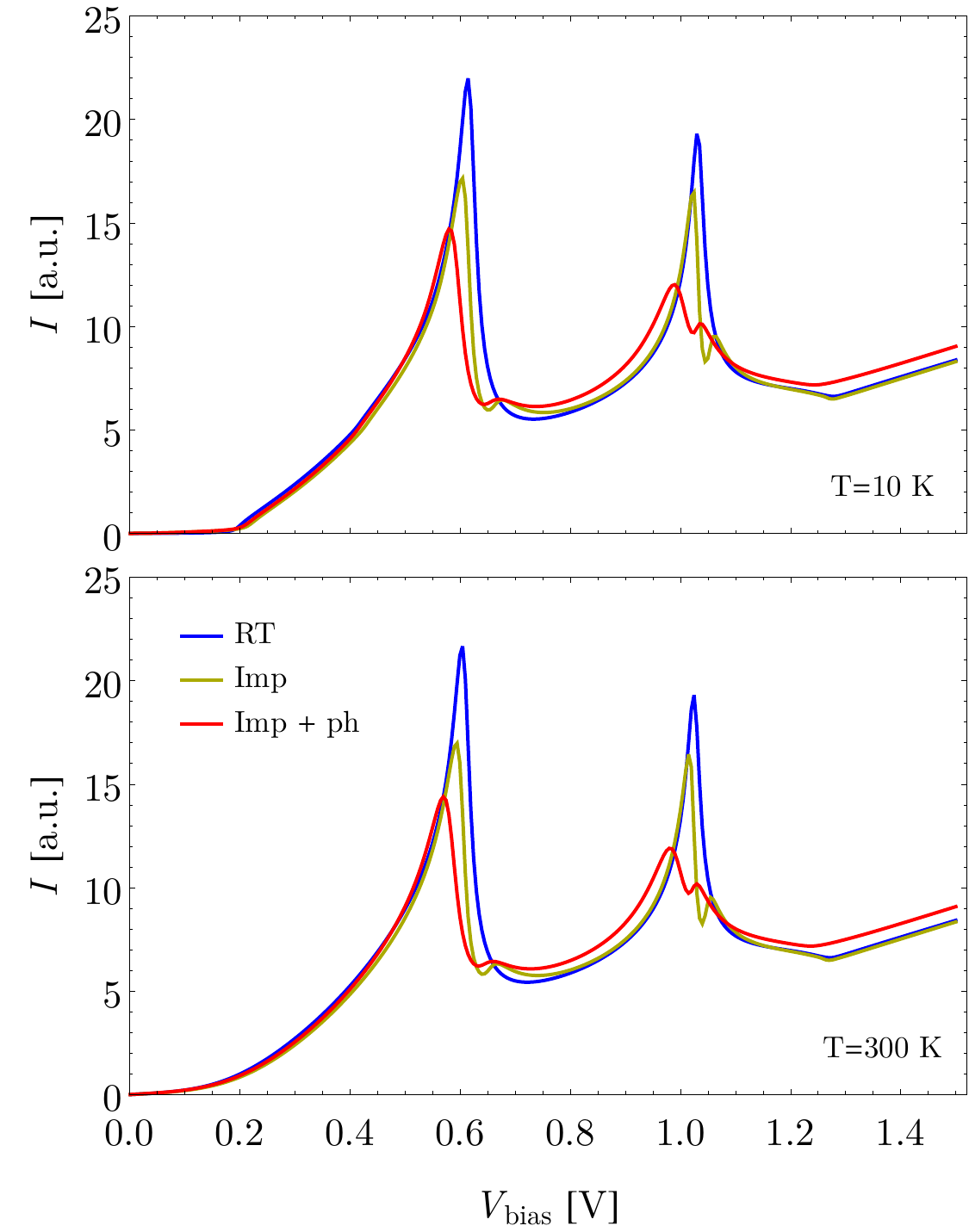}
\par\end{centering}

\protect\caption{\label{fig:I-V-scattering}I-V curves at constant $V_{\text{gate}}=0$
in a graphene-hBN-graphene device with rotations angles $\theta_{\text{tg}}=1^{\circ}$
and $\theta_{\text{hBN}}=1.5^{\circ}$, considering different sources
of scattering in the graphene layers: (RT) constant relaxation time
of $\gamma=3$ meV; (Imp) scattering by resonant scatterers treated
within the SCBA with an impurity concentration of $n_{\text{imp}}=10^{-4}$
impurities per graphene unit cell; (Imp+RT) scattering by resonant
scatterers and graphene in-plane optical phonons also with $n_{\text{imp}}=10^{-4}$. }
\end{figure}

\begin{figure}
\begin{centering}
\includegraphics[width=8cm]{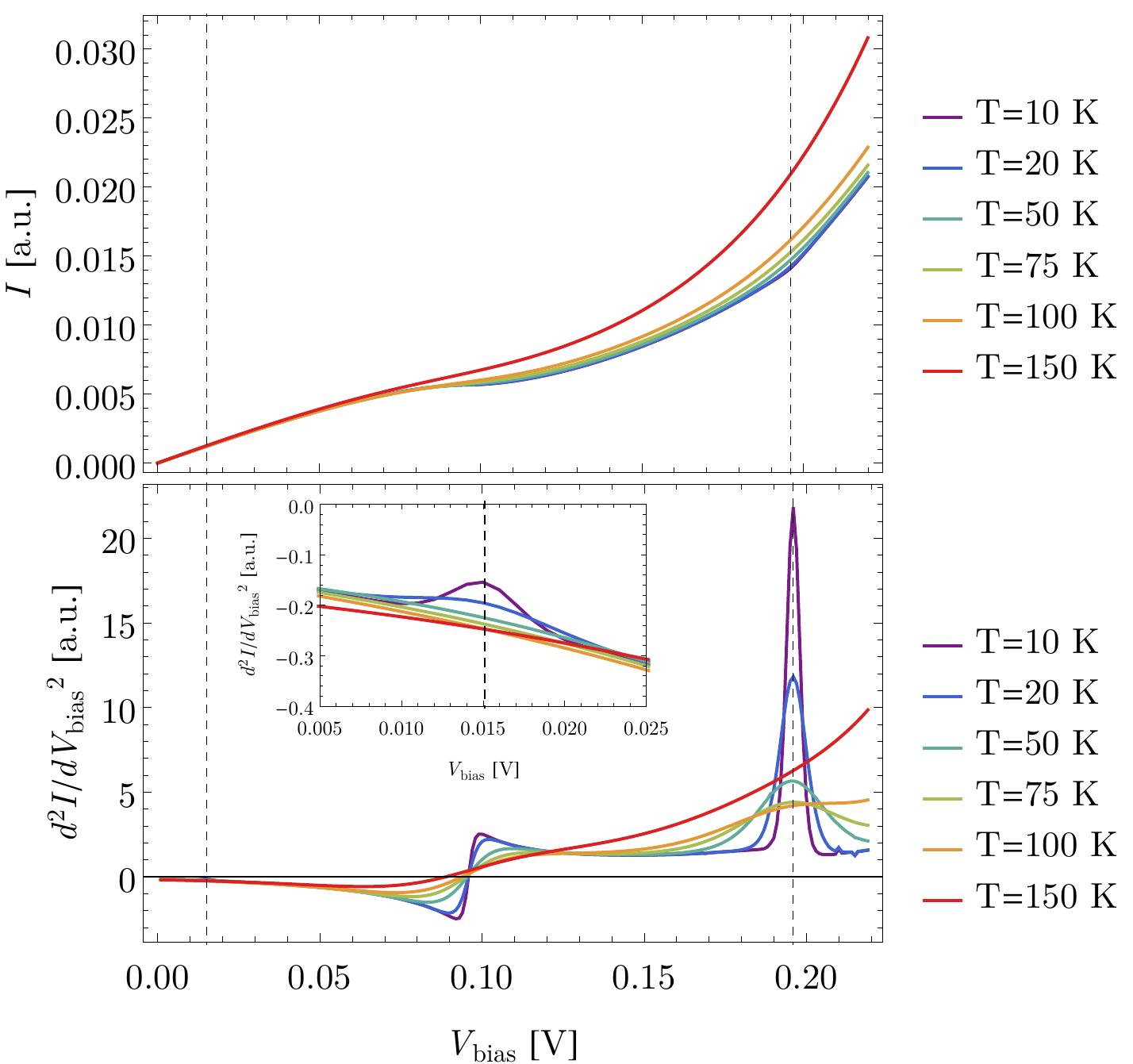}
\par\end{centering}

\protect\caption{\label{fig:I-V-assisted}I-V curve and $d^{2}I/dV_{\text{bias}}^{2}$
as a function of bias voltages at a constant $V_{\text{gate}}=10$
V for different temperatures and for rotation angles $\theta_{\text{tg}}=2^{\circ}$
and $\theta_{\text{hBN}}=3^{\circ}$, including effects of scattering
by out-of-plane breathing phonons of hBN, $\omega_{\text{ZB}}^{\text{hBN}}=15$
meV\cite{Michel2011}, and of the in-plane graphene phonons, $\omega_{\Gamma\text{O}}^{\text{g}}=196$
meV\cite{Wirtz_2004} (represented by the vertical dashed lines).
Processes involving spontaneous emission of phonons open up new tunneling
channels that appear as peaks in $d^{2}I/dV_{\text{bias}}^{2}$ at
low temperature. The inset zooms in the small peak due to the hBN
out-of-planes breathing phonon. We point out that the feature that
occurs around $V_{\text{bias}}\sim0.1$ V is not due to phonons, but
due to the tunneling density of states structure.}

\end{figure}

In Fig.~\ref{fig:I-V-scattering} we show the vertical current as
a function of bias voltage taking into account the effect of scattering
of graphene electrons by resonant scatterers (treated within the SCBA,
see Appendix~\ref{sec:Ressonant-impurties-SCBA}) and in-plane graphene
electrons (see Appendix~\ref{sec:Graphene-electron-phonon}). For
comparison we also show current computed used a constant relaxation
time. The main difference between modeling electron scattering with
a constant relaxation rate or considering scattering by resonant scatters,
is that for resonant scatters the electron decay rate has a strong
dependence in energy, behaving as $\omega^{-1}$. Therefore, for higher
bias voltages (when the graphene Fermi levels are higher), the electron
lifetime is larger. This is manifest in Fig.~\ref{fig:I-V-scattering},
where it is seen that when assuming a constant relaxation rate the
second peak in the I-V current is considerably smaller than the first
one, while for resonant scatterers both peaks are roughly the same
height. Inclusion of phonons, makes again the peak at higher bias
voltage smaller due to the fact that the decay rate due to scattering
with graphene in-plane optical phonons increases with frequency. Also
notice that inclusion of resonant disorder and phonons leads to a
small splitting of the peaks in the I-V current. This splitting is
due to real part of the self-energy due to both resonant scatterers
and phonons. Apart from increasing graphene electron's decay rate
and as such providing an additional broadening of peaks in the I-V
current, phonons do not play a relevant role for the high bias I-V
characteristics of a graphene-hBN-graphene device. This changes if
one focus on small bias. At very low temperature, the spontaneous
emission of optical phonons becomes possible whenever $V_{\text{bias}}>\omega_{O\text{ph}}$,
where $\omega_{O\text{ph}}$ is the optical phonon frequency, opening
up new tunneling channels for electrons. Although for small electron-phonon
coupling, this phonon assisted contribution to the current is small
the opening up of a new tunneling channel can be observed in the derivatives
of the current with respect to the bias, as can be seen in Fig.~\ref{fig:I-V-assisted}.
The features in $d^{2}I/dV_{\text{bias}}^{2}$ are only significant
at low temperature, being smoothed out at higher temperatures due
to the smearing of the Fermi occupation factors in graphene. We point
out however, that the features due to phonons are a small contribution
can be overridden due to features in the coherent current induced
by the rotation between different layers (shown in Fig.~\ref{fig:Density-plot}),
even if we treat the phonons as dispersionless leading to a complete
degradation of electron momentum conservation. We also note in passing,
that tunneling assisted by emission of multiple phonons is also possible
(see Eqs.~(\ref{eq:Incoherent_n_order})-(\ref{eq:Transmission_backward_n_ph}))
which would open up new scattering channels when $n\omega_{O\text{ph}}>V_{\text{bias}}$,
where $n$ is the number of phonons. These would lead to additional
peaks in $d^{2}I/dV_{\text{bias}}^{2}$ but would be instead suppressed
by higher powers of the electron-phonon coupling.

\section{Conclusions\label{sec:Conclusions}}

This works provides another example of the extreme sensitivity of
the properties of vdW structures to the rotational alignment of the
different constitutive layers. We have seen how this additional degree
of freedom can be exploited in order to create devices displaying
multiple regions of negative differential conductance. The development
of devices that display multiple NDC regions is relevant for the development
of multivalued logic devices \cite{Sen_1987,Sen_1988}, which showcases
another possible application of vdW structures. We have studied in
detail the effect of the rotational alignment between the boron nitride
slab and the graphene layers in the vertical current of a graphene-hBN-graphene
vdW structure for small rotational misalignment, which have so far
not been observed\cite{MTC14}. We have seen now the transference
of momentum by the hBN crystalline structure to the tunneling electrons
gives origin to additional peaks in the I-V characteristics of this
device, followed by regions of negative differential conductance.
These additional peaks are however extremely sensitive to the rotation
angle between the graphene layers and the hBN slab, and rotational
angles as small as $3^{\circ}$ can already push these additional
peaks to bias voltages higher than $1.5$ V. Therefore, the observation
of multiple NDC in graphene-hBN-graphene devices requires a control
of the rotational angle between the different layers with a precision
of $\lesssim1^{\circ}$, something which is within experimental reach
\cite{Ponomarenko_2013,Woods_2014,MTC14}. We expect that the possible
reconstruction of graphene spectrum due to the periodic potential
induced by hBN for small rotational angles should not affect in a
qualitative way the occurrence of multiple NDC regions in graphene-hBN-graphene
devices, provided the applied bias voltage is much larger than the
width of the region where the spectrum reconstruction is significant.
However, a more quantitative treatment of these effects is required.

We have also analyzed the effect of treating graphene as being the
source and drain contacts of the graphene-hBN-graphene device, or
by treating them as part of the device and taking the source and drain
as being external metallic contacts. We have seen that, provided the
metallic contacts do not significantly spoil translation invariance
of graphene (as expected if the contact is deposited only over a small
region of the graphene layer), and in the non-interacting case, both
approaches are equivalent. In the presence of interactions both approaches
are equivalent to lowest order in the graphene-hBN coupling.

Finally, we have studied, in a unified way, the effect of scattering
by disorder and phonon scattering in the vertical current of graphene-hBN-graphene
devices. Starting from a NEGF formalism we derived the contribution
to the current due to phonon (or disorder) assisted tunneling processes.
We have seen now scattering by short range disorder or dispersionless
phonons leads to a complete degradation of electron momentum conservation
in the graphene-to-graphene tunneling process and how spontaneous
emission of phonons at lower temperature appear as sharp features
in the derivatives of the current with respect to the bias voltage
at the energy of the phonons. These features can however be hidden
by features due to the rotational alignment between the different
layers. We have focused on the effect of graphene in-plane optical
phonons and hBN optical out-of-plane breathing phonons. We have not
considered the effect of vibrations at the graphene-hBN interface,
as these would require the description of phonons in incommensurate
structures something which will be focus of future work. 
\begin{acknowledgments}
B. Amorim acknowledges financial support from Funda\c{c}\~{a}o para a Ci\^{e}ncia e a Tecnologia (Portugal), through 
Grant No. SFRH/BD/78987/2011. 
R.M. Ribeiro and  N.M.R. Peres acknowledge the financial support of EC under Graphene Flagship (Contract No. CNECT-ICT-604391). 
N. M. R. Peres acknowledges financial support from the FCT project EXPL-FIS-NAN-1728-2013.
\end{acknowledgments}

\appendix

\section{\label{sec:Capacitor}Thomas-Fermi modeling of electrostatic doping}

\begin{figure}
\begin{centering}
\includegraphics[width=8cm]{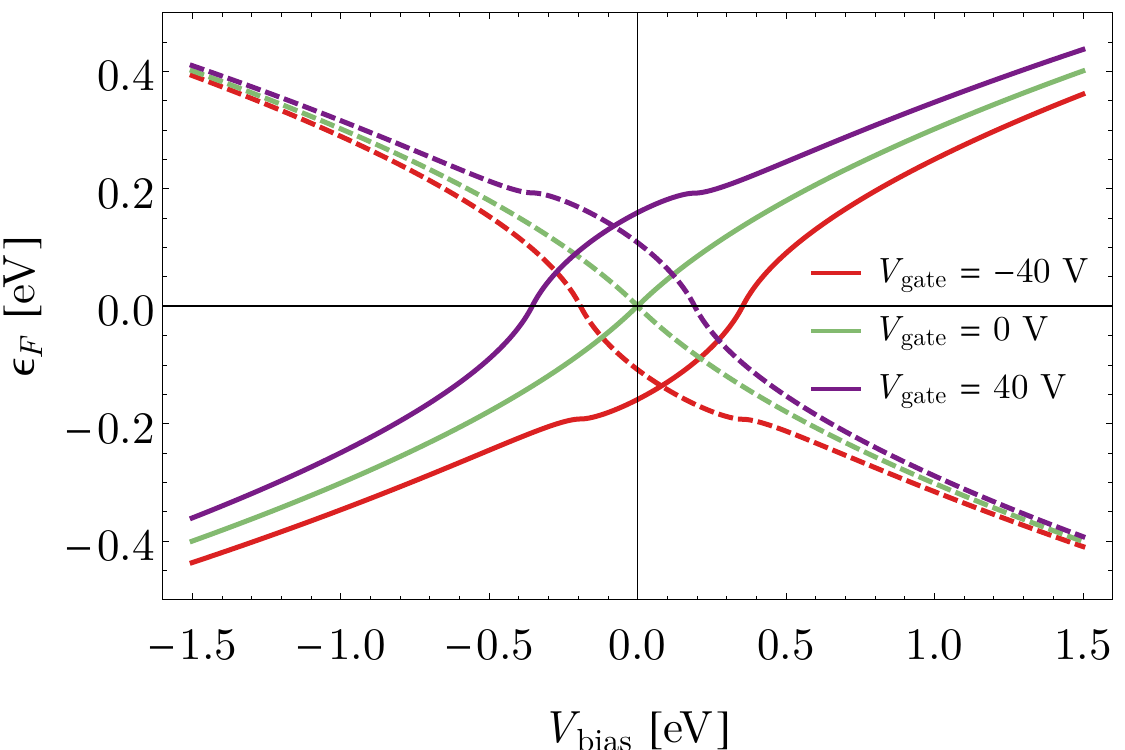}
\par\end{centering}

\protect\caption{\label{fig:gating}Computed Fermi levels for the bottom and top graphene
layers as a function of bias voltage for different gate voltages obtained
by solving Eqs.~(\ref{eq:gating_bg}) and (\ref{eq:gating_tg}).
We assume that the following parameters $d_{\text{SiO}_{2}}=285$
nm, $d_{\text{hBN}}=40$ nm for the thickness of the back gate dielectric,
with out-of-plane dielectric constants $\bar{\epsilon}_{\text{SiO}_{2}}=3.9$
and $\bar{\epsilon}_{\text{hBN}}=5.09$\cite{Geick_1966}. We assumed
that the distance between the two graphene layers are separated by
4 monolayers of hBN, which corresponds to a distance between the graphene
layers of $d\simeq1.6$ nm.}
\end{figure}

We wish to model the charging of an a graphene-hBN-graphene device
by application of a gate, $V_{\text{gate}}$, and bias, $V_{\text{bias}}$,
voltages. The graphene-hBN-graphene structure is formed by $\mathcal{N}$
hBN monolayers, sandwiched between two graphene layers. The graphene-hBN-graphene
structure is on top of a dielectric spacer (typically hBN/SiO$_{2}$)
separating the structure from a back gate, typically a highly doped
Si layer. We treat each layer forming the graphene-hBN-graphene structure
as a 2D film with a two dimensional charge density given by $\rho_{\ell}$,
$\ell=-1,...,\mathcal{N}+1$, where $\ell=-1$ indexes the Si layer,
$\ell=0$ and $\ell=\mathcal{N}+1$ are, respectively, the bottom
and top graphene layers and $\ell=1,...,\mathcal{N}$ index the layers
of hBN slab. Layers $\ell-1$ and $\ell$ are separated by a distance
$d_{\ell}$ and we assume that this is filled with a dielectric with
relative constant along the $z$ direction given by given by $\bar{\epsilon}_{\ell}$.
Applying Gauss's law around each plate, and assuming charge neutrality,
$\sum_{\ell=-1}^{\mathcal{N}}\rho_{\ell}=0$, we obtain
\begin{eqnarray}
\bar{\epsilon}_{0}E_{0} & = & \rho_{-1}/\epsilon_{0}.\\
\bar{\epsilon}_{\ell+1}E_{\ell+1}-\bar{\epsilon}_{\ell}E_{\ell} & = & \rho_{\ell}/\epsilon_{0},\,\ell=0,...,\mathcal{N},\\
-\bar{\epsilon}_{\mathcal{N}+1}E_{\mathcal{N}+1} & = & \rho_{\mathcal{N}+1}/\epsilon_{0},
\end{eqnarray}
where $E_{\ell}$ is the electric field along the $z$ direction,
between layers $\ell-1$ and $\ell$, and $\epsilon_{0}$ is vacuum\textquoteright s
permittivity. From these equations we can write
\begin{equation}
\bar{\epsilon}_{\ell}E_{\ell}=\frac{1}{\epsilon_{0}}\sum_{k=-1}^{\ell-1}\rho_{k},\,\ell=0,...,\mathcal{N}+1,
\end{equation}
and the stored electrostatic energy is given by
\begin{eqnarray}
U_{EM} & = & \sum_{\ell=0}^{\mathcal{N}+1}\frac{1}{2}\epsilon_{0}d_{\ell}\bar{\epsilon}_{\ell}E_{\ell}^{2}\nonumber \\
 & = & \frac{1}{2}\sum_{\ell,\ell^{\prime}=0}^{\mathcal{N}+1}\rho_{\ell}\left(\sum_{k=0}^{\min(\ell,\ell^{\prime})}\frac{d_{k}}{\epsilon_{0}\bar{\epsilon}_{k}}\right)\rho_{\ell^{\prime}}.
\end{eqnarray}
where we have used the charge neutrality condition in order to eliminate
the charge in the Si gate $\rho_{-1}$. This is nothing more than
the Hartree energy for a layered material. We split the charge density
of each layer into a contribution from charge carriers and another
from charged impurities, $\rho_{\ell}=-en_{\ell}+en_{\ell}^{\text{imp}}$,
where $n_{\ell}$ is the charge carrier concentration ($n_{\ell}>0$
for electron doping) and $n_{\ell}^{\text{imp}}$ is the concentration
of charged impurities ($n_{\ell}^{\text{imp}}>0$ for positively charged
impurities). Including the effects of a gate voltage, $V_{\text{gate}}$,
applied between the $\ell=-1$ and the $\ell=0$ layers and a bias
voltage between the $\ell=\mathcal{N}+1$and the $\ell=0$ layers,
we obtain a Thomas-Fermi functional
\begin{eqnarray}
\Phi & = & \frac{1}{2}\sum_{\ell,\ell^{\prime}=1}^{\mathcal{N}+1}n_{\ell}\left(\sum_{k=0}^{\min(\ell,\ell^{\prime})}\frac{e^{2}d_{k}}{\epsilon_{0}\bar{\epsilon}_{k}}\right)n_{\ell^{\prime}}-\sum_{\ell=0}^{\mathcal{N}+1}n_{\ell}e\phi_{\ell}^{\text{imp}}\nonumber \\
 & - & eV_{\text{gate}}\sum_{\ell=0}^{\mathcal{N}+1}n_{\ell}+eV_{\text{bias}}\sum_{\ell=0}^{\mathcal{N}+1}\frac{\ell}{\mathcal{N}+1}n_{\ell},
\end{eqnarray}
where 
\begin{equation}
e\phi_{\ell}^{\text{imp}}=\sum_{\ell^{\prime}=0}^{\mathcal{N}+1}\sum_{k=1}^{\min(\ell,\ell^{\prime})}\frac{e^{2}d_{k}}{\epsilon_{0}\bar{\epsilon}_{k}}n_{\ell^{\prime}}^{\text{imp}},
\end{equation}
is the potential created by the charged impurities. The Hartree potential
felt by electrons in layer $\ell$ is then given by
\begin{eqnarray}
V_{\ell}^{H} & = & -\frac{\pd\Phi}{\pd n_{\ell}}\nonumber \\
 & = & V_{\text{gate}}-V_{\text{bias}}\frac{\ell}{\mathcal{N}+1}n_{\ell}+e\phi_{\ell}^{\text{imp}}\nonumber \\
 & - & \sum_{\ell^{\prime}=0}^{\mathcal{N}+1}\left(\sum_{k=0}^{\min(\ell,\ell^{\prime})}\frac{e^{2}d_{k}}{\epsilon_{0}\bar{\epsilon}_{k}}\right)n_{\ell^{\prime}},\,\ell=0,...,\mathcal{N}+1.\label{eq:Hartree_potential}
\end{eqnarray}
Now, we assume that the vertical current flowing between the two graphene
layers is small enough, such that we can assume that these are in
a near equilibrium state. Furthermore, we employ the Thomas-Fermi
approximation, in which the local Fermi level for each layer is given
by $\epsilon_{\text{F},\ell}=V_{\ell}^{H}$, where $\epsilon_{\text{F},\ell}$
is a function of the local carrier density. This together with Eq.~(\ref{eq:Hartree_potential})
becomes a system of non-linear equations in the carrier density /
local Fermi level. 

It can be checked, that due to the large band gap of hBN, most charge
density will be accumulated in the graphene layers. As such we approximate
$n_{\ell}=0,$ for $\ell=1,...,\mathcal{N}$ and therefore the $\mathcal{N}+2$
equations are reduced to two 
\begin{eqnarray}
\epsilon_{\text{F,bg}} & = & eV_{\text{gate}}-\left(n_{\text{bg}}+n_{\text{tg}}\right)C_{\text{t}}^{-1}+e\phi_{\text{bg}}^{\text{imp}}\label{eq:gating_bg}\\
\epsilon_{\text{F,tg}} & = & eV_{\text{gate}}-eV_{\text{bias}}-n_{\text{tg}}C_{\text{t}}^{-1}-n_{\text{bg}}C_{\text{b}}^{-1}+e\phi_{\text{tg}}^{\text{imp}}\label{eq:gating_tg}
\end{eqnarray}
where the capacitances are given by (taking into account the series
capacitances of a hBN/SiO$_{2}$ spacer with $d_{\text{hBN}}$ the
hBN thickness and $d_{\text{SiO}_{2}}$the SiO$_{2}$ thickness) 
\begin{align}
C_{\text{b}}^{-1} & =\frac{e^{2}d_{0}}{\epsilon_{0}\bar{\epsilon}_{0}}=\frac{e^{2}d_{\text{SiO}_{2}}}{\epsilon_{0}\bar{\epsilon}_{\text{SiO}_{2}}}+\frac{e^{2}d_{\text{hBN}}}{\epsilon_{0}\bar{\epsilon}_{\text{hBN}}}\nonumber \\
C_{\text{t}}^{-1} & =\sum_{\ell=0}^{\mathcal{N}+1}\frac{e^{2}d_{\ell}}{\epsilon_{0}\bar{\epsilon}_{\ell}}\\
 & =\frac{e^{2}d_{\text{SiO}_{2}}}{\epsilon_{0}\bar{\epsilon}_{\text{SiO}_{2}}}+\frac{e^{2}d_{\text{hBN}}}{\epsilon_{0}\bar{\epsilon}_{\text{hBN}}}+\frac{e^{2}d}{\epsilon_{0}\bar{\epsilon}_{\text{hBN}}},
\end{align}
and $d$ is the distance between the two graphene layers. The terms
$e\phi_{\text{bg}/\text{tg}}^{\text{imp}}$ are the potentials induced
by the charged impurities in the bottom/top graphene layer that can
be tuned to account for intrinsic doping of the graphene layers (acting
as an offset in the measurement of $V_{\text{gate}}$ and $V_{\text{bias}}$).
We finally point out that in the case where the hBN layers have no
charge carrier, then the Hartree potential within the hBN slab is
given from Eq.~(\ref{eq:Hartree_potential}) in terms of $\epsilon_{\text{F,bg}/\text{t}g}$
as
\begin{align}
V_{\ell}^{H} & =\epsilon_{\text{F,tg}}-e\phi_{\text{tg}}^{\text{imp}}+e\phi_{\ell}^{\text{imp}}\nonumber \\
 & -\frac{\ell}{\mathcal{N}+1}\left(\epsilon_{\text{F,tg}}+eV_{\text{bias}}-\epsilon_{\text{F,bg}}-e\phi_{\text{tg}}^{\text{imp}}+e\phi_{\text{bg}}^{\text{imp}}\right)
\end{align}
which in the absence of impurities reduces to the expression given
in Sec.~\ref{sub:Model-Hamiltonian}. The solutions of Eqs.~(\ref{eq:gating_bg})-(\ref{eq:gating_tg})
for a particular device are shown in Fig.~\ref{fig:gating}

\section{\label{sec:Interlayer-hopping-twisted}Interlayer hopping Hamiltonian
between non-commensurate layers}

We describe the graphene-boron nitride coupling using the general
theory of coupling between non-commensurate layers of Refs.~\onlinecite{Bistritzer2010,Koshino2015}.
We wish to describe the coupling between two 2D crystals, labeled
as $\ell$ and $\ell^{\prime}$, with Bravais lattices spanned by
$\left\{ \bm{a}_{1,\ell},\bm{a}_{2,\ell}\right\} $ and $\left\{ \bm{a}_{1,\ell^{\prime}},\bm{a}_{2,\ell^{\prime}}\right\} $,
respectively. In a tight-binding representation the interlayer hopping
between layers $\ell$ and $\ell^{\prime}$ can be written as
\begin{equation}
T_{\ell,\ell^{\prime}}=-\sum_{n,a,m,b}t\left({\bf R}_{n,a,\ell},{\bf R}_{m,b,\ell^{\prime}}\right)c_{n,a,\ell}^{\dagger}c_{m,b,\ell^{\prime}},
\end{equation}
where the indices $n,m$ run over Bravais lattice sites, $a,b$ run
over orbitals/sublattice sites, $c_{n,\alpha,\ell}^{\dagger}$ creates
an electron state in layer $\ell$ at position ${\bf R}_{n,a,\ell}=n_{1}\bm{a}_{1,\ell}+n_{2}\bm{a}_{2,\ell}+\bm{\tau}_{a,\ell},$
and orbital/sublattice $a$, with $\bm{\tau}_{a,\ell}$ a sublattice
vector, and $t\left({\bf R}_{n,a,\ell},{\bf R}_{m,b,\ell^{\prime}}\right)$
are hopping terms. Assuming that the hopping $t\left({\bf R}_{n,a,\ell},{\bf R}_{m,b,\ell^{\prime}}\right)$
only depends on ${\bf R}_{n,a,\ell}-{\bf R}_{m,b,\ell^{\prime}}$
it is possible to write it in Fourier components as\cite{Koshino2015}
\begin{multline*}
t\left({\bf R}_{n,a,\ell},{\bf R}_{m,b,\ell^{\prime}}\right)=\sqrt{A_{\text{cell},\ell}A_{\text{cell},\ell^{\prime}}}\times\\
\times\int\frac{d^{2}\bm{q}}{\left(2\pi\right)^{2}}t_{a,b}^{\ell,\ell^{\prime}}\left(\bm{q}\right)e^{i\bm{q}\cdot\left({\bf R}_{n,a,\ell}-{\bf R}_{m,b,\ell^{\prime}}\right),}
\end{multline*}
where $A_{\text{cell},\ell/\ell^{\prime}}$ is the area of the unit
cell of layer $\ell/\ell^{\prime}$. If we express, $c_{n,a,\ell}^{\dagger}$
and $c_{m,b,\ell^{\prime}}$ in a Bloch basis 
\begin{eqnarray}
c_{n,a,\ell}^{\dagger} & = & \frac{1}{\sqrt{N_{\ell}}}\sum_{\bm{k}}e^{-i\bm{k}\cdot{\bf R}_{n,a,\ell}}c_{\bm{k},a,\ell}^{\dagger},\\
c_{m,b,\ell^{\prime}}^{\dagger} & = & \frac{1}{\sqrt{N_{\ell^{\prime}}}}\sum_{\bm{k}}e^{-i\bm{k}\cdot{\bf R}_{m,b,\ell^{\prime}}}c_{\bm{k},b,\ell^{\prime}}^{\dagger},
\end{eqnarray}
where $N_{\ell/\ell^{\prime}}$ is the number of unit cells in layer
$\ell/\ell^{\prime}$, such that$N_{\ell}A_{\text{cell},\ell}=N_{\ell^{\prime}}A_{\text{cell},\ell^{\prime}}$,
the interlayer Hamiltonian becomes
\begin{align}
T_{\ell,\ell^{\prime}} & =-\sum_{\substack{\bm{k},{\bf G}_{n,\ell}\\
\bm{k}^{\prime},{\bf G}_{m,\ell^{\prime}}
}
}e^{i\bm{\tau}_{a,\ell}\cdot{\bf G}_{n,\ell}}t_{a,b}^{\ell,\ell^{\prime}}\left(\bm{k}+{\bf G}_{n,\ell}\right)\times\nonumber \\
 & \times e^{-i\bm{\tau}_{b,\ell^{\prime}}\cdot{\bf G}_{m,\ell^{\prime}}}c_{\bm{k},a,\ell}^{\dagger}c_{\bm{k}^{\prime},b,\ell^{\prime}}\delta_{\bm{k}+{\bf G}_{n,\ell},\bm{k}^{\prime}+{\bf G}_{m,\ell^{\prime}}}\label{eq:Interlayer_general}
\end{align}
where ${\bf G}_{n,\ell/\ell^{\prime}}$ are reciprocal lattice vectors
of the 2D crystal $\ell/\ell^{\prime}$. The Kronecker-$\delta$ imposes
that in a interlayer hoping process, momentum is conserved modulo
any combination of reciprocal lattice vectors of both layers. In general,
$t_{a,b}^{\ell,\ell^{\prime}}\left(\bm{q}\right)$ will decay for
large values of $\left|\bm{q}\right|$, and therefore only the processes
with smallest $\left|\bm{k}+{\bf G}_{n,\ell}\right|$ need be considered.

We now specialize to the case where $\ell^{\prime}$ is a graphene
layer and $\ell$ is a boron nitride layer. The graphene unit cell
contains two carbon atoms in the unit cell, A and B, while boron nitride
contains one boron atom, B, and one nitrogen atom, N, in the unit
cell, see Fig.~\ref{fig:Schematic}. We will focus on low energy
states, which lie close to the Dirac points, $\pm\bm{K}_{\text{g}}$,
of the graphene layer. Considering only the three most relevant processes
coupling the graphene and boron nitride layers, we must consider processes
involving ${\bf G}_{n,\text{\text{g}}}=\bm{0},\bm{b}_{2,\text{g}},-\bm{b}_{1,g}$
and ${\bf G}_{n,\text{\text{hBN}}}=\bm{0},\bm{b}_{2,\text{hBN}},-\bm{b}_{1,\text{hBN}}$
for states close to the $\bm{K}_{\text{g}}$ point and processes involving
${\bf G}_{n,\text{\text{g}}}=\bm{0},-\bm{b}_{2,\text{g}},\bm{b}_{1,\text{g}}$
and ${\bf G}_{n,\text{\text{hBN}}}=\bm{0},-\bm{b}_{2,\text{hBN}},\bm{b}_{1,\text{hBN}}$
for states close to the $-\bm{K}_{\text{g}}$ point. It is also assumed
that the momentum dependence of $t_{a,b}^{\ell,\ell^{\prime}}\left(\bm{k}\right)$
is weak such that we can approximate $t_{a,b}^{\ell,\ell^{\prime}}\left(\bm{k}+\bm{K}_{\pm}+{\bf G}_{n,\text{\text{g}}}\right)\simeq t_{a,b}^{\ell,\ell^{\prime}}\left(\bm{K}\right)$,
setting :$t_{\text{B},\text{A}}^{\text{\text{hBN}},\text{g}}\left(\bm{K}\right)=t_{\text{B},\text{B}}^{\text{\text{hBN}},\text{g}}\left(\bm{K}\right)=t_{\text{B},\text{C}}$
and $t_{\text{N},\text{A}}^{\text{\text{hBN}},\text{g}}\left(\bm{K}\right)=t_{\text{N},\text{B}}^{\text{\text{hBN}},\text{g}}\left(\bm{K}\right)=t_{\text{N},\text{C}}$. 

In order to describe the coupling between the bottom and top graphene
layers to a slab formed by $\mathcal{N}$ hBN monolayers, we notice
that the products of unit cell basis vectors and reciprocal lattice
vectors that appears in Eq.~(\ref{eq:Interlayer_general}) can be
written for the bottom graphene layer as $\text{\ensuremath{\bm{\tau}}}_{\text{hBN},\text{B}1}\cdot{\bf G}_{n,\text{hBN}}=\text{\ensuremath{\bm{\tau}}}_{\text{bg},\text{A}}\cdot{\bf G}_{n,\text{bg}}=0$
and $\text{\ensuremath{\bm{\tau}}}_{\text{hBN},\text{N1}}\cdot{\bf G}_{n,\text{hBN}}=\text{\ensuremath{\bm{\tau}}}_{\text{bg},\text{B}}\cdot{\bf G}_{n,\text{bg}}=\pm n2\pi/3$
(for states close to $\pm\bm{K}_{\text{g}}$ point). For the coupling
between the top graphene layer and the $\mathcal{N}^{\text{th}}$
hBN layer, one must consider separately the cases when the hBN slab
is formed by and even or odd number of layers. For an odd number of
layers, in the $\mathcal{N}$th layer the boron and nitrogen atoms
occupy the same positions as in the 1st layer and therefore we still
have $\text{\ensuremath{\bm{\tau}}}_{\text{hBN},\text{B}\mathcal{N}}\cdot{\bf G}_{n,\text{hBN}}=\text{\ensuremath{\bm{\tau}}}_{\text{tg},\text{A}}\cdot{\bf G}_{n,\text{tg}}=0$
and $\text{\ensuremath{\bm{\tau}}}_{\text{hBN},\text{N}\mathcal{N}}\cdot{\bf G}_{n,\text{hBN}}=\text{\ensuremath{\bm{\tau}}}_{\text{tg},\text{B}}\cdot{\bf G}_{n,\text{tg}}=\pm n2\pi/3$.
If we have an even number of hBN layers, then in the $\mathcal{N}^{\text{th}}$
layer, the boron an nitrogen atoms switch positions compared to the
1st layer, and one obtains instead $\text{\ensuremath{\bm{\tau}}}_{\text{hBN},\text{N}\mathcal{N}}\cdot{\bf G}_{n,\text{hBN}}=\text{\ensuremath{\bm{\tau}}}_{\text{tg},\text{A}}\cdot{\bf G}_{n,\text{tg}}=0$
and $\text{\ensuremath{\bm{\tau}}}_{\text{hBN},\text{B}\mathcal{N}}\cdot{\bf G}_{n,\text{hBN}}=\text{\ensuremath{\bm{\tau}}}_{\text{tg},\text{B}}\cdot{\bf G}_{n,\text{tg}}=\pm n2\pi/3\mp n2\pi/3$.
With these approximations, one obtains Eq.~(\ref{eq:Hamiltonian_graphene_hBN})
of the main text.

\section{\label{sec:Ananlitic-expression-TDoS}Analytic expression for the
tunneling density of states}

In this appendix we provide an analytic expression for Eq.~(\ref{eq:TDoS}).
First, we notice that Eq.~(\ref{eq:TDoS}) can be written in a the
graphene sublattice basis as
\begin{multline}
\text{TDoS}_{n,m}(\omega_{\text{bg}},\omega_{\text{tg}})=\\
=\int\frac{d^{2}\bm{k}}{\left(2\pi\right)^{2}}\text{tr}\left[\bm{R}_{-\frac{2\pi}{3}}^{n}\cdot\bm{A}_{\text{bg,}\bm{k}}(\omega_{\text{bg}})\cdot\bm{R}_{\frac{2\pi}{3}}^{n}\cdot\right.\\
\left.\cdot\bm{J}\cdot\bm{R}_{-\frac{2\pi}{3}}^{m}\cdot\bm{A}_{\text{tg},\bm{k}+\bm{\mathcal{Q}}_{n,m},T}(\omega_{\text{tg}})\cdot\bm{R}_{\frac{2\pi}{3}}^{m}\cdot\bm{J}\right],
\end{multline}
where $\text{tr}\left\{ ...\right\} $ is the trace over graphene
sublattice indices, $\bm{J}$ is a $2\times2$ matrix of ones, and
we have written the spectral function in the sublattice basis as
\begin{equation}
\bm{A}_{\text{bg}/\text{tg},\bm{k}}(\omega)=i\left[\bm{G}_{\bm{k},\theta_{\text{bg}/\text{tg}}}\left(\omega_{\text{bg/tg}}^{+}\right)-\bm{G}_{\bm{k},\theta_{\text{bg}/\text{tg}}}\left(\omega_{\text{bg/tg}}^{-}\right)\right],
\end{equation}
where the graphene retarded/advanced electron Green\textasciiacute s
function in the sublattice space is given by
\begin{equation}
\bm{G}_{\bm{k},\theta}\left(\omega_{\text{bg/tg}}^{\pm}\right)=\frac{\omega_{\text{bg/tg}}^{\pm}\Id+v_{F}\hbar\bm{k}\cdot\bm{\sigma}_{\theta}}{\left(\omega_{\text{bg/tg}}^{\pm}\right)^{2}-\left(v_{F}\hbar\right)^{2}\left|\bm{k}\right|^{2}},
\end{equation}
with $\bm{\sigma}_{\theta}=\left(\cos\theta\sigma_{x}-\sin\theta\sigma_{y},\sin\theta\sigma_{x}+\cos\theta\sigma_{y}\right)$.
In the limit of an infinite electron lifetime, we have $\omega_{\text{bg}/\text{tg}}^{\pm}=\omega_{\text{bg}/\text{tg}}\pm i0^{+}$.
In the presence of perturbations that induce a momentum independent
self-energy that is diagonal in the sublattice basis (such as short
range diagonal disorder or scattering by in-plane optical phonons),
we make the replacement $\omega_{\text{bg}/\text{tg}}^{\pm}\rra\omega_{\text{bg}/\text{tg}}-\re\Sigma_{\text{bg}/\text{tg}}(\omega_{\text{bg}/\text{tg}})\pm i\gamma_{\text{bg}/\text{tg}}(\omega_{\text{bg}/\text{tg}})$,
where $\gamma_{\text{bg}/\text{tg}}(\omega_{\text{bg}/\text{tg}})=-\im\Sigma_{\text{bg}/\text{tg}}^{R}(\omega_{\text{bg}/\text{tg}})$
is the broadening factor. In the presence of the external metallic
contacts and disorder/phonon scattering, we obtain $\gamma_{\text{bg}/\text{tg}}=\left(\Gamma_{\text{b}/\text{t}}+\Gamma_{\text{e-ph},\text{bg/tg}}\right)/2$
. In terms of Green's functions, and noticing that the matrices $\bm{R}_{\pm\frac{2\pi}{3}}^{n}$
perform a rotation of the electronic Green's functions, $\text{TDoS}_{n,m}(\omega)$
can be written as
\begin{multline}
\text{TDoS}_{n,m}(\omega_{\text{bg}},\omega_{\text{tg}})=i^{2}\sum_{s,s^{\prime}=\pm1}\int\frac{d^{2}\bm{k}}{\left(2\pi\right)^{2}}ss^{\prime}\times\\
\times\text{tr}\left[\bm{G}_{\bm{k},\theta_{\text{bg}}+n\frac{2\pi}{3}}\left(\omega_{\text{bg}}^{s}\right)\cdot\bm{J}\cdot\right.\\
\left.\cdot\bm{G}_{\bm{k}+\bm{\mathcal{Q}}_{n,m},\theta_{\text{tg}}+m\frac{2\pi}{3}}\left(\omega_{\text{tg}}^{s^{\prime}}\right)\cdot\bm{J}\right]
\end{multline}
Performing the trace over the sublattice degrees of freedom we get
\begin{multline}
\text{TDoS}_{n,m}(\omega_{\text{bg}},\omega_{\text{tg}})=i^{2}\sum_{s,s^{\prime}=\pm1}\int\frac{d^{2}\bm{k}}{\left(2\pi\right)^{2}}ss^{\prime}\times\\
\times\frac{2\left(\omega_{\text{bg}}^{s}+v_{F}\hbar\bm{k}\cdot\hat{\bm{K}}_{\text{bg},n}\right)}{\left(\omega_{\text{bg}}^{s}\right)^{2}-\left(v_{F}\hbar\right)^{2}\left|\bm{k}\right|^{2}}\\
\times\frac{2\left(\omega_{\text{tg}}^{s^{\prime}}+v_{F}\hbar\left(\bm{k}+\bm{\mathcal{Q}}_{n,m}\right)\cdot\hat{\bm{K}}_{\text{tg},m}\right)}{\left(\omega_{\text{tg}}^{s^{\prime}}\right)^{2}-\left(v_{F}\hbar\right)^{2}\left|\bm{k}+\bm{\mathcal{Q}}_{n,m}\right|^{2}}\label{eq:TDoS_alt}
\end{multline}
The advantage of this form, with respect to Eq.~(\ref{eq:TDoS}),
is that Eq.~(\ref{eq:TDoS_alt}) is analytic in $\bm{k}$ and as
such, contour integration methods can be used to compute the integrals.
In order to make analytic progress, in the first term of the previous
expression we take the limit $\gamma_{\text{bg}}\rra0$, such that
$\omega_{\text{bg}}^{s^{\prime}}\rra\omega_{\text{bg}}=\omega+\epsilon_{\text{F,bg}}$
and
\begin{multline}
i\sum_{s=\pm1}s\frac{\omega_{\text{bg}}^{s}+v_{F}\hbar\bm{k}\cdot\hat{\bm{K}}_{\text{bg},n}}{\left(\omega_{\text{bg}}^{s}\right)^{2}-\left(v_{F}\hbar\right)^{2}\left|\bm{k}\right|^{2}}\rra\\
\rra2\pi\frac{\omega_{\text{bg}}+v_{F}\hbar\bm{k}\cdot\hat{\bm{K}}_{\text{bg},n}}{2v_{F}\hbar\left|\bm{k}\right|}\times\\
\times\sum_{s=\pm1}s\delta\left(\omega_{\text{bg}}-sv_{F}\hbar\left|\bm{k}\right|\right)
\end{multline}
We use the $\delta$-function to perform the integration over $\left|\bm{k}\right|$,
obtaining
\begin{multline}
\text{TDoS}_{n,m}(\omega_{\text{bg}},\omega_{\text{tg}})\simeq\\
\simeq i\frac{\omega_{\text{bg}}}{\left(v_{F}\hbar\right)^{2}}\int\frac{d\theta_{\bm{k}}}{2\pi}\left(\frac{\omega_{\text{bg}}+v_{F}\hbar\bm{k}\cdot\hat{\bm{K}}_{\text{bg},n}}{v_{F}\hbar\left|\bm{k}\right|}\right)\Biggr|_{\left|\bm{k}\right|=\frac{\left|\omega_{\text{bg}}\right|}{v_{F}\hbar}}\times\\
\times\sum_{s^{\prime}=\pm1}s^{\prime}\frac{2\left(\omega_{\text{tg}}^{s^{\prime}}+v_{F}\hbar\left(\bm{k}+\bm{\mathcal{Q}}_{n,m}\right)\cdot\hat{\bm{K}}_{\text{tg},m}\right)}{\left(\omega_{\text{tg}}^{s^{\prime}}\right)^{2}-\left(v_{F}\hbar\right)^{2}\left|\bm{k}+\bm{\mathcal{Q}}_{n,m}\right|^{2}}.\label{eq:TDoS_oneInt}
\end{multline}
The remaining integration over the angular variable $\theta_{\bm{k}}$
can be performed using contour integration methods. Performing a change
of variables $z=e^{i\theta_{\bm{k}}}$ such that
\begin{eqnarray}
\cos\theta_{\bm{k}} & = & \frac{z+z^{-1}}{2},\\
\sin\theta_{\bm{k}} & = & \frac{z-z^{-1}}{2i},
\end{eqnarray}
Eq.~(\ref{eq:TDoS_oneInt}) can be written as an integral over the
$z$ variable around the unit circle in the complex plane\begin{widetext}
\begin{multline}
\text{TDoS}_{n,m}(\omega_{\text{bg}},\omega_{\text{tg}})\simeq i\frac{\omega_{\text{bg}}}{\left(v_{F}\hbar\right)^{2}}\varointctrclockwise_{\left|z\right|=1}\frac{dz}{2\pi i}\frac{1}{z}\left(\frac{\omega_{\text{bg}}+v_{F}\hbar\left|\bm{k}\right|\left(\frac{z+z^{-1}}{2}\hat{K}_{\text{bg},n}^{x}+\frac{z-z^{-1}}{2i}\hat{K}_{\text{bg},n}^{y}\right)}{v_{F}\hbar\left|\bm{k}\right|}\right)\Biggr|_{\left|\bm{k}\right|=\frac{\left|\omega_{\text{bg}}\right|}{v_{F}\hbar}}\times\\
\times\sum_{s^{\prime}=\pm1}s^{\prime}\frac{2\left(\omega_{\text{tg}}^{s^{\prime}}+v_{F}\hbar\left|\bm{k}\right|\left(\frac{z+z^{-1}}{2}\hat{K}_{\text{tg},m}^{x}+\frac{z-z^{-1}}{2i}\hat{K}_{\text{tg},m}^{y}\right)+v_{F}\hbar\bm{\mathcal{Q}}{}_{n,m}\cdot\hat{\bm{K}}_{\text{bg},m}\right)}{\left(\omega_{\text{tg}}^{s^{\prime}}\right)^{2}-\left|\bm{k}\right|^{2}-\left|\bm{\mathcal{Q}}{}_{n,m}\right|^{2}-2v_{F}\hbar\left|\bm{k}\right|\left|\bm{\mathcal{Q}}_{n,m}\right|\left(\frac{z+z^{-1}}{2}\cos\theta_{\bm{\mathcal{Q}}_{n,m}}+\frac{z-z^{-1}}{2i}\sin\theta_{\bm{\mathcal{Q}}_{n,m}}\right)},\label{eq:TDoS_oneInt-1}
\end{multline}
\end{widetext} with $\theta_{\bm{\mathcal{Q}}_{n,m}}$ the angle
of the vector $\bm{\mathcal{Q}}_{n,m}$ with the reference $x$ axis.
The integrand has a double pole at $z=0$ and two simple poles at
$z=e^{i\theta_{\bm{\mathcal{Q}}{}_{m,n}}}w_{\gtrless,\text{tg}}^{s}$,
with
\begin{align}
w_{\gtrless,\text{tg}}^{s} & =\mathcal{C}_{\text{tg}}^{s}\mp i\mathcal{S}_{\text{tg}}^{s},\\
\mathcal{C}_{\text{tg}}^{s} & =\frac{\left(\omega_{\text{tg}}+si\gamma_{\text{tg}}\right)^{2}-\left(v_{F}\hbar\right)^{2}\left(\left|\bm{\mathcal{Q}}_{n,m}\right|^{2}+\left|\bm{k}\right|^{2}\right)}{2\left(v_{F}\hbar\right)^{2}\left|\bm{\mathcal{Q}}{}_{n,m}\right|\left|\bm{k}\right|},\label{eq:CoSine_like}\\
\mathcal{S}_{\text{tg}}^{s} & =\sgn\left(\omega_{\text{tg}}^{2}-\gamma_{\text{tg}}^{2}-\left(v_{F}\hbar\right)^{2}\left(\left|\bm{\mathcal{Q}}_{n,m}\right|^{2}+\left|\bm{k}\right|^{2}\right)\right)\times\nonumber \\
 & \times i\sqrt{\left(\mathcal{C}_{\text{tg}}^{s}\right)^{2}-1},\label{eq:Sine_like}
\end{align}
defined such that $\left|w_{<,\text{tg}}\right|<1$ and $w_{>,\text{tg}}=w_{<,\text{tg}}^{-1}$.
The contour integration around the unit circle can be performed analytically
collecting the residues at $z=e^{i\theta_{\pm\bm{\mathcal{Q}}_{n,m}}}w_{<,\text{tg}}^{s}$
and $z=0$. Notice that we have made the approximation $\gamma_{\text{bg}}\rra0$.
In general, both $\gamma_{\text{bg}}$ and $\text{\ensuremath{\gamma}}_{\text{tg}}$
will be non-zero. The simplest way to that this into account is to
symmetrize Eq.~(\ref{eq:TDoS_alt}) with respect to the bottom and
the top graphene layer an then taking the limit $\gamma_{\text{bg}}\rra0$
in the first term and $\gamma_{\text{tg}}\rra0$ in the second. The
final symmetrized result is given by\begin{widetext}
\begin{multline}
\text{TDoS}_{n,m}(\omega_{\text{bg}},\omega_{\text{tg}})\simeq\frac{\omega_{\text{bg}}}{\left(v_{F}\hbar\right)^{3}\left|\bm{\mathcal{Q}}_{n,m}\right|}\times\\
\times\left[\frac{-1}{\mathcal{S}_{\text{tg}}^{+}}\left(\frac{\omega_{\text{tg}}^{+}+\left|\omega_{\text{bg}}\right|\left(\mathcal{C}_{\text{tg}}^{+}X_{n,m}^{\text{tg}}+\mathcal{S}_{\text{tg}}^{+}Y_{n,m}^{\text{tg}}\right)+v_{F}\hbar\bm{\mathcal{Q}}_{n,m}\cdot\hat{\bm{K}}_{\text{tg},m}}{\left|\omega_{\text{bg}}\right|}\right)\left(\frac{\omega_{\text{bg}}+\left|\omega_{\text{bg}}\right|\left(\mathcal{C}_{\text{tg}}^{+}X_{n,m}^{\text{bg}}+\mathcal{S}_{\text{tg}}^{+}Y_{n,m}^{\text{bg}}\right)}{\left|\omega_{\text{bg}}\right|}\right)\right.\\
+\frac{1}{\mathcal{S}_{\text{tg}}^{-}}\left(\frac{\omega_{\text{tg}}^{-}+\left|\omega_{\text{bg}}\right|\left(\mathcal{C}_{\text{tg}}^{-}X_{n,m}^{\text{tg}}+\mathcal{S}_{\text{tg}}^{-}Y_{n,m}^{\text{tg}}\right)+v_{F}\hbar\bm{\mathcal{Q}}_{n,m}\cdot\hat{\bm{K}}_{\text{tg},m}}{\left|\omega_{\text{bg}}\right|}\right)\left(\frac{\omega_{\text{bg}}+\left|\omega_{\text{bg}}\right|\left(\mathcal{C}_{\text{tg}}^{-}X_{n,m}^{\text{bg}}+\mathcal{S}_{\text{tg}}^{-}Y_{n,m}^{\text{bg}}\right)}{\left|\omega_{\text{bg}}\right|}\right)\\
+\left.\frac{2\gamma_{\text{tg}}\left(X_{n,m}^{\text{bg}}+iY_{n,m}^{\text{bg}}\right)}{v_{F}\hbar\left|\bm{\mathcal{Q}}_{n,m}\right|}\left(\frac{v_{F}\hbar\left|\bm{\mathcal{Q}}_{n,m}\right|+\omega_{\text{tg}}\left(X_{n,m}^{\text{tg}}+iY_{n,m}^{\text{tg}}\right)}{\left|\omega_{\text{bg}}\right|}\right)\right]\Biggr|_{\left|\bm{k}\right|=\left|\omega_{\text{bg}}\right|/\left(v_{F}\hbar\right)}\\
+\frac{1}{2}\frac{\omega_{\text{tg}}}{\left(v_{F}\hbar\right)^{3}\left|\bm{\mathcal{Q}}_{n,m}\right|}\times\\
\times\left[\frac{-1}{\mathcal{S}_{\text{bg}}^{+}}\left(\frac{\omega_{\text{bg}}^{+}-\left|\omega_{\text{tg}}\right|\left(\mathcal{C}_{\text{bg}}^{+}X_{n,m}^{\text{bg}}+\mathcal{S}_{\text{bg}}^{+}Y_{n,m}^{\text{bg}}\right)-v_{F}\hbar\bm{\mathcal{Q}}_{n,m}\cdot\hat{\bm{K}}_{\text{bg},n}}{\left|\omega_{\text{tg}}\right|}\right)\left(\frac{\omega_{\text{tg}}-\left|\omega_{\text{tg}}\right|\left(\mathcal{C}_{\text{bg}}^{+}X_{n,m}^{\text{tg}}+\mathcal{S}_{\text{bg}}^{+}Y_{n,m}^{\text{tg}}\right)}{\left|\omega_{\text{tg}}\right|}\right)\right.\\
+\frac{1}{\mathcal{S}_{\text{bg}}^{-}}\left(\frac{\omega_{\text{bg}}^{-}-\left|\omega_{\text{tg}}\right|\left(\mathcal{C}_{\text{bg}}^{-}X_{n,m}^{\text{bg}}+\mathcal{S}_{\text{bg}}^{-}Y_{n,m}^{\text{bg}}\right)-v_{F}\hbar\bm{\mathcal{Q}}_{n,m}\cdot\hat{\bm{K}}_{\text{bg},n}}{\left|\omega_{\text{tg}}\right|}\right)\left(\frac{\omega_{\text{tg}}-\left|\omega_{\text{tg}}\right|\left(\mathcal{C}_{\text{bg}}^{-}X_{n,m}^{\text{tg}}+\mathcal{S}_{\text{bg}}^{-}Y_{n,m}^{\text{tg}}\right)}{\left|\omega_{\text{tg}}\right|}\right)\\
+\left.\frac{2\gamma_{\text{bg}}\left(-X_{n,m}^{\text{tg}}-iY_{n,m}^{\text{tg}}\right)}{v_{F}\hbar\left|\bm{\mathcal{Q}}_{n,m}\right|}\left(\frac{v_{F}\hbar\left|\bm{\mathcal{Q}}_{n,m}\right|-\omega_{\text{bg}}\left(X_{n,m}^{\text{bg}}+iY_{n,m}^{\text{bg}}\right)}{\left|\omega_{\text{tg}}\right|}\right)\right]\Biggr|_{\left|\bm{k}\right|=\left|\omega_{\text{tg}}\right|/\left(v_{F}\hbar\right)},\label{eq:TDoS_analytic_sym}
\end{multline}
\end{widetext}where we have introduced the quantities
\begin{equation}
\begin{array}{cc}
X_{n,m}^{\text{bg}}=\hat{\bm{\mathcal{Q}}}_{n,m}\cdot\hat{\bm{K}}_{\text{bg},n},\quad & Y_{n,m}^{\text{bg}}=\hat{\bm{\mathcal{Q}}}_{n,m}\times\hat{\bm{K}}_{\text{bg},n},\\
X_{n,m}^{\text{tg}}=\hat{\bm{\mathcal{Q}}}_{n,m}\cdot\hat{\bm{K}}_{\text{tg},m},\quad & Y_{n,m}^{\text{tg}}=\hat{\bm{\mathcal{Q}}}_{n,m}\times\hat{\bm{K}}_{\text{tg},m},
\end{array}
\end{equation}
and the quantities $\mathcal{C}_{\text{tg}}^{\pm}$ and $\mathcal{S}_{\text{tg}}^{\pm}$
given by Eqs.~(\ref{eq:CoSine_like}) and (\ref{eq:Sine_like}) with
the replacements $\omega_{\text{\text{tg}}}\rra\omega_{\text{bg}}$
and $\gamma_{\text{\text{tg}}}\rra\gamma_{\text{bg}}.$ It was checked
that Eq.~(\ref{eq:TDoS_analytic_sym}) provides a very good approximation
to the numeric evaluation of Eq.~(\ref{eq:TDoS_alt}) when both $\gamma_{\text{bg}}$
and $\gamma_{\text{tg}}$ are non-zero, if the broadening function
for each layer is assumed to the the sum of the broadening factors
of both layers, i.e., performing the replacement $\gamma_{\text{bg}},\gamma_{\text{tg}}\rra\gamma_{\text{bg}}+\gamma_{\text{tg}}$. 

In the limit of infinite electron lifetime in both layers $\gamma_{\text{bg}/\text{tg}}\rra0$,
we obtain
\begin{eqnarray}
\mathcal{C}_{\text{tg}}^{s} & = & \frac{\omega_{\text{tg}}^{2}-\omega_{\text{bg}}^{2}-\left(v_{F}\hbar\right)^{2}\left|\bm{\mathcal{Q}}_{n,m}\right|^{2}}{2\left(v_{F}\hbar\right)\left|\bm{\mathcal{Q}}{}_{n,m}\right|\left|\omega_{\text{bg}}\right|},\\
\mathcal{S}_{\text{tg}}^{s} & = & -s\sgn(\omega_{\text{tg}})\sqrt{1-\left(\mathcal{C}_{\text{tg}}^{s}\right)^{2}},
\end{eqnarray}
and $\mathcal{S}_{\text{bg}}^{s}/\mathcal{C}_{\text{bg}}^{s}$ are
obtained by replacing $\omega_{\text{bg}}\leftrightarrow\omega_{\text{tg}}$.
and $\text{TDoS}_{n,m}(\omega_{\text{bg}},\omega_{\text{tg}})$ simplifies
to \begin{widetext}
\begin{multline}
\text{TDoS}_{n,m}(\omega_{\text{bg}},\omega_{\text{tg}})=\frac{\omega_{\text{tg}}}{\left(v_{F}\hbar\right)^{3}\left|\bm{\mathcal{Q}}_{n,m}\right|}\times\\
\times\left[\frac{-1}{\mathcal{S}_{\text{tg}}^{+}}\left(\frac{\omega_{\text{tg}}+\left|\omega_{\text{bg}}\right|\left(\mathcal{C}_{\text{tg}}^{+}X_{n,m}^{\text{tg}}+\mathcal{S}_{\text{tg}}^{+}Y_{n,m}^{\text{tg}}\right)+v_{F}\hbar\bm{\mathcal{Q}}_{n,m}\cdot\hat{\bm{K}}_{\text{tg},m}}{\left|\omega_{\text{bg}}\right|}\right)\left(\frac{\omega_{\text{bg}}+\left|\omega_{\text{bg}}\right|\left(\mathcal{C}_{\text{tg}}^{+}X_{n,m}^{\text{bg}}+\mathcal{S}_{\text{tg}}^{+}Y_{n,m}^{\text{bg}}\right)}{\left|\omega_{\text{bg}}\right|}\right)\right.\\
+\left.\frac{1}{\mathcal{S}_{\text{tg}}^{-}}\left(\frac{\omega_{\text{tg}}+\left|\omega_{\text{bg}}\right|\left(\mathcal{C}_{\text{tg}}^{-}X_{n,m}^{\text{tg}}+\mathcal{S}_{\text{tg}}^{-}Y_{n,m}^{\text{tg}}\right)+v_{F}\hbar\bm{\mathcal{Q}}_{n,m}\cdot\hat{\bm{K}}_{\text{tg},m}}{\left|\omega_{\text{bg}}\right|}\right)\left(\frac{\omega_{\text{bg}}+\left|\omega_{\text{bg}}\right|\left(\mathcal{C}_{\text{tg}}^{-}X_{n,m}^{\text{bg}}+\mathcal{S}_{\text{tg}}^{-}Y_{n,m}^{\text{bg}}\right)}{\left|\omega_{\text{bg}}\right|}\right)\right].
\end{multline}
\end{widetext} We notice that, in this limit, $\text{TDoS}_{n,m}(\omega_{\text{bg}},\omega_{\text{tg}})$
is only non-zero when $4\left(v_{F}\hbar\right)^{2}\left|\bm{\mathcal{Q}}_{n,m}\right|^{2}\omega_{\text{bg}}^{2}>\left(\omega_{\text{tg}}^{2}-\omega_{\text{bg}}^{2}-\left(v_{F}\hbar\right)^{2}\left|\bm{\mathcal{Q}}_{n,m}\right|^{2}\right)^{2}$.

We finally study how the spinorial character of graphene's wavefunction
manifests in the form of $\text{TDoS}_{n,m}(\omega_{\text{bg}},\omega_{\text{tg}})$.
If we set the wavefunction overlap factors $\Upsilon_{\bm{k},\lambda}^{\text{bg}/\text{tg},n}$
to $1$ in Eq.~(\ref{eq:TDoS}), then instead of Eq.~(\ref{eq:TDoS_alt})
we would obtain
\begin{multline}
\text{TDoS}_{n,m}^{\text{scalar}}(\omega_{\text{bg}},\omega_{\text{tg}})=i^{2}\sum_{s,s^{\prime}=\pm1}\int\frac{d^{2}\bm{k}}{\left(2\pi\right)^{2}}ss^{\prime}\times\\
\times\frac{2\omega_{\text{bg}}^{s}}{\left(\omega_{\text{bg}}^{s}\right)^{2}-\left(v_{F}\hbar\right)^{2}\left|\bm{k}\right|^{2}}\\
\times\frac{2\omega_{\text{tg}}^{s^{\prime}}}{\left(\omega_{\text{tg}}^{s^{\prime}}\right)^{2}-\left(v_{F}\hbar\right)^{2}\left|\bm{k}+\bm{\mathcal{Q}}_{n,m}\right|^{2}}.
\end{multline}
In order to evaluate $\text{TDoS}_{n,m}^{\text{scalar}}(\omega_{\text{bg}},\omega_{\text{tg}})$,
we proceed as previously. the only difference is that when performing
the integration over the unit circle in the complex variable $z$,
there is no double pole at $z=0$, and the contour integration only
collects the contribution from $z=e^{i\theta_{\pm\bm{\mathcal{Q}}{}_{m,n}}}w_{<,\text{tg}/\text{bg}}^{s}$.
Symmetrizing the result, this leads to
\begin{multline}
\text{TDoS}_{n,m}^{\text{scalar}}(\omega_{\text{bg}},\omega_{\text{tg}})=\frac{1}{\left(v_{F}\hbar\right)^{3}\left|\bm{\mathcal{Q}}_{n,m}\right|}\times\\
\times\frac{1}{2}\left[\left(\frac{\omega_{\text{tg}}^{-}}{\mathcal{S}_{\text{tg}}^{-}}-\frac{\omega_{\text{tg}}^{+}}{\mathcal{S}_{\text{tg}}^{+}}\right)+\left(\frac{\omega_{\text{bg}}^{-}}{\mathcal{S}_{\text{bg}}^{-}}-\frac{\omega_{\text{bg}}^{+}}{\mathcal{S}_{\text{bg}}^{+}}\right)\right].\label{eq:TDoS_scalar}
\end{multline}

\section{\label{sec:Ressonant-impurties-SCBA}Resonant impurities within the
SCBA}

\begin{figure}
\begin{centering}
\includegraphics[width=8cm]{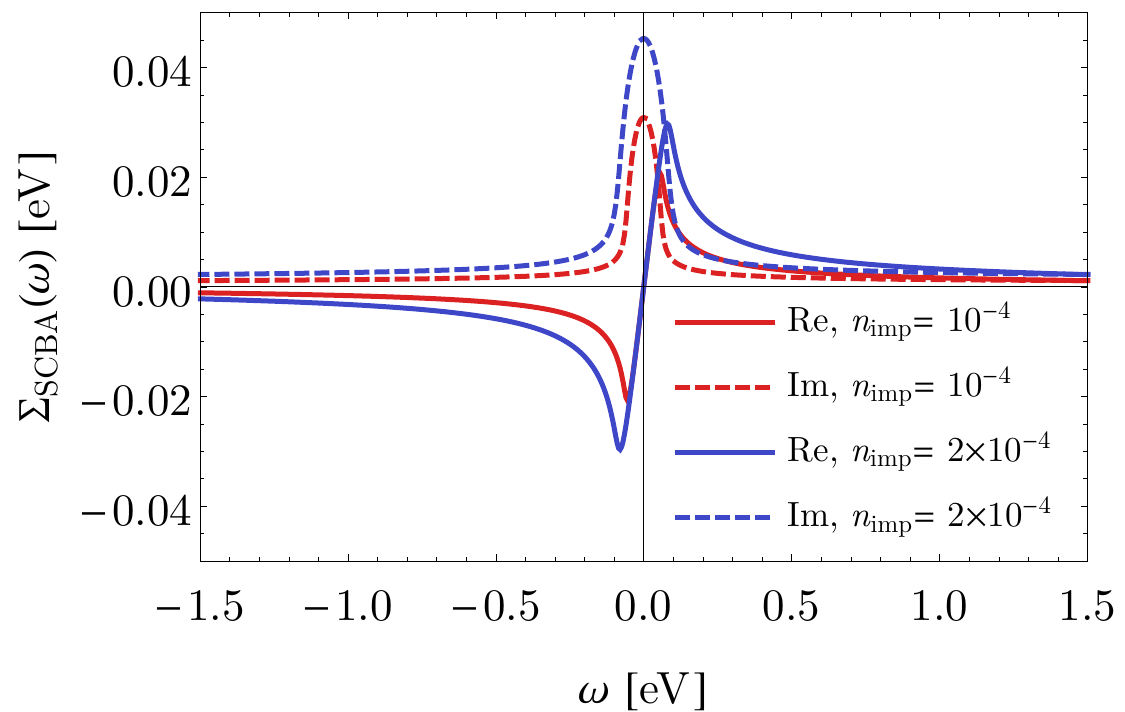}
\par\end{centering}

\protect\caption{\label{fig:Self_energy_SCBA}Real and (minus) imaginary parts of the
retarded self-energy for graphene electrons due to resonant impurities
treated within the SCBA, for two different impurity concentrations
(number of impurities per graphene unit cell).}

\end{figure}

We consider the effect of resonant impurities, such as vacancies,
in the properties of graphene. We focus on this kind of impurities
due to the possibility for analytical progress and due to the fact
that this model for impurities correctly predicts a transport lifetime
in graphene that depends on the Fermi energy as $\tau_{\text{tr}}(\epsilon_{F})\propto\epsilon_{F}$
\cite{Peres_Rev2010}. Resonances due to short range disorder cannot
be taken into account by treating then within a Gaussian approximation.
A way to overcome this limitation is to employ the T-matrix, which
properly takes into account multiple scatterings by the same impurity
in the limit of low impurity concentration. Using the T-matrix within
the non-crossing approximation, the self-consistent Born approximation
(SCBA) for the Green's function of an isolated graphene layer reads 

\begin{equation}
\overline{\bm{G}_{\bm{k}}^{0,R}}(\omega)=\bm{G}_{\bm{k}}^{0,R}(\omega)+\bm{G}_{\bm{k}}^{0,R}(\omega)\cdot\bm{\Sigma}_{\text{imp}}^{R}(\omega)\cdot\overline{\bm{G}_{\bm{k}}^{0,R}}(\omega),
\end{equation}
where matrices the have indices in the sublattice space, a bar denotes
disorder averaging and $\bm{\Sigma}_{\text{imp}}^{R}(\omega)=n_{\text{imp}}\bm{T}^{R}(\omega)$
is the impurity self-energy, where $n_{\text{imp}}$ is the impurity
concentration (number of impurities by graphene unit cell) and $\bm{T}^{R}(\omega)$
is the T-matrix for a single $\delta$-like impurity with strength
$u$. For an impurity potential diagonal in the sublattice basis,
the T-matrix is also diagonal with equal components, given by 
\begin{equation}
T^{R}(\omega)=\frac{u}{1-u\overline{G_{1}^{R}}(\omega)}\stackrel[u\rra\infty]{}{=}-\frac{1}{\overline{G_{1}^{R}}(\omega)},\label{eq:impurity_self_energy}
\end{equation}
where we have taken the limit $u\rra\infty$ in order to describe
vacancies and defined 
\begin{equation}
\overline{G_{1}^{R}}(\omega)=\int\frac{d^{2}\bm{k}}{\left(2\pi\right)^{2}}\left[\overline{G_{\bm{k}}^{0,R}}(\omega)\right]_{\ph AA}^{A}
\end{equation}
In the Dirac cone approximation, graphene Green's function in the
sublattice basis and taking into account a finite electron life (induce
by the metallic contact) is given by 
\begin{align}
\left[G_{\bm{k}}^{0,R}(\omega)\right]_{\ph ab}^{a} & =\frac{1}{\omega-\lambda v_{F}\hbar\left|\bm{k}\right|+i\gamma_{\text{c}}-\Sigma^{R}(\omega)}\times\nonumber \\
 & \times\frac{1}{2}\left[\delta_{\ph ab}^{a}+\lambda\frac{\bm{k}}{\left|\bm{k}\right|}\cdot\bm{\sigma}_{\ph ab}^{a}\right],
\end{align}
with $a$, $b$ indices running over the A, B sublattice sites and
$\gamma_{\text{c}}$ is the lifetime induced by the metallic contacts,
$\gamma_{\text{c}}=\Gamma_{\text{bg}/\text{tg}}/2$ (assuming the
metallic contacts couple equally to all graphene states and do not
spoil translational invariance of graphene). For resonant impurities,
the self-energy is momentum independent. Writing it as $\Sigma_{\text{imp}}^{R}(\omega)=\Sigma_{\text{imp}}(\omega)-i\gamma_{\text{imp}}(\omega)$,
we can evaluate $G_{1}^{R}(\omega)$ analytically obtaining
\begin{eqnarray}
\overline{G_{1}^{R}}(\omega) & = & \frac{g_{1}\left(\omega-\Sigma_{\text{imp}}(\omega),\gamma_{\text{imp}}(\omega)+\gamma_{\text{c}}\right)}{4\pi\left(v_{F}\hbar\right)^{2}}\nonumber \\
 & - & i\frac{g_{2}\left(\omega-\Sigma_{\text{imp}}(\omega),\gamma_{\text{imp}}(\omega)+\gamma_{\text{c}}\right)}{4\pi\left(v_{F}\hbar\right)^{2}},\label{eq:G_scba}
\end{eqnarray}
where the functions $g_{1}$ and $g_{2}$ are given by
\begin{align}
g_{1}(\omega,\eta) & =-\frac{\omega}{2}\left[\log\left(\frac{\left(\Lambda_{E}-\omega\right)^{2}+\eta^{2}}{\omega^{2}+\eta^{2}}\right)+\left(\omega\rra-\omega\right)\right]\nonumber \\
+\eta & \left[\arctan\left(\frac{\Lambda_{E}-\omega}{\eta}\right)+\arctan\left(\frac{\omega}{\eta}\right)-\left(\omega\rra-\omega\right)\right]\label{eq:scba_g_1}\\
g_{2}(\omega,\eta) & =\frac{\eta}{2}\left[\log\left(\frac{\left(\Lambda_{E}-\omega\right)^{2}+\eta^{2}}{\omega^{2}+\eta^{2}}\right)+\left(\omega\rra-\omega\right)\right]\nonumber \\
+\omega & \left[\arctan\left(\frac{\Lambda_{E}-\omega}{\eta}\right)+\arctan\left(\frac{\omega}{\eta}\right)-\left(\omega\rra-\omega\right)\right]\label{eq:scba_g_2}
\end{align}
with $\Lambda_{E}\simeq v_{F}\hbar\left(4\pi/\left(\sqrt{3}a_{\text{g}}^{2}\right)\right)^{1/2}$
a high energy cutoff. In terms of $g_{1}$ and $g_{2}$ the self-energy
is given by
\begin{eqnarray}
\Sigma_{\text{imp}}(\omega) & = & -\lambda_{\text{imp}}\frac{g_{1}\left(\omega^{\prime},\gamma^{\prime}\right)}{g_{1}^{2}\left(\omega^{\prime},\gamma^{\prime}\right)+g_{2}^{2}\left(\omega^{\prime},\gamma^{\prime}\right)},\\
\gamma_{\text{imp}}(\omega) & = & \lambda_{\text{imp}}\frac{g_{2}\left(\omega^{\prime},\gamma^{\prime}\right)}{g_{1}^{2}\left(\omega^{\prime},\gamma^{\prime}\right)+g_{2}^{2}\left(\omega^{\prime},\gamma^{\prime}\right)}.\label{eq:gama_scba}
\end{eqnarray}
where we have defined $\omega^{\prime}=\omega-\Sigma_{\text{imp}}(\omega)$,
$\gamma^{\prime}=\gamma_{\text{c}}+\gamma_{\text{imp}}(\omega)$ and
$\lambda_{\text{imp}}=4\pi\left(v_{F}\hbar\right)^{2}n_{\text{imp}}$
is a constant characterizing the scattering by resonant disorder.
Eqs.~(\ref{eq:scba_g_1})-(\ref{eq:gama_scba}) form a set of equations
that can be easily solved. The solution for self-energy is shown in
Fig.~\ref{fig:Self_energy_SCBA}.

\section{\label{sec:Graphene-electron-phonon}Graphene electron self-energy
due to in-plane optical phonons}

\begin{figure}
\begin{centering}
\includegraphics[width=8cm]{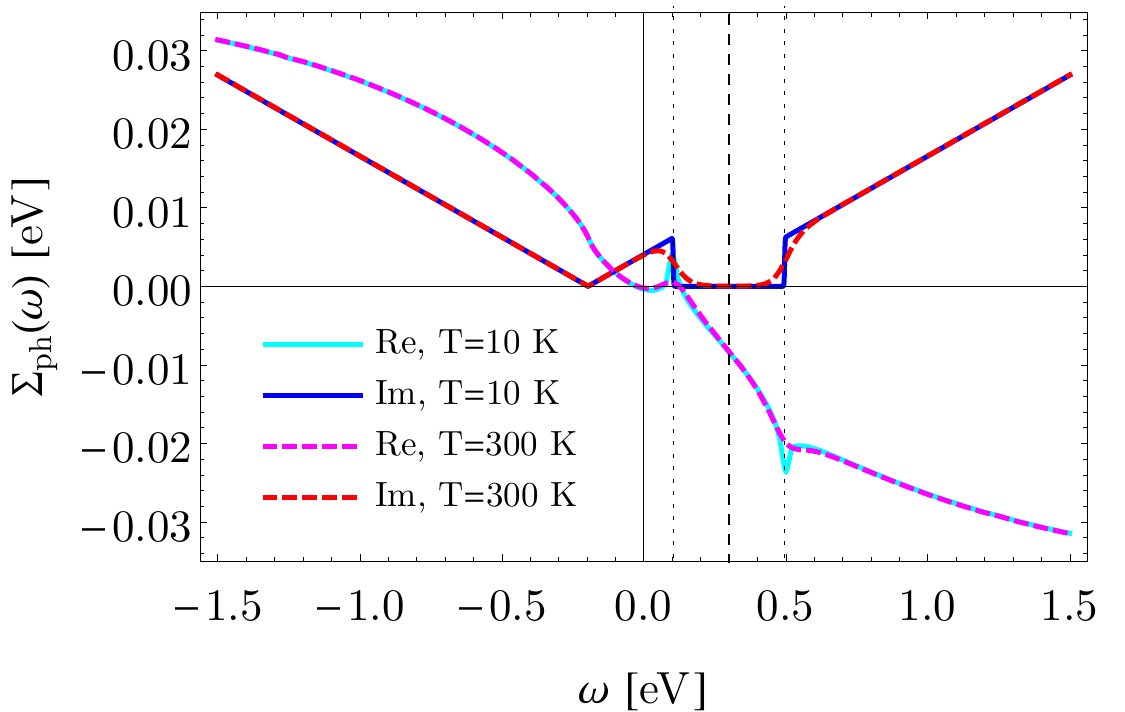}
\par\end{centering}

\protect\caption{\label{fig:Self_energy_phonon}Real and (minus) imaginary parts of
the self-energy for graphene electrons due to scattering by in-plane
optical phonons for two different temperatures for doped graphene
with $\epsilon_{\text{F}}=0.3$ eV. The zero of energy corresponds
to the Dirac point. The dashed vertical line marks $\omega=\epsilon_{\text{F}}$
and the dotted lines mark $\omega=\epsilon_{\text{F}}\pm\omega_{\Gamma\text{O}}^{\text{g}}$.}

\end{figure}

Electron-phonon interaction in graphene can be modeled by starting
from a nearest neighbour tight-binding Hamiltonian for the electrons
and assuming that the lattice distortions due to phonons lead to a
modulation of hopping integrals\cite{Suzuura2002}. For graphene longitudinal
and transverse in-plane phonons close to the $\Gamma$ point and electrons
close to the $\bm{K}$ point the obtained electron-phonon interaction
Hamiltonian is given by
\begin{equation}
H_{\text{g,\text{e-ph}}}=\frac{g_{\Gamma\text{O}}^{\text{g}}}{\sqrt{N}}\sum_{\substack{\bm{k},\bm{q}\\
\zeta=LO,TO
}
}\bm{c}_{\bm{k}+\bm{q},\text{g}}^{\dagger}\left(\vec{\sigma}\times\vec{\epsilon}_{\bm{q},\zeta}\right)\bm{c}_{\bm{k},\text{g}}\phi_{\bm{q},\zeta},
\end{equation}
wheres
\begin{equation}
g_{\Gamma\text{O}}^{\text{g}}=\frac{3}{2}\left(-\frac{d\log t}{d\log a_{\text{CC}}}\right)\frac{t}{a_{\text{CC}}}\sqrt{\frac{\hbar}{\mu_{\text{g}}\omega_{\Gamma\text{O}}^{\text{g}}}},
\end{equation}
is the electron-phonon coupling constant, with $-d\log t/d\log a_{\text{CC}}\simeq3$
describing the change in the nearest neighbour hopping, $t$, with
the distance, $a_{\text{CC}}$; $\mu_{\text{g}}=m_{\text{C}}/2$ is
the reduced mass of the phonon mode, with $m_{\text{C}}$ the carbon
atom mass; and $\omega_{\Gamma\text{O}}^{\text{g}}$ is the phonon
dispersion for the longitudinal/transverse in-plane optical phonon
mode (which are degenerate at $\Gamma$ and and assume we approximate
them as dispersionless). The polarization vectors for the longitudinal
and transverse mode can be written as $\vec{\epsilon}_{\bm{q},LO}=\left(1,0\right)$
and $\vec{\epsilon}_{\bm{q},TO}=\left(0,1\right)$. With these approximations
we obtain the momentum independent electron-phonon interaction matrices
\begin{eqnarray}
\bm{M}_{\text{LO}}^{\text{g}} & = & -g_{\Gamma\text{O}}^{\text{g}}\sigma_{y},\\
\bm{M}_{\text{TO}}^{\text{g}} & = & g_{\Gamma\text{O}}^{\text{g}}\sigma_{x}.
\end{eqnarray}
Assuming the graphene layer is in thermal equilibrium and to lowest
order in the electron-phonon interaction, the self energy is diagonal
in sublattice space and given by 
\begin{multline}
\Sigma_{\text{ph}}^{R}(\omega)=A_{\text{cell}}\left(g_{\Gamma\text{O}}^{\text{g}}\right)^{2}\times\\
\times\sum_{\lambda,s}\int\frac{d^{2}\bm{q}}{\left(2\pi\right)^{2}}s\frac{1+b(s\omega_{\Gamma\text{O}}^{\text{g}})-f\left(\epsilon_{\bm{q},\lambda}-\epsilon_{\text{F}}\right)}{\omega-\epsilon_{\bm{q},\lambda}-s\omega_{\Gamma\text{O}}^{\text{g}}+i0^{+}}.
\end{multline}
The imaginary part can be computed for pristine graphene at finite
temperature as
\begin{multline}
-\im\Sigma_{\text{ph}}^{R}(\omega)=\\
=\left(g_{\Gamma\text{O}}^{\text{g}}\right)^{2}\left[1+b(\omega_{\Ga\text{O}}^{\text{g}})-f\left(\omega-\omega_{\Gamma\text{O}}^{\text{g}}-\epsilon_{\text{F}}\right)\right]\times\\
\times\frac{A_{\text{cell}}\left|\omega-\omega_{\Gamma\text{O}}^{\text{g}}\right|}{2\left(v_{F}\hbar\right)^{2}}+\\
+\left(g_{\Gamma\text{O}}^{\text{g}}\right)^{2}\left[b(\omega_{\Ga\text{O}}^{\text{g}})-f\left(\omega+\omega_{\Gamma\text{O}}^{\text{g}}-\epsilon_{\text{F}}\right)\right]\times\\
\times\frac{A_{\text{cell}}\left|\omega+\omega_{\Gamma\text{O}}^{\text{g}}\right|}{2\left(v_{F}\hbar\right)^{2}},
\end{multline}
where $\omega$ and the Fermi energy, $\epsilon_{\text{F}}$, are
both measured from the Dirac cone. From this, the real part can be
efficiently obtained using the Kramers-Kronig relation
\begin{equation}
\re\Sigma_{\text{ph}}^{R}(\omega)=-\int\frac{d\nu}{\pi}\frac{\im\Sigma_{\text{ph}}^{R}(\omega-\nu)-\im\Sigma_{\text{ph}}^{R}(\omega+\nu)}{\nu}.
\end{equation}
The computed self-energy is shown in Fig.~\ref{fig:Self_energy_phonon}.

\section{\label{sec:Vertex-corrections}Vertex corrections for resonant impurities}

\begin{figure}
\begin{centering}
\includegraphics[width=8cm]{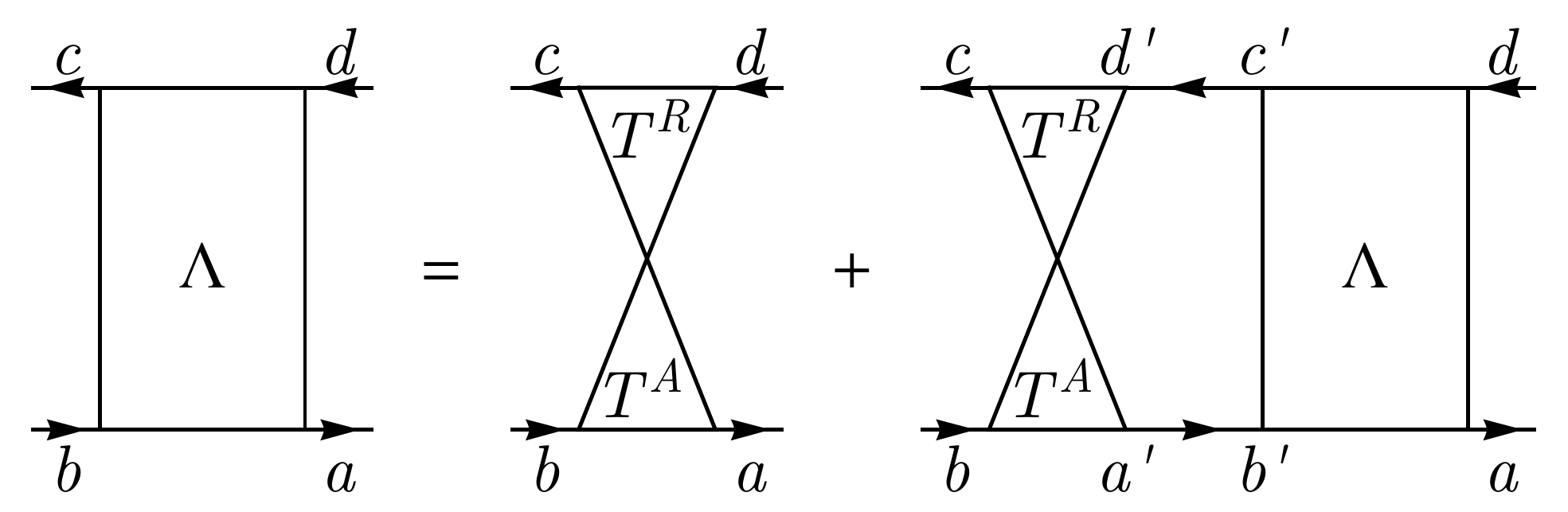}
\par\end{centering}

\protect\caption{\label{fig:BS_eq_scba}Diagrammatic representation of the Bethe-Salpeter
equation, Eq.~(\ref{eq:Bethe-Salpeter_eq})}
\end{figure}

In this Appendix, we provide an alternative derivation of Eq.~(\ref{eq:Current_lowestorder}),
for the vertical current in a graphene-hBN-graphene device taking
into account disorder in the graphene layers, employing approach (B).
Instead of describing disorder as an interaction, we will start from
Eq.~(\ref{eq:transmission_B}) and perform disorder averages of it.
Just as in Appendix.~\ref{sec:Ressonant-impurties-SCBA} we will
consider scattering by resonant disorder. This model will both serve
as a concrete example for elastic scattering of the general results
present in Sec.~\ref{sec:Incoherent-tunnelling} regarding the equivalences
of approaches (A) and (B) and will also show the formal equivalence
between the contributions to the current arising from Eq.~(\ref{eq:Incoherent_current})
and vertex corrections. Just as in Sec.~\ref{sec:Coherent-tunnelling}
we will assume for simplicity that the external metallic contacts
couple to all graphene states and that graphene electronic states
are still well describe by Bloch states. With these approximations,
we write 
\begin{equation}
\bm{\Gamma}_{\text{b}/\text{t}}=\Gamma_{\text{b}/\text{t}}\bm{I}_{\text{bg}/\text{tg}}.
\end{equation}
Performing an averaging of Eq.~(\ref{eq:transmission_B}) with respect
to disorder in the bottom and top graphene layers, assuming that these
are uncorrelated, and to lowest order in the graphene-hBN coupling
we obtain
\begin{eqnarray}
\overline{\mathcal{T}} & = & \Gamma_{\text{b}}\Gamma_{\text{t}}\Tr\left[\overline{\bm{G}_{\text{bg}}^{0,A}\cdot\bm{I}_{\text{bg}}\cdot\bm{G}_{\text{bg}}^{0,R}}\cdot\bm{\mathcal{T}}_{\text{bg},\text{tg}}\cdot\right.\label{eq:Transmission_average}\\
 &  & \left.\cdot\overline{\bm{G}_{\text{tg}}^{0,R}\cdot\bm{I}_{\text{tg}}\cdot\bm{G}_{\text{tg}}^{0,A}}\cdot\mathcal{\bm{\mathcal{T}}}_{\text{tg},\text{bg}}\right].
\end{eqnarray}
The disorder averaged product of Green's functions is not just the
product of average Green's function, as the averaging procedure establishes
correlations between the two functions. From now on, we will employing
a notation where an upper indice represents an out-going electronic
state and a lower indice represents an incoming state, with repeated
indices being summed over. With this convention, the average of the
product of two Green's functions, in sublattice space, can be written
as (suppressing the frequency argument and the bg/tg indice)
\begin{multline}
\overline{\left[G_{\bm{k}}^{0,A}\right]_{\ph ab}^{a}\delta_{\ph cc}^{b}\left[G_{\bm{k}}^{R,0}\right]_{\ph cd}^{c}}=\left[\overline{G_{\bm{k}}^{0,A}}\right]_{\ph ab}^{a}\delta_{\ph cc}^{b}\left[\overline{G_{\bm{k}}^{R,0}}\right]_{\ph cd}^{c}+\\
+\left[\overline{G_{\bm{k}}^{0,A}}\right]_{\ph aa^{\prime}}^{a}\Lambda_{\ph{a^{\prime}}b^{\prime}\ph{c^{\prime}}d^{\prime}}^{a^{\prime}\ph{b^{\prime}}c^{\prime}\ph d^{\prime}}\left[G_{2}^{AR}\right]_{\ph{b^{\prime}}b\ph cc^{\prime}}^{b^{\prime}\ph bc}\delta_{\ph cc}^{b}\left[\overline{G_{\bm{k}}^{0,R}}\right]_{\ph{d^{\prime}}d}^{d^{\prime}},\label{eq:Impurity_average_2_GreenFunc}
\end{multline}
where the second term are vertex corrections, we have define the quantity
\begin{equation}
\left[G_{2}^{AR}\right]_{\ph ab\ph cd}^{a\ph bc}=\int\frac{d^{2}\bm{p}}{\left(2\pi\right)^{2}}\left[\overline{G_{\bm{p}}^{A}}\right]_{\ph ab}^{a}\left[\overline{G_{\bm{p}}^{R}}\right]_{\ph cd}^{c},
\end{equation}
and $\Lambda_{\ph ab\ph cd}^{a\ph bc\ph d}$ is a 4-point function,
which obeys a Bethe-Salpeter equation (see Fig.~\ref{fig:BS_eq_scba})
\begin{equation}
\Lambda_{\ph ab\ph cd}^{a\ph bc\ph d}=U_{\ph ab\ph cd}^{a\ph bc\ph d}+U_{\hphantom{a^{\prime}}b\hphantom{c}d^{\prime}}^{a^{\prime}\hphantom{b}c}\left[G_{2}^{AR}\right]_{\hphantom{b^{\prime}}a^{\prime}\hphantom{d^{\prime}}c^{\prime}}^{b^{\prime}\hphantom{a^{\prime}}d^{\prime}}\Lambda_{\ph ab^{\prime}\ph{c^{\prime}}d}^{a\ph{b^{\prime}}c^{\prime}\ph d},\label{eq:Bethe-Salpeter_eq}
\end{equation}
where $U_{\ph ab\ph cd}^{a\ph bc\ph d}$ is an irreducible 4-point
function, which within the T-matrix and non-crossing approximation
for resonant impurities is given by
\begin{equation}
U_{\ph ab\ph cd}^{a\ph bc\ph d}=n_{\text{imp}}\left|T_{\text{imp}}^{R}\right|^{2}\delta_{\ph ab}^{a}\delta_{\ph cd}^{c}.
\end{equation}
The quantity $\left[G_{2}^{AR}\right]_{\ph ab\ph cd}^{a\ph bc}$ can
be evaluated analytically yielding
\begin{equation}
\left[G_{2}^{AR}\right]_{\hphantom{a}b\hphantom{c}d}^{a\hphantom{b}c}=L_{1}\left(\omega^{\prime},\gamma^{\prime}\right)\delta_{\ph ab}^{a}\delta_{\ph cd}^{c}+L_{\text{2}}\left(\omega^{\prime},\gamma^{\prime}\right)\frac{1}{2}\bm{\sigma}_{\ph ab}^{a}\cdot\bm{\sigma}_{\ph cd}^{c},
\end{equation}
where
\begin{eqnarray}
L_{1}\left(\omega,\eta\right) & = & \frac{1}{8\pi\left(v_{F}\hbar\right)^{2}}\left(\frac{1}{\eta}g_{2}(\omega,\eta)+\frac{1}{\omega}g_{1}(\omega,\eta)\right),\label{eq:BS_L_1}\\
L_{2}\left(\omega,\eta\right) & = & \frac{1}{8\pi\left(v_{F}\hbar\right)^{2}}\left(\frac{1}{\eta}g_{2}(\omega,\eta)-\frac{1}{\omega}g_{1}(\omega,\eta)\right),\label{eq:BS_L_2}
\end{eqnarray}
with the functions $g_{1}$and $g_{2}$ defined by Eqs.~(\ref{eq:scba_g_1}),
(\ref{eq:scba_g_2}) and where we have written $\omega^{\prime}=\omega-\Sigma_{\text{imp}}$
and $\gamma^{\prime}=\gamma_{\text{imp}}+\gamma_{\text{c}}$ as in
Appendix.~\ref{sec:Ressonant-impurties-SCBA}. The Bethe-Salpeter
equation for $\Lambda_{\ph ab\ph cd}^{a\ph bc\ph d}$ is now a simple
problem of linear algebra. Solving Eq.~(\ref{eq:Bethe-Salpeter_eq}),
yields the non-zero components of $\Lambda_{\ph ab\ph cd}^{a\ph bc\ph d}$
in the sublattice basis\begin{widetext}
\begin{eqnarray}
\Lambda_{\hphantom{A}A\hphantom{A}A}^{A\hphantom{A}A\hphantom{A}}=\Lambda_{\hphantom{B}B\hphantom{B}B}^{B\hphantom{B}B\hphantom{B}} & = & \frac{n_{\text{imp}}\left|T^{R}\right|^{2}\left(1-L_{1}n_{\text{imp}}\left|T^{R}\right|^{2}\right)}{\left[1-\left(L_{1}-L_{2}\right)n_{\text{imp}}\left|T^{R}\right|^{2}\right]\left[1-\left(L_{1}+L_{2}\right)n_{\text{imp}}\left|T^{R}\right|^{2}\right]},\\
\Lambda_{\hphantom{A}A\hphantom{B}B}^{A\hphantom{A}B\hphantom{B}}=\Lambda_{\hphantom{B}B\hphantom{A}A}^{B\hphantom{B}A\hphantom{A}} & = & \frac{n_{\text{imp}}\left|T^{R}\right|^{2}}{1-L_{1}n_{\text{imp}}\left|T^{R}\right|^{2}},\\
\Lambda_{\hphantom{A}B\hphantom{B}A}^{A\hphantom{B}B\hphantom{B}}=\Lambda_{\hphantom{A}A\hphantom{A}B}^{B\hphantom{B}A\hphantom{B}} & = & \frac{L_{2}n_{\text{imp}}^{2}\left|T^{R}\right|^{4}}{\left[1-\left(L_{1}-L_{2}\right)n_{\text{imp}}\left|T^{R}\right|^{2}\right]\left[1-\left(L_{1}+L_{2}\right)n_{\text{imp}}\left|T^{R}\right|^{2}\right]}.
\end{eqnarray}
\end{widetext}, where we have omitted the frequency arguments of
$L_{1/2}$. Using the fact that $\left[G_{2}^{AR}\right]_{\ph{b^{\prime}}b\ph cc^{\prime}}^{b^{\prime}\ph bb}=\left(L_{1}(\omega)+L_{2}(\omega)\right)\delta_{\ph{b^{\prime}}c\prime}^{b^{\prime}}$,
the vertex correction contribution in Eq.~(\ref{eq:Impurity_average_2_GreenFunc})
can be written as
\begin{multline}
\Lambda_{\ph{a^{\prime}}b^{\prime}\ph{c^{\prime}}d^{\prime}}^{a^{\prime}\ph{b^{\prime}}c^{\prime}\ph d^{\prime}}\left[G_{2}^{AR}\right]_{\ph{b^{\prime}}b\ph cc^{\prime}}^{b^{\prime}\ph bc}=\\
=\frac{n_{\text{imp}}\left|T^{R}\right|^{2}\left(L_{1}+L_{2}\right)}{1-\left(L_{1}+L_{2}\right)n_{\text{imp}}\left|T^{R}\right|^{2}}\delta_{b}^{c}
\end{multline}
Expressing $T^{R}$ and $L_{1/2}$ in terms of $g_{1}$ and $g_{2}$,
and using Eqs.~(\ref{eq:gama_scba}) it can be seen that the quantity
$\left(L_{1}+L_{2}\right)n_{\text{imp}}\left|T^{R}\right|^{2}$ can
be written as the ratio
\begin{equation}
\left(L_{1}+L_{2}\right)n_{\text{imp}}\left|T^{R}\right|^{2}=\frac{\gamma_{\text{imp}}}{\gamma_{\text{imp}}+\gamma_{\text{c}}}.
\end{equation}
Therefore, Eq.~(\ref{eq:Impurity_average_2_GreenFunc}) can be written
as 
\begin{align}
\overline{\left[G_{\bm{k}}^{0,A}\right]_{\ph ab}^{a}\left[G_{\bm{k}}^{R,0}\right]_{\ph cd}^{b}} & =\left[\overline{G_{\bm{k}}^{0,A}}\right]_{\ph ab}^{a}\left[\overline{G_{\bm{k}}^{R,0}}\right]_{\ph cd}^{b}\nonumber \\
 & +\frac{\gamma_{\text{imp}}}{\gamma_{\text{c}}}\left[\overline{G_{\bm{k}}^{0,A}}\right]_{\ph ab}^{a}\left[\overline{G_{\bm{k}}^{R,0}}\right]_{\ph cd}^{b},
\end{align}
Therefore, the product of a retarded and an advanced Green function
is related to the spectral function as 
\begin{equation}
\left[\overline{G_{\bm{k}}^{0,A}}(\omega)\right]_{\ph ab}^{a}\left[\overline{G_{\bm{k}}^{R,0}}(\omega)\right]_{\ph cd}^{b}=\frac{1}{\gamma_{\text{imp}}+\gamma_{\text{c}}}\left[\overline{A_{\bm{k}}^{0}}(\omega)\right]_{\ph ab}^{a},
\end{equation}
and therefore, the contributions from vertex corrections (incoherent
contributions) due to impurities adds to the contribution coming from
the product of two average Green's functions (coherent contribution),
in such a way that Eq.~(\ref{eq:Transmission_average}) reduces to
Eq.~(\ref{eq:transmission_A}) of the main text. This result is a
particular case of the more general discussion of Sec.~\ref{sub:Scattering-graphene},
which is not limited to elastic scattering.


%

\end{document}